\documentclass[fleqn,usenatbib,usedcolumn]{mnras}
\usepackage[british]{babel}
\usepackage{graphicx}
\usepackage{color,soul}
\usepackage{mathtools}
\usepackage{newtxtext}
\usepackage[slantedGreek]{newtxmath}
\usepackage[T1]{fontenc}
\usepackage{bm}
\usepackage{amsmath}
\usepackage{tablefootnote}
\usepackage{scrextend}
\usepackage{hyperref}
\usepackage{booktabs}
\usepackage[inline]{enumitem}
\usepackage{relsize}
\usepackage{todonotes}
\usepackage{cuted}

\addto\extrasbritish{%
}

\let\mr\mathrm
\graphicspath{{Figures/}}

\newcommand{\sTheta}{\bm{\upTheta}} %set Theta
\newcommand{\sPhi}{\bm{\upPhi}}     %set Phi
\newcommand{\sSigma}{\bm{\upSigma}} %set Sigma
\begin{document}
\title[High fidelity Bayesian calibration I]{A Bayesian approach to high fidelity interferometric calibration I: mathematical formalism}
\author[Sims et al.]{Peter H. Sims,$^{1,2}$\thanks{E-mail: peter.sims@mail.mcgill.ca} Jonathan C. Pober,$^3$ and Jonathan L. Sievers$^{1,2}$ \\
$^1$McGill Space Institute, McGill University, 3550 University Street, Montreal, QC H3A 2A7, Canada \\
$^2$Department of Physics, McGill University, 3600 University Street, Montreal, QC H3A 2T8, Canada \\
$^3$Department of Physics, Brown University, Providence, RI 02912, USA \\ 
}

\maketitle
\label{firstpage}
\begin{abstract}
High fidelity radio interferometric data calibration that minimises spurious spectral structure in the calibrated data is essential in astrophysical applications, such as 21 cm cosmology, which rely on knowledge of the relative spectral smoothness of distinct astrophysical emission components to extract the signal of interest. Existing approaches to radio interferometric calibration have been shown to impart spurious spectral structure to the calibrated data if the sky model used to calibrate the data is incomplete. In this paper, we introduce \textsc{BayesCal}: a novel solution to the sky-model incompleteness problem in interferometric calibration, designed to enable high fidelity data calibration. The \textsc{BayesCal} data model supplements the a priori known component of the forward model of the sky with a statistical model for the missing and uncertain flux contribution to the data, constrained by a prior on the power in the model. We demonstrate how the parameters of this model can be marginalised out analytically, reducing the dimensionality of the parameter space to be sampled from and allowing one to sample directly from the posterior probability distribution of the calibration parameters. Additionally, we show how physically motivated priors derived from theoretical and measurement-based constraints on the spectral smoothness of the instrumental gains can be used to constrain the calibration solutions. In a companion paper, we apply this algorithm to simulated observations with a HERA-like array and demonstrate that it enables up to four orders of magnitude suppression of power in spurious spectral fluctuations relative to standard calibration approaches.
\end{abstract}


%%%%%%%%%%%%%%%%%%%%%%%%%%%%%%%%%%%%%%%%%%%%%%%%%%%%%%%%%
\section{Introduction}
\label{Sec:Introduction}
%%%%%%%%%%%%%%%%%%%%%%%%%%%%%%%%%%%%%%%%%%%%%%%%%%%%%%%%%

Observations of the 21 cm hyperfine line radiation emitted by neutral hydrogen in the high redshift Universe have the potential to provide precision constraints on cosmological parameters during the Cosmic Dark Ages (e.g. \citealt{2004PhRvL..92u1301L, 2008PhRvD..78b3529M}), to probe directly the initial stages of structure formation and characterise the properties of the first stars, proto-galaxies and accreting black holes during Cosmic Dawn (CD) and the Epoch of Reionization (EoR) (e.g. \citealt{2012MNRAS.424..762D, 2013MNRAS.431..621M, 2014MNRAS.439.3262M, 2015MNRAS.449.4246G}), and, at lower redshifts, to constrain the dark energy equation of state (e.g. \citealt{2014SPIE.9145E..4VN, 2020PASP..132f2001L}). However, to achieve this, the cosmological signals must be extracted from data containing astrophysical foreground emission that, depending on the frequency range, angular scale and field observed, is 3--6 orders of magnitude brighter.

Numerous approaches to separating the cosmological signal from the foregrounds have been discussed in the literature, but a unifying theme amongst them is the use of knowledge of the intrinsic spectral structure of the constituent astrophysical signals. This knowledge is employed either \begin{enumerate*} \item implicitly, when using general signal separation techniques that discriminate between the components through the assumed spectral smoothness of the foreground emission relative to the cosmological signal, which is expected to fluctuate more rapidly due to inhomogeneities in the ionization, temperature and density of hydrogen in the IGM along the line of sight (e.g. \citealt{2012ApJ...756..165P, 2012MNRAS.423.2518C, 2013MNRAS.429..165C, 2016ApJ...818..139T, 2016MNRAS.462.3069S, 2018MNRAS.478.3640M, 2019MNRAS.484.4152S}), or \item explicitly, when jointly fitting models for the foregrounds and the cosmological signal to the calibrated data (e.g. \citealt{2019MNRAS.488.2904S}).\end{enumerate*}

The observable of interest in radio interferometry is the spatial autocorrelation function of the electric field due to the sky brightness distribution, evaluated at the baseline separations of the antennas in the array. The physical observables are the cross-correlations of the voltages at the correlator induced by the electric field at the antennas. These two quantities would be equal if the voltages at the correlator were equal to the electric field evaluated at the antenna locations; however, the antenna signal chain scales the amplitude and shifts the phase of the voltage relative to the incident electric field. Thus, before interferometric data can be analysed and separation of the cosmological signal from the foregrounds attempted, the effect of the complex instrumental transfer function on the correlated antenna voltages must be modelled and removed via data calibration. 

Considering the vector electric field, $\bm{E}$, due to the far-field sky brightness distribution above the instrument and assuming all transformations along the signal chain between the sources of emission producing the observed brightness distribution and the antenna, and between the antenna and the instrument correlator\footnote{For generality, we refer to the signal propagation effects between the antenna and the correlator; however, signal propagation effects can be restricted to the pre-digitisation subset of the signal chain by digitisation of the voltage stream prior to transmission to the correlator.}, are linear, the frequency-, time- and polarisation-dependent vectors describing the  incident electric field and uncalibrated induced voltage at the correlator can be related via, 
\begin{equation}
\label{Eq:AntennaVoltage}
\bm{v} =  \int\limits_{4\pi} \mathbfss{J}\bm{E} ~\mathrm{d}\Omega \ . 
\end{equation}
Here $\mathbfss{J}$ is the Jones matrix of the antenna and its elements are complex scale factors, known as antenna gains, that describe the (position-dependent) amplification and phase imparted to the incident vector electric field by the signal chain,  $\mathrm{d}\Omega$ is the differential solid angle, and the integral is over the sky\footnote{Here we treat the integral as the limit of the contributions to the measured voltage from a large number of individual point sources.}. During calibration, the complex gains of per-antenna Jones matrix models, $\mathbfss{J}^\mathrm{m}$, are fit for. The inverses of the model Jones matrices are then used to calibrate the data.

The sky brightness distribution determining the incident electric field at the antenna, $\bm{E}$, and correspondingly, the measured interferometric data formed from the cross-correlation of the voltages of pairs of antennas in the array can be divided into two categories: \begin{enumerate*} \item an a priori known component comprised of catalogued point sources and measured diffuse emission (each with associated uncertainties) and \item an a priori unknown component associated with sources below the completeness level of catalogues in the field of interest and with the uncertainties associated with measurements of low frequency diffuse radio emission and, to a lesser extent, with the locations and flux-densities of catalogued point sources.\end{enumerate*}

Existing approaches to determining the antenna dependent complex gains during interferometric calibration rely on either \begin{enumerate*}\item fitting a forward model of the sky to the data, or \item the presence of redundancy in the baseline layout of the array to solve for the majority of the calibration degrees of freedom, followed by fitting a forward model of the sky to the data to solve for the remaining calibration terms \end{enumerate*}. However, both approaches produce biased results if the sky model used to calibrate the data is incomplete (e.g. \citealt{2016MNRAS.461.3135B} and \citealt{2017MNRAS.470.1849E}, and \citealt{2019ApJ...875...70B}, respectively). Ultimately, this occurs because, when fitting the data with the product of gain parameters and a fixed and incomplete forward model of the sky defined exclusively by the a priori known component of the sky brightness distribution, the gain parameters provide the only degrees of freedom for absorbing the effect of the sky-model incompleteness. As a result, the fitted model Jones matrices do not match the true Jones matrices of the antennas in the array ($\mathbfss{J}^\mathrm{m} \ne \mathbfss{J}$). Furthermore, due to the inherent chromaticity of the interferometer's sky response, even if the component of sky emission in the data that is missing in the incomplete forward model of the sky is spectrally smooth, that will not be true of the difference between the true visibilities that would be obtained if $\mathbfss{J}$ were the identity matrix and the incomplete calibration model visibilites (e.g. \citealt{2017MNRAS.470.1849E}). The spectral structure in this difference translates to spectral fluctuations in the gain solutions and these spectral fluctuations impart spurious spectral structure into the calibrated data. 

For a given level of calibration model incompleteness, approaches to mitigating spectral fluctuations in the frequency-dependent biases of the gain solutions discussed in the literature have focused on \begin{enumerate*} \item downweighting longer baselines, which are  more chromatic, during calibration (e.g. \citealt{2017MNRAS.470.1849E}) \item penalising deviations of the spectral structure\footnote{In the context of direction-dependent calibration, penalising deviations of the spatial structure of the model gains from spherical harmonics has also recently been explored (\citealt{2022MNRAS.510.2718Y}).} of the model gains from a low-order polynomial description (e.g. \citealt{2015MNRAS.449.4506Y}), \item modelling the spectral structure in the gains with low-order polynomials (e.g. \citealt{2016MNRAS.461.3135B}), or \item filtering of spectral structure in the gain solutions below a threshold spectral scale once they have been fit (e.g. \citealt{2020ApJ...890..122K}). \end{enumerate*}

These approaches have been shown to suppress power in calibration systematics but have some limitations. For example, the increasing contribution to the visibility data from the more uncertain diffuse sky brightness distribution on shorter baselines, which makes the calibration sky model on these baselines lengths less complete. Additionally, in the context of the latter three approaches, constraining the gain solutions, while valuable when one has a priori knowledge regarding the level of spectral structure in the instrumental gains, will redistribute bias rather than eliminating it if the underlying cause of sky-model incompleteness bias in the gain solutions is not addressed in parallel. This is undesirable if one aims to recover the 21 cm signal on the full range of spectral scales accessible in the data. More generally, for any choice of spectral window in which one wishes to estimate the signal, using this form of gain prior creates an additional avenue for introducing spurious spectral structure into the gain solutions, and correspondingly into the calibrated data, if the gains are overly constrained and are unable to fit real instrumental structure. In this context, if the real instrument deviates from the prior, applying a hard (zero-uncertainty) prior on the amplitude of spectral fluctuations in the gains on given spectral scales poses a risk for two reasons. Firstly, in terms of bias since in this case the prior dominates the likelihood, irregardless of how much better a fit of the calibration model with deviations from the prior would fit the data. Secondly, in terms of error propagation between calibration parameters and calibrated visibilities, since the prior implies zero uncertainty on the calibration solutions in the context of covariant propagation of uncertainties on the calibration parameters through to the calibrated data. 

In this paper, we introduce a new Bayesian calibration formalism (\textsc{BayesCal}), designed to minimise spurious spectral structure imparted to the calibrated data via these two avenues, while making optimal use of our a priori knowledge of the expected level of spectral structure in the gain amplitude solutions. \textsc{BayesCal} addresses the sky-model incompleteness problem directly by supplementing the a priori known component of the calibration forward model of the sky with a statistical model for the a priori unknown component of the sky brightness distribution. In principle, when calibrating the data with this model, one could jointly fit for the calibration solutions and the most probable sky-based parameters of this statistical model. However, in general, this means a significant increase in the dimensonality of the parameter space to be sampled from. We demonstrate how this limitation can be overcome by analytically marginalising over the fitted sky-model parameters of our statistical model, allowing us to sample directly from the posterior probability distribution of the calibration parameters. Next, we describe how theoretical and measurement-based constraints (derived from, for example, electric and electromagnetic co-simulation and reflectometry measurements of the receiver system) can be used to construct physically motivated priors on the spectral smoothness of the instrumental gains and how, by reparametrising the instrumental gains in terms of Fourier modes, these priors are seamlessly integrated into the calibration framework.

As our primary case study, we derive the posterior probability distribution for the direction independent, absolute calibration parameters of a redundantly calibrated array in the \textsc{BayesCal} calibration framework. In a companion paper (hereafter, paper II) we demonstrate, on simulated observations, that sampling from this \textsc{BayesCal} posterior enables recovery of significantly higher fidelity calibration solutions relative to calibration using an incomplete sky model in the construction of the calibration model, yielding up to four orders of magnitude additional suppression of power in spurious spectral fluctuations in the calibration solutions. Depending on the level of completeness of the a priori known component of the calibration model in the field considered, we show in paper II, in the context of 21 cm cosmology, that the overall fidelity of these calibration solutions is sufficiently high for foreground systematics imparted by imperfections in the calibration solutions to be at least a factor of 5 below a fiducial 10 mK RMS EoR signal on large spectral scales ($\sim 9~\mathrm{MHz}$, which corresponds to $k_{\parallel} \simeq 0.06~h\mathrm{Mpc}^{-1}$ in the $160 \le \nu \le 169~\mathrm{MHz}$ spectral band considered there) and more than an order of magnitude below the fiducial signal on intermediate and small spectral scales ($ \le 4.5~\mathrm{MHz}$, which corresponds to $k_{\parallel} \gtrsim 0.11~h\mathrm{Mpc}^{-1}$).

Recently, calibration approaches have been developed that relax the assumption of explicit redundancy by calculating the expected covariance of baselines (\citealt{2017arXiv170101860S, 2021MNRAS.503.2457B}). This enables one to jointly constrain the gain solutions using the partial redundancy in the baseline layout of the array and via fitting a forward model of the sky to the data. Additionally, \citet{2017arXiv170101860S} and \citet{2021MNRAS.503.2457B} consider techniques that allow one to take advantage of partial knowledge of the sky, such as point sources with known positions but unknown fluxes. In \citet{2021MNRAS.503.2457B} it is shown that these approaches improve calibration in the presence of random model error, as well as reducing calibration error from missing faint sources in the sky model. In \citet{2017arXiv170101860S}, application of these techniques yields a factor of five reduction in gain phase errors relative to an equivalent application of partially-redundant calibration that neglects point source information and comparable errors in the spectral structure of redundant calibration gain amplitudes.

Our focus with \textsc{BayesCal}, on developing a new data model to address the sky-model incompleteness problem and incorporating physically derived priors on the spectral smoothness of the instrumental gains, is complementary to that of these approaches for making optimal use of data constraints deriving from a combination of partial baseline redundancy and knowledge of the sky. We discuss in \autoref{Sec:BayesCalForSkyBasedCalibration} the generalisation to non-redundant interferometric arrays of our primary case study, deriving the posterior probability distribution for the absolute calibration parameters of a redundantly calibrated array in the \textsc{BayesCal} calibration framework. This provides a route to combining the benefits of these approaches, and a more thorough exploration of this topic is an interesting direction for future work.

The remainder of the paper is organised as follows. In \autoref{Sec:InterferometricCalibration}, we give an overview of current approaches to interferometric calibration. In \autoref{Sec:BayesCal}, we cast interferometric calibration as a Bayesian parameter estimation problem and present the \textsc{BayesCal} data model and calibration framework. We summarise in \autoref{Sec:SummaryAndConclusions}. For readers' convenience, Table \ref{Tab:VariablesList1} lists the variables used in this paper.

%%%%%%%%%%%%%%%%%%%%%%%%%%%%%%%%%%%%%%%%%%%%%%%%%%%%%%%%%
\section{Interferometric Calibration}
\label{Sec:InterferometricCalibration}
%%%%%%%%%%%%%%%%%%%%%%%%%%%%%%%%%%%%%%%%%%%%%%%%%%%%%%%%%

In order to introduce the notation used and to provide context when deriving the \textsc{BayesCal} calibration model in \autoref{Sec:BayesCal}, in this section, we start by outlining the general interferometric calibration problem (\autoref{Sec:CalibrationFormalism}), before discussing its specialisation to direction-independent sky-based interferometric calibration (\autoref{Sec:SkyBasedCalibration}) and describing how this approach can be supplemented by constraints on the data deriving from array redundancy (\autoref{Sec:RedundantCalibration}).

\subsection{Calibration formalism}
\label{Sec:CalibrationFormalism}

For a generic radio interferometer, the Measurement Equation (\citealt{1996A&AS..117..137H, 2011A&A...527A.106S}; hereafter, HBS96 and Sm11, respectively), for a pair of antennas $p,q$ observing a continuous brightness distribution, allows one to construct a `polarised visibility vector', $\bm{V}_{pq}$, as,
\begin{equation}
\label{Eq:MEq}
\bm{V}_{pq} = \bm{n}_{pq} + \iint\limits_{lm}\frac{\mr{d}l\mr{d}m}{n}\,  \mathbfss{J}_{pq}(\bm{\hat{l}})\bm{C}(\bm{\hat{l}})  \ . 
\end{equation}
Here, the coherency vector, $\bm{C}$, and $\bm{V}_{pq}$ are $4\times 1$ column vectors encoding the four correlations of the two components of the electric field incident on the antennas, measured in a given polarisation basis, and the four correlations of the two voltage signals per antenna associated with those electric field components, respectively; $l,m$ and $n = \sqrt{1-l^2-m^2}$ are the direction cosines of the unit vector, $\bm{\hat{l}}$, pointing from the antenna to the source; $\mathbfss{n}_{pq}$ is a $4\times 1$ column vector describing the noise on the data. $\mathbfss{J}_{pq}$ is a $4 \times 4$ matrix describing signal propagation effects on the voltage correlations measured by the interferometer and can be written as,
\begin{equation}
\label{Eq:Jpq}
\mathbfss{J}_{pq} = \mathbfss{J}_{p}  \otimes \mathbfss{J}_{q}^{*} \ ,
\end{equation}
where $\mathbfss{J}_{p}$ and $\mathbfss{J}_{q}$ are $2\times 2$ Jones matrices that describe the cumulative product of all propagation effects along the signal paths of antennas $p$ and $q$ to the correlator and $\otimes$ and $(.)^{*}$ denote the Kronecker product and complex conjugation, respectively. Each of the terms in \autoref{Eq:MEq} is a function of time, frequency and polarisation; here, for brevity, we leave these dependencies implicit, but we return to the explicit polarisation structure of $\bm{V}_{pq}$ shortly. 

For concreteness, writing the vector electric field of the sky brightness distribution visible to the antennas in an equatorial basis, $\bm{E} = [E_{\delta}(\bm{\hat{l}}), E_{\alpha}(\bm{\hat{l}})]^{T}$, and the declination (Dec) and right ascension (RA) equatorial basis vectors $\hat{\bm{e}}_{\delta}$ and $\hat{\bm{e}}_{\alpha}$, respectively, as $2\times 1$ column vectors, the coherency vector can be expressed in terms of Stokes parameters, $I$, $Q$, $U$, and $V$, as (e.g. HBS96, \citealt{2018ApJ...869...79M, 2019ApJ...882...58K}), 
\begin{align}
\label{Eq:CoherencyVector}
\bm{C} &= \langle  E_{\delta} E_{\delta}^{*} \rangle \hat{\bm{e}}_{\delta} \otimes \hat{\bm{e}}_{\delta} + \langle E_{\delta} E_{\alpha}^{*} \rangle \hat{\bm{e}}_{\delta} \otimes \hat{\bm{e}}_{\alpha} \\ \nonumber
&+ \langle E_{\alpha} E_{\delta}^{*} \rangle \hat{\bm{e}}_{\alpha} \otimes \hat{\bm{e}}_{\delta} + \langle E_{\alpha} E_{\alpha}^{*} \rangle \hat{\bm{e}}_{\alpha} \otimes \hat{\bm{e}}_{\alpha} \\ 
&= \begin{pmatrix}
I + Q  \\ \nonumber
U + iV  \\ \nonumber
U - iV  \\ \nonumber
I - Q
\end{pmatrix} \ .
\end{align}
Here, we have left the direction and frequency dependence of the electric field components and Stokes parameters implicit. 

Interferometric calibration, in its most general form, is the process of fitting for the elements of $\mathbfss{J}_{p}$ and $\mathbfss{J}_{q}$, given a model for the instrument and the data. Jones matrices can be decomposed into chains of products of Jones matrices, where each matrix in the chain encodes particular physical propagation effects. Decomposing the cumulative product of all propagation effects along the signal paths of antenna $p$ to the correlator into a chain comprised of sky-based propagation effects described by $\mathbfss{J}^\mathrm{sky}_{p}(\bm{l})$, antenna-based propagation effects described by $\mathbfss{J}^\mathrm{ant}_{p}(\bm{\hat{l}}) = \mathbfss{G}_{p} \mathbfss{E}_{p}(\bm{\hat{l}})$ (where $\mathbfss{G}_{p}$ and $\mathbfss{E}_{p}(\bm{\hat{l}})$ encode direction-independent and direction-dependent antenna-based propagation effects, respectively) and a phase term $K_{p}(\bm{\hat{l}})$, we can write,
\begin{align}
\label{Eq:JonesDecomposition}
\mathbfss{J}_{p}(\bm{\hat{l}}) &= \mathbfss{J}^\mathrm{ant}_{p}(\bm{\hat{l}}) \mathbfss{J}^\mathrm{sky}_{p}(\bm{\hat{l}}) K_{p}(\bm{\hat{l}}) \\ 
&= \mathbfss{G}_{p} \mathbfss{E}_{p}(\bm{\hat{l}}) \mathbfss{J}^\mathrm{sky}_{p}(\bm{\hat{l}}) \mr{e}^{ -2\pi i \frac{\bm{x}_{p}^{T} \bm{\hat{l}}}{\lambda} }. \nonumber
\end{align}
Here, $\bm{x}_{p}$ and $\mathbfss{E}_{p}(\bm{\hat{l}})$ describe the position vector and the voltage beam pattern of antenna $p$, respectively, $K_{p}(\bm{\hat{l}})$ describes the geometrical delay as a function of direction on the sky associated with antenna $p$, $\lambda$ is the wavelength of the incoming radiation and $\mathbfss{G}_{p}$ describes the direction independent signal propagation effects, such as amplification and phase variation due to propagation through the electronics and cables, between the feed and the correlator. $\mathbfss{J}^\mathrm{sky}_{p}(\bm{\hat{l}})$ encodes direction-dependent sky-based signal propagation effects, such as direction-dependent Faraday rotation of the electric field in the Earth's ionosphere.

Full direction-independent gain calibration requires solving for the complex gains of the four visibility correlations encoded in $\mathbfss{G}_{p}$ on a per-antenna basis, as a function of frequency and time. In arrays with long baselines or a large field-of-view (FoV), $\mathbfss{J}^\mathrm{sky}(\bm{\hat{l}})$ of antennas $p$ and $q$, situated under, or observing through, patches of sky with differing ionospheric conditions, will vary according to those conditions, necessitating direction-dependent calibration even if antennas $p$ and $q$ have identical physical construction and environments. However, in this paper, we will assume that the voltage beam patterns associated with all antennas in the array are identical for a given instrumental correlation and we will focus on array configurations that are sufficiently compact and have a sufficiently narrow FoV for direction-dependent sky-based signal propagation effects to be small (i.e. $\mathbfss{J}^\mathrm{sky}_{p}(\bm{\hat{l}}) \simeq \mathbfss{I}$, with $\mathbfss{I}$ the identity matrix). Additionally, we assume $\mathbfss{E}(\bm{\hat{l}})$ is known a priori. We correspondingly limit our focus to direction-independent gain calibration. The adaption of the Bayesian calibration framework presented in \autoref{Sec:BayesCal} to calibration of arrays when these assumptions are relaxed will be considered in future work. 

In the limit that direction-dependent sky-based signal propagation effects are small,
we can re-write \autoref{Eq:Jpq} as,
\begin{align}
\label{Eq:MEqexpanded2}
\bm{V}_{pq} &= \bm{n}_{pq} + \mathbfss{G}_{pq} \iint\limits_{lm}\frac{\mr{d}l\mr{d}m}{n}\, \mathbfss{P}_{pq}(\bm{\hat{l}})\bm{C}^\mathrm{S}(\bm{\hat{l}}) \mr{e}^{-2\pi i \frac{\bm{b}_{pq}^{T} \bm{\hat{l}}}{\lambda}}
\ .
\end{align}
Here, $\mathbfss{G}_{pq} = \mathbfss{G}_{p}  \otimes \mathbfss{G}_{q}^{*}$ is a $4 \times 4$ direction-independent gain matrix. In general, $\mathbfss{G}_{pq}$ is a dense matrix in which off-diagonal terms describe cross-coupling of the induced voltages along the signal path between the antenna feeds and the correlator. The off-diagonal terms in $\mathbfss{G}$ are commonly refered to as $\mathbfss{D}$-terms because a dense $\mathbfss{G}$ matrix can be split out into a chain comprised of a diagonal $\mathbfss{G}$ matrix and a feed-error matrix, $\mathbfss{D}$, that encodes cross-coupling effects (e.g. HBS96). Interferometric arrays aim to minimise cross-coupling and in this paper we will assume that cross-coupling effects are negligible. In this case, $\mathbfss{D}$ is an identity matrix and $\mathbfss{G}_{pq}$ is diagonal with diagonal elements: $(g_{p}^\mathrm{n})(g_{q}^\mathrm{n})^{*}$, $(g_{p}^\mathrm{n})(g_{q}^\mathrm{e})^{*}$, $(g_{p}^\mathrm{e})(g_{q}^\mathrm{n})^{*}$, $(g_{p}^\mathrm{e})(g_{q}^\mathrm{e})^{*}$, where $g_{p}^\mathrm{n}$ and $g_{p}^\mathrm{e}$ are time and frequency dependent complex scalars describing the voltage's amplification and phase shift between the feed and the correlator, in the north and east oriented feed of antenna $p$, respectively. $\mathbfss{P}_{pq} = \mathbfss{E}_{pq}\mathbfss{T}^\mathrm{S}$ is the polarised primary beam matrix encoding the transformation between the Stokes coherency vector, $\bm{C}^\mathrm{S} = (I, Q, U, V)^{T}$ and the true visibilities that would be measured if $\mathbfss{G}_{pq}$ was described by the identity matrix. $\mathbfss{E}_{pq} = \mathbfss{E}_{p}  \otimes \mathbfss{E}_{q}^{*}$ is the polarised primary beam matrix encoding the transformation between the coherency vector, $\bm{C}$, given by the cross-correlation of the electric field written in an equatorial basis (see \autoref{Eq:CoherencyVector}), and the true visibilities. $\mathbfss{T}^\mathrm{S}$ is the Stokes transformation matrix relating the two representations of the coherency vector and defined such that $\bm{C} = \mathbfss{T}^\mathrm{S} \bm{C}^\mathrm{S}$ (e.g. HBS96).

In general, one can individually or jointly calibrate the four visibility correlations of the polarised visibility vector. Here, we assume that the antennas $p$ and $q$ utilise linearly polarised orthogonal dual-feeds and, without loss of generality, we consider the two feeds to be orientated in the north-south and east-west directions, respectively. Writing the electric field incident on antenna $p$ in a topocentric basis, with north and east $2 \times 1$  column basis vectors, $\hat{\bm{e}}_\mathrm{n}$ and $\hat{\bm{e}}_\mathrm{e}$, aligned with the antenna feeds, the induced voltage at the correlator of the interferometer due to the vector electric field incident on antenna $p$, is given by the voltage vector,
\begin{align}
\label{Eq:VoltageNE}
\bm{v}_{p} &= \iint\limits_{lm}\frac{\mr{d}l\mr{d}m}{n}\,  \mathbfss{J}_{p}(\bm{\hat{l}}) \bm{E}(\bm{\hat{l}})  \mathrm{d}\Omega \\
& = (v_{p}^\mathrm{n}, v_{p}^\mathrm{e})^{T} \nonumber 
\ ,
\end{align}
and the polarised visibility vector given by the cross-correlation of $\bm{v}_{p}$ and $\bm{v}_{q}$ can be written as,
\begin{align}
\label{Eq:PolarisedVisibilityVector}
\bm{V}_{pq} &= \langle (v_{p}^\mathrm{n})(v_{q}^{\mathrm{n}})^{*} \rangle \hat{\bm{e}}_\mathrm{n} \otimes \hat{\bm{e}}_\mathrm{n} + \langle (v_{p}^\mathrm{n})(v_{q}^{\mathrm{e}})^{*} \rangle \hat{\bm{e}}_\mathrm{e} \otimes \hat{\bm{e}}_\mathrm{n} \\ \nonumber
&+ \langle (v_{p}^\mathrm{e})(v_{q}^{\mathrm{n}})^{*} \rangle \hat{\bm{e}}_\mathrm{n} \otimes \hat{\bm{e}}_\mathrm{e} + \langle (v_{p}^\mathrm{e})(v_{q}^{\mathrm{e}})^{*} \rangle \hat{\bm{e}}_\mathrm{e} \otimes \hat{\bm{e}}_\mathrm{e} \\ 
&= \begin{pmatrix}
V^\mathrm{nn} \\ \nonumber
V^\mathrm{ne} \\ \nonumber
V^\mathrm{en} \\ \nonumber
V^\mathrm{ee}
\end{pmatrix} \ ,
\end{align}
where, $\langle . \rangle$ denotes an average over a small interval in time and frequency.

In this paper, we consider calibration of an individual component of $\bm{V}_{pq}$ which, without loss of generality, we will take to be $V^\mathrm{nn}$. The three remaining components of the polarised visibility vector, corresponding to the east-west feed correlations, $V^\mathrm{ee}$, the east-west with north-south cross-correlations, $V^\mathrm{ne}$, and the north-south with east-west cross-correlations, $V^\mathrm{en}$, are neglected; however, in general, the calibration approach we describe can be used to individually calibrate any of the four components.

In the topocentric basis of \autoref{Eq:PolarisedVisibilityVector}, we can write the elements of the polarised primary beam matrix as,
\begin{align}
\label{Eq:PolarisedPrimaryBeamMatrixElements}
\mathbfss{P} =  
\begin{pmatrix}
P_{\mathrm{nn},I} & P_{\mathrm{nn},Q} & P_{\mathrm{nn},U} & P_{\mathrm{nn},V}\\ 
P_{\mathrm{ne},I} & P_{\mathrm{ne},Q} & P_{\mathrm{ne},U} & P_{\mathrm{ne},V}\\ 
P_{\mathrm{en},I} & P_{\mathrm{en},Q} & P_{\mathrm{en},U} & P_{\mathrm{en},V}\\ 
P_{\mathrm{ee},I} & P_{\mathrm{ee},Q} & P_{\mathrm{ee},U} & P_{\mathrm{ee},V} 
\end{pmatrix}
\ .
\end{align}
Here, $P_{i,j}$ encodes the effective primary beam of the baselines coupling Stokes parameter $j \in [I,Q,U,V]$, on the sky, to instrumental correlation $i \in [\mathrm{nn}, \mathrm{ne}, \mathrm{en}, \mathrm{ee}]$ and we have left implicit the dependence of the primary beam on the choice of constituent antennas, $p$ and $q$, for notational simplicity.

Expanding \autoref{Eq:MEqexpanded2}, we can write $V^\mathrm{nn}$ in the frame of the instrument, in which the brightness distribution on the celestial sphere is rotating overhead, as,
\begin{multline}
\label{Eq:MEqSingePol}
V^\mathrm{nn}(\bm{u}_{pq},\nu, t) = n(\nu, t) + (g_{p}^\mathrm{n}(\nu, t))(g_{q}^\mathrm{n}(\nu, t))^{*} \\ 
\times \iint\limits_{lm}\frac{\mr{d}l\mr{d}m}{n}\,  
[P_{\mathrm{nn},I}(\bm{\hat{l}}, \nu) I(\bm{\hat{l}}, \nu, t) + P_{\mathrm{nn},Q}(\bm{\hat{l}}, \nu) Q(\bm{\hat{l}}, \nu, t) \\
+ P_{\mathrm{nn},U}(\bm{\hat{l}}, \nu) U(\bm{\hat{l}}, \nu, t) + P_{\mathrm{nn},V}(\bm{\hat{l}}, \nu) V(\bm{\hat{l}}, \nu, t)] \mr{e}^{ -2\pi i \bm{u}_{pq}^{T}(\nu) \bm{\hat{l}}(\nu) } \ .
\end{multline}
Here, we have expressed the visibilities in terms of the time-stationary $uvw$-coordinates $\bm{u}_{pq} = \bm{b}_{pq}/\lambda = (u_{pq},v_{pq},w_{pq})$ of the baseline in the frame of the instrument; $n(\nu, t)$ describes the noise on $V^\mathrm{nn}(\bm{u}_{pq},\nu, t)$, which we will assume is Gaussian and white; $P_{\mathrm{nn},j}(\bm{\hat{l}}, \nu)$, with $j \in [I,Q,U,V]$ are the elements of the polarised primary beam matrix that $V^\mathrm{nn}$ is dependent on, and we have included explicit dependencies on observing frequency $\nu$ and time $t$. Going forward, all calculated visibilities refer to correlations of pairs of north-south oriented feeds, so for brevity we omit the `$\mathrm{nn}$' and `$\mathrm{n}$' superscripts on subsequent visibilities and gains associated with these visibilities, respectively.

\subsection{Sky Based Calibration}
\label{Sec:SkyBasedCalibration}

In the following sections, it will be helpful to write a discretised version of \autoref{Eq:MEqSingePol} for a set of baselines as,
\begin{equation}
\label{Eq:MEqSingePolShorthand}
\bm{V}^\mathrm{obs} = \bm{n} + \mathbfss{G}\bm{V}^\mathrm{true} \ ,
\end{equation}
where we will refer to $\mathbfss{G}\bm{V}^\mathrm{true}$ and $\bm{n}$ as the signal and noise components of the observed visibilities, respectively, and $\bm{V}^\mathrm{obs}$, $\bm{n}$ and $\bm{V}^\mathrm{true}$ are given by the concatenation over baselines of the observed  cross-correlation visibilities, noise on the data and `true cross-correlation visibilities' (corresponding to the integral in \autoref{Eq:MEqSingePol}), respectively, each vectorised over a discrete set of frequencies and times. $\mathbfss{G}$ is a diagonal matrix encoding the antenna, frequency and time dependent instrumental gains, and, for a single frequency and time, it has diagonal elements $G_{ii} = g_{p}g_{q}^{*}$. Here $i$ is an index over visibilities which runs over the $N_\mathrm{vis} = N_t N_\nu N_\mathrm{ant}(N_\mathrm{ant}-1)/2$ cross-correlations between the signals from the $N_\mathrm{ant}$ antennas in the array, with $p$ and $q$ the antennas associated with the $i$th visibility, such that $\bm{V}^\mathrm{obs}$ has elements,
\begin{equation}
\label{Eq:MEqSingePolShorthandElement}
V^\mathrm{obs}_{i} = n_{pq} + g_{p}g_{q}^{*} V^\mathrm{true}_{i}\, .
\end{equation}

Simultaneous calibration of multiple frequencies is achieved by writing $\bm{V}^\mathrm{obs} = ((\bm{V}^\mathrm{obs}_{\nu_{1}})^{T}, (\bm{V}^\mathrm{obs}_{\nu_{2}})^{T}, \cdots (\bm{V}^\mathrm{obs}_{\nu_{N_{\nu}}})^{T})^{T}$, with $\bm{V}^\mathrm{obs}_{\nu_{i}}$ the observed visibilities in frequency channel $\nu_{i}$, where $\nu_{i}$ is an index over the $N_{\nu}$ channels in the data set, and writing $\bm{V}^\mathrm{true}$ equivalently; then $\mathbfss{G}$ becomes a block-diagonal matrix comprised of $N_{\nu}$ blocks, where the $\nu_{i}$-th block describes the instrumental gains in the $\nu_{i}$-th channel. Simultaneous calibration of multiple time integration is similarly achieved by concatenating the observed and true visibilities over the $N_{t}$ time integrations in the data set. Simultaneous calibration of multiple frequency channels and time integrations is achieved by  writing $\bm{V}^\mathrm{obs}_{\nu_{i}} = ((\bm{V}^\mathrm{obs}_{\nu_{i}, t_{1}})^{T}, (\bm{V}^\mathrm{obs}_{\nu_{i}, t_{2}})^{T}, \cdots, (\bm{V}^\mathrm{obs}_{\nu_{i}, t_{N_{\nu}}})^{T})^{T}$, with $\bm{V}^\mathrm{obs}_{\nu_{i}, t_{j}}$ the observed visibilities in frequency channel $\nu_{i}$ and time integration $t_{j}$, where $t_{j}$ is an index over the $N_{t}$ channels in the data set; then $\mathbfss{G}$ becomes a block-diagonal matrix comprised of $N_{\nu}$ block-diagonal matrices where the $\nu_{i}$-th block-diagonal matrix is comprised of $N_{t}$ block-diagonal matrices and where the $t_{j}$-th block describes the instrumental gains in the $\nu_{i}$-th channel and $t_{j}$-th time integration. Finally, the procedure described above can also be simply generalised to calibrate multiple instrumental correlations (e.g. $V^\mathrm{nn}$, $V^\mathrm{ne}$, $V^\mathrm{en}$, $V^\mathrm{ee}$) simultaneously.

In a non-redundant array, the data calibration problem requires solving for model gain parameters $g_{p}^\mathrm{m}$ given a model for the signal component of the observed visibilities,
\begin{equation}
\label{Eq:MEqSingePolShorthandModel}
\bm{V}^\mathrm{model} = \mathbfss{G}^\mathrm{m}\bm{V}^\mathrm{sim} \ ,
\end{equation}
where $\mathbfss{G}^\mathrm{m}$ is a diagonal matrix with diagonal elements, $G^\mathrm{m}_{ii} = g_{p}^\mathrm{m}g_{q}^{\mathrm{m}*}$, with $p$ and $q$ the antennas associated with the $i$th visibility. $\bm{V}^\mathrm{sim}$ is a vector of simulated visibilities encoding our best estimate of the true (gain-free) visibilites derived from our a priori knowledge of the antenna beam, array layout and emission sources in the sky under observation, $\bm{V}^\mathrm{model}$ is a vector of model visibilities, and the elements of \autoref{Eq:MEqSingePolShorthandModel} are ordered such that,
\begin{equation}
\label{Eq:MEqSingePolShorthandModelElement}
V^\mathrm{model}_{i} = g_{p}^\mathrm{m}g_{q}^{\mathrm{m}*}V^\mathrm{sim}_{i}\, .
\end{equation}

In practice, $\bm{V}^\mathrm{sim}$ will generally be a model for the dominant astrophysical components contributing to $\bm{V}^\mathrm{true}$. Common approximations for $\bm{V}^\mathrm{sim}$ include: \begin{enumerate}
\item In the simplest case, when a single point source dominates the observed field, a forward model for the visibilities that derive from only that source may be used.
\item More usually, in current approaches to calibration in 21 cm cosmology, this will be improved on by including in the calibration model many thousands of the brightest sources in a field and, in place of the point source approximation, using more accurate models for extended sources.
\item The calibration model can be further improved by including a model for diffuse emission from the Galaxy, the power from which is inversely proportional to baseline length and dominates the measured visibilities on short $\mathcal{O}(10~\mathrm{m})$ baselines. Incorporating a model for diffuse emission is particularly important for obtaining robust calibration solutions that minimise spurious spectral structure in the data  when short baselines are used for the absolute calibration of the instrument.
\end{enumerate} 

Each of the above approximations will inevitably be, at some level, an incomplete and imperfect description of the foregrounds due to unmodelled sources below the flux-density limits of current surveys and uncertainties on the sky brightness distributions associated with the models. Thus, one can improve upon models based on these approximations by additionally fitting for a statistical model for the difference between the true emission on the sky and the approximations described above. We will describe a physically motivated approach to achieving this in \autoref{Sec:BayesCal}.

\subsubsection{Constructing a sky-based calibration likelihood}
\label{Sec:ConstructingASkyBasedCalibrationLikelihood}

To fit for the gain parameters in a purely sky-based calibration framework, one must define a model for the noise on the data, which, in many cases of interest (see \autoref{Sec:DataLikelihood} for further discussion), can be approximated as being drawn from a zero-mean, statistically homogeneous, complex Gaussian random distribution, uncorrelated between different visibilities, with covariance matrix $\mathbfss{N}$ given by,
\begin{equation}
\label{Eq:DataCovarianceMatrix}
N_{ij} = \left< n_in_j^{*}\right> = \delta_{ij}\sigma_{j}^{2} \ ,
\end{equation}
where $\left< .. \right>$ represents the expectation value and $\sigma_{j}^{2}$ is the variance of the complex noise on visibility $j$. 

We can, therefore, write a Gaussian likelihood for $\bm{V}^\mathrm{obs}$ and our model of the data, constructed from the set of gain parameters $\bm{g}^\mathrm{m}$, encoded in the calibration matrix, $\mathbfss{G}^\mathrm{m}$, as,
\begin{align}
\label{Eq:BasicVisLike}
\mathrm{Pr}(\bm{V}^\mathrm{obs} \;|\; \bm{g}^\mathrm{m}) &= \frac{1}{\pi^{N_{\mathrm{vis}}}\mathrm{det}(\mathbfss{N})} \nonumber \\
&\times\exp\left[-\left(\bm{V}^\mathrm{obs} - \mathbfss{G}^\mathrm{m}\bm{V}^\mathrm{sim}\right)^\dagger\mathbfss{N}^{-1}\left(\bm{V}^\mathrm{obs} - \mathbfss{G}^\mathrm{m}\bm{V}^\mathrm{sim}\right)\right] \ .
\end{align} 
Here, our data and model vector, $\bm{V}^\mathrm{obs}$ and $\mathbfss{G}^\mathrm{m}\bm{V}^\mathrm{sim}$, respectively, are each of length $N_{\mathrm{vis}}$.  The gain parameters can then be estimated by finding the maximum likelihood solution to \autoref{Eq:BasicVisLike}, which can be achieved using, for example, gradient descent optimization \citep{Gilbert1989}. An efficient linear least squares fitting algorithm for the more general case of direction-dependent sky-based calibration is described in \citet{2014A&A...571A..97S}.

\autoref{Eq:BasicVisLike} can be re-written as (e.g. \citealt{FoSTP}),
\begin{multline}
\label{Eq:BasicVisLikeReal}
\mathrm{Pr}(\bm{V}^\mathrm{obs} \;|\; \bm{g}^\mathrm{m}) = \mathrm{Pr}(\bm{V}^\mathrm{obs, r} \;|\; \bm{g}^\mathrm{m, r}) = \frac{1}{\sqrt{(2\pi)^{2N_{\mathrm{vis}}}\mathrm{det}(\mathbfss{N}^\mathrm{r})}} \\
\times\exp\left[-\frac{1}{2}\left(\bm{V}^\mathrm{obs, r} - \bm{V}^\mathrm{model, r}\right)^T\mathbfss{N}^{-1}\left(\bm{V}^\mathrm{obs, r} - \bm{V}^\mathrm{model, r}\right)\right] \ .
\end{multline} 
Here vectors with r superscripts are defined such that for a complex column vector $\bm{x}$ of length $N_x$, $\bm{x}^\mathrm{r} = [\mathrm{Re}(\bm{x})^{T}, \mathrm{Im}(\bm{x})^{T}]^{T}$ is a real vector of length $2N_x$ and,
\begin{align}
\label{Eq:RealCovarianceMatrix}
\mathbfss{N}^\mathrm{r} =  
\begin{pmatrix}
\mathrm{cov}(\mathrm{Re}(\bm{x})) & 0 \\ 
0 & \mathrm{cov}(\mathrm{Im}(\bm{x})) \\ 
\end{pmatrix}
\ ,
\end{align}
and we have assumed that the real and imaginary noise components are uncorrelated, $\mathrm{cov}(\mathrm{Re}(\bm{x})) = \mathrm{cov}(\mathrm{Im}(\bm{x}))$, and have the same form as \autoref{Eq:DataCovarianceMatrix} but with $\sigma_{j}^{2}$ now representing the variance of the real component of the noise on visibility $j$.

Converting between Equations \ref{Eq:BasicVisLike} and \ref{Eq:BasicVisLikeReal} can be computationally convenient when an efficient optimisation algorithm is available in either the complex or real domain. However, for brevity, we quote only the complex form going forward.

In the limit of a high fidelity model for the true visibilities, \autoref{Eq:BasicVisLike} and \autoref{Eq:BasicVisLikeReal} provide unbiased estimators of the gain solutions. Thus, in this limit and subject to the noise on the data, the maximum likelihood gain parameters will yield a model calibration matrix equal to the true calibration matrix of the data set (i.e. if $\bm{V}^\mathrm{sim}=\bm{V}^\mathrm{true}$,  $\lim\limits_{\bm{n} \to \bm{0}} \mathbfss{G}^\mathrm{m} = \mathbfss{G}$). However, since the antenna gains are the only degrees of freedom in \autoref{Eq:BasicVisLike} and \autoref{Eq:BasicVisLikeReal}, their maximum likelihood solutions will deviate from the true instrumental gains when there are imperfections in $\bm{V}^\mathrm{sim}$, due either to the sky model from which $\bm{V}^\mathrm{sim}$ is derived being incomplete (\citealt{2016MNRAS.461.3135B}) or because of imperfections in the instrument model used to propagate the calibration sky model to the simulated interferometric visibilites.

For 21 cm cosmology applications, even low-level spurious spectral structure can produce spectrally fluctuating foreground systematics in excess of the signal of interest, preventing unbiased extraction of the 21 cm cosmological signal from the data. Redundant calibration can go some way towards mitigating this issue by reducing the number of calibration degrees of freedom that $\bm{V}^\mathrm{sim}$ is used to solve for. Nevertheless, it still requires using $\bm{V}^\mathrm{sim}$ to derive the full set of calibration solutions and, thus, does not eliminate the problem entirely (e.g. \citealt{2019ApJ...875...70B}). We discuss this approach to calibration in the next section.

\subsection{Redundant Calibration}
\label{Sec:RedundantCalibration}

In the sky-based calibration framework described above, the only constraint on the gains derives from one's simulated model for the sky. When an array has a redundant antenna layout, such that it measures visibilities derived from the cross-correlation of the voltage responses from sets of antennas with identical baseline separation vectors and beam patterns, the fact that $V^\mathrm{true}$ is identical\footnote{However, see e.g. \citet{2018AJ....156..285J}, \citet{2019MNRAS.487..537O} and \citet{2021MNRAS.506.2066C} for biases introduced in the relative calibration solutions if there are non-redundancies in an array that is assumed to be redundant.} on these redundant baselines provides an additional constraint on the model gain parameters that is independent of the sky (e.g. \citealt{1992ExA.....2..203W, 2010MNRAS.408.1029L}).

To make use of this additional constraint, one writes the data model as a function of redundant antenna gain parameters, $h_{p}^\mathrm{m}$, and true visibility parameters, $V^\mathrm{red}_{\alpha}$ (e.g. \citealt{2019ApJ...875...70B}),
\begin{equation}
\label{Eq:MEqSingePolShorthandRedundantModel}
V^\mathrm{model}_{\alpha,j} = h_{p}^\mathrm{m}h_{q}^\mathrm{m*}V^\mathrm{red}_{\alpha}\, .
\end{equation}
Here, $\alpha$ indexes unique baseline types in the array, $V^\mathrm{model}_{\alpha,j}$ is our model for the $j$th redundant visibility in the set of visibilities with the $\alpha$th unique baseline, $h_{p}^\mathrm{m}$ is the redundant gain parameter associated with antenna $p$  and $V^\mathrm{red}_{\alpha}$ is a parameter representing the true visibility measured by the baselines in the $\alpha$th unique baseline group. The change of variable (from $g$ to $h$) for the antenna gain parameter in \autoref{Eq:MEqSingePolShorthandRedundantModel}, foreshadows the fact that fully calibrating an array using redundant calibration is a two stage process that requires: \begin{enumerate}                                                                                               \item \textit{'relative' calibration}, in which baseline redundancy is used to derive redundant gain parameters, $h_{p}^\mathrm{m}$, that are equal to the general direction-independent gain parameter, $g_{p}^\mathrm{m}$, solved for in \autoref{Eq:MEqSingePolShorthandModel}, to within a frequency, time and polarisation dependent complex degeneracy factor. This complex degeneracy factor is described by a set of degenerate calibration parameters that cannot be solved for using baseline redundancy.
\item \textit{'absolute' calibration}, in which the degenerate calibration parameters associated with relative calibration are solved for with reference to a sky model.
\end{enumerate}

We describe approaches for solving for the gain parameters associated with (i) and (ii) in Sections \ref{Sec:ConstructingARelativeCalibrationLikelihood} and \ref{Sec:ConstructingAnAbsoluteCalibrationLikelihood}, respectively.

\subsubsection{Constructing a relative calibration likelihood}
\label{Sec:ConstructingARelativeCalibrationLikelihood}

We can write a Gaussian likelihood for $\bm{V}^\mathrm{obs}$ and the data model in \autoref{Eq:MEqSingePolShorthandRedundantModel} as,
\begin{multline}
\label{Eq:RedundantVisLike}
\mathrm{Pr}(\bm{V}^\mathrm{obs} \;|\; \bm{h}^\mathrm{m}, \bm{V}^\mathrm{red,set}) = \frac{1}{\pi^{N_{\mathrm{vis}}}\mathrm{det}(\mathbfss{N})} \\
\times\exp\left[-\left(\bm{V}^\mathrm{obs} - \mathbfss{H}^\mathrm{m}\bm{V}^\mathrm{red}\right)^\dagger\mathbfss{N}^{-1}\left(\bm{V}^\mathrm{obs} - \mathbfss{H}^\mathrm{m}\bm{V}^\mathrm{red}\right)\right] \ .
\end{multline}
Here, we have parametrised the model in terms of a set of redundant gain parameters, $\bm{h}$, encoded in the redundant calibration matrix, $\mathbfss{H}^\mathrm{m}$, and a set of unique redundant visibility parameters, $\bm{V}^\mathrm{red,set}$ defined such that $V^\mathrm{red,set}_{\alpha}$ is the estimate of the true visibility value measured by baselines in the $\alpha$-th redundant-baseline group. $\mathbfss{H}^\mathrm{m}$ is a diagonal matrix with elements $H_{ij}^\mathrm{m} = \delta_{ij}h_{p}^\mathrm{m}h_{q}^{\mathrm{m}*}$, with $p$ and $q$ the antennas associated with the $i$th visibility.
$\bm{V}^\mathrm{red}$ is a vector of length $N_\mathrm{vis} = \sum\limits_{\alpha=1}^{N_\mathrm{red}} N_{\alpha}$, where $N_\mathrm{vis}$ is the total number of observed visibility data points, as defined below \autoref{Eq:MEqSingePolShorthand}. $\bm{V}^\mathrm{red}$ derives from $\bm{V}^\mathrm{red,set}$ via $\bm{V}^\mathrm{red} = ((\bm{V}^\mathrm{red}_{1})^{T}, (\bm{V}^\mathrm{red}_{2})^{T}, \cdots, (\bm{V}^\mathrm{red}_{N_\mathrm{red}})^{T})^{T}$, where $N_\mathrm{red}$ is the number of redundant baseline groups in the array. Here, $\bm{V}^\mathrm{red}_{\alpha}$ consists of $N_{\alpha}$ elements, each equal to $V^\mathrm{red,set}_{\alpha}$, where $N_{\alpha}$ is the number of baselines in redundant baseline group $\alpha$.

The parameters in \autoref{Eq:MEqSingePolShorthandRedundantModel} are degenerate (see e.g. \citealt{2010MNRAS.408.1029L}). If this were not the case, assuming $N_\mathrm{vis} \ge N_\mathrm{ant} + N_\mathrm{red}$, where $N_\mathrm{ant}$ is the number of antennas in the array and $N_\mathrm{red}$ is the number of unique redundant visibilities in the array, \autoref{Eq:MEqSingePolShorthandRedundantModel} alone could be used to derive a set of maximum likelihood estimates for the model parameters. One must place additional constrains on either $\bm{h}^\mathrm{m}$ or $\bm{V}^\mathrm{red,set}$ to break this degeneracy. The degeneracy between $\bm{h}^\mathrm{m}$ and $\bm{V}^\mathrm{red,set}$ has the functional form, $f_\mathrm{ant}(A, \sPhi, \psi) = A^{-1/2}\mr{e}^{-i(\bm{x}_{p}^{T} \sPhi - \psi)}$ and $f_\mathrm{vis}(A, \sPhi) = A\mr{e}^{i\bm{b}_{i}^{T} \sPhi}$ for the redundant gains and redundant visibility parameter estimates, respectively. Here, $A$ is a real amplitude and $\psi$ is an absolute gain phase. For a coplanar\footnote{
For a non-coplanar redundant array, $\bm{x}_{p} = (x,y,z)$ and an additional degenerate phase, $\phi_{n}$, which corresponds to a translation of the phase center of the observation in direction cosine $n$, must be solved for. In this case, one must fit for redundant gains using constraints deriving from degeneracy between baselines in 3D. In this paper we focus on calibration of a coplanar array. However, the calibration formalisms discussed in the paper generalise in a straightforward manner to the 3D via a redefinition of $\sPhi = (\phi_{l}, \phi_{m}, \phi_{n})$. Given this definition, one then fits for the value of $\phi_{n}$, jointly with the degenerate gain parameters for a coplanar array, as a function of time, frequency and polarisation; however, the functional form of the degeneracies and the manner in which they can be fit is otherwise unchanged. 
} array, $\bm{x}_{p} = (x,y)$ is the position vector of antenna $p$, $\bm{b}_{i} = \bm{x}_{p} - \bm{x}_{q}$ is the  baseline vector $i$ between antennas $p$ and $q$, $\sPhi = (\phi_{l}, \phi_{m})$ corresponds to the `tip-tilt' phase angles that shift the apparent phase center of the sky in $l$ and $m$, respectively. Each of these degeneracy parameters are themselves functions of frequency, time and, in the general case, polarisation; however, for brevity, we omit writing this dependence explicitly.

To illustrate the existence of this parameter degeneracy, one can define redundant antenna gains and redundant visibility parameters equal to their counterparts in \autoref{Eq:MEqSingePolShorthandRedundantModel} scaled by $f_\mathrm{ant}(A, \sPhi, \psi)$ and $f_\mathrm{vis}(A, \sPhi)$, respectively,
\begin{align}
\label{Eq:DegenerateParameterDefinitions}
\bar{h}_{p}^\mathrm{m} &= A^{-1/2}\mr{e}^{-i(\bm{x}_{p}^{T} \sPhi - \psi)} h_{p}^\mathrm{m} \\ 
\bar{V}^\mathrm{red}_{\alpha} &= A\mr{e}^{i\bm{b}_{i}^{T} \sPhi}V^\mathrm{red}_{\alpha} \ . \nonumber
\end{align}
Substituting $\bar{h}_{p}^\mathrm{m}$ and $\bar{V}^\mathrm{red}_{\alpha}$ for $h_{p}^\mathrm{m}$ and $V^\mathrm{red}_{\alpha}$ when calculating the model visibilites corresponding to these redundant gain parameters, we have:
\begin{align}
\label{Eq:DegenerateParameterSubstitution}
\bar{V}^\mathrm{model}_{\alpha,j} &= \bar{h}_{p}^\mathrm{m} \bar{h}_{q}^\mathrm{m*}\bar{V}^\mathrm{red}_{\alpha} \\  \nonumber
 &= A^{-1/2}\mr{e}^{-i(\bm{x}_{p}^{T} \sPhi - \psi)} h_{p}^\mathrm{m}A^{-1/2}\mr{e}^{+i(\bm{x}_{q}^{T} \sPhi + \psi)} h_{q}^\mathrm{m} A\mr{e}^{\bm{b}_{i}^{T} \sPhi}V^\mathrm{red}_{\alpha} \\ \nonumber
 &= h_{p}^\mathrm{m}h_{q}^\mathrm{m*}V^\mathrm{red}_{\alpha} \\ \nonumber
 &= V^\mathrm{model}_{\alpha,j}\ , \nonumber
\end{align}
where, in going from the second to third line, we have used the fact that $\bm{b}_{i} = \bm{x}_{p} - \bm{x}_{q}$.

One way to break this degeneracy, for the purposes of relative calibration, is by fixing, to an arbitrary choice of values, the gain amplitude of a reference antenna and the gain phases of two reference antennas\footnote{Eliminating the absolute amplitude and tip-tilt phase degeneracies can also be achieved in an equivalent manner by fixing the average gain amplitude, phase and phase gradient between antennas.} (e.g. \citealt{2018ApJ...863..170L, 2020arXiv200308399D}). For any given choice of reference gain amplitude and phases, there exists a corresponding unique solution to \autoref{Eq:DegenerateParameterSubstitution} in the form of a redundant gain parameter estimate per antenna, per frequency and per time and an estimate of $\bm{V}^\mathrm{red}$ for each unique redundant baseline in the array. Maximum likelihood estimates for these parameters can be derived using non-linear optimization algorithms such as dual annealing (\citealt{SimulatedAnnealing} and \citealt{2020NatMe..17..261V} for its implementation\footnote{https://docs.scipy.org/doc/scipy/reference/optimize.html} in \textsc{scipy}). Alternatively, a frequently used approach is to linearise the equation and split it into its real and imaginary parts. A set of maximum likelihood parameters can then be derived by iteratively solving the corresponding sets of linearised equations (see e.g. \citealt{2010MNRAS.408.1029L, 2016ApJ...826..181D, 2017MNRAS.470.1849E, 2018ApJ...863..170L, 2018MNRAS.477.5670D,2018AJ....156..285J}). However, regardless of the approach to solving the equation, the redundant gain parameters thus derived are only correct to within a factor $f_\mathrm{ant}(A, \sPhi, \psi)$ determined by the difference between the true amplitude and phases of the reference antennas and the arbitrary choice of values used for these parameters during relative calibration. Therefore, full calibration of the data requires subsequent fitting for the true values of $A, \sPhi, \psi$ to calibrate out this degeneracy factor in the relatively calibrated data. This, in turn, must be done with reference to a calibration model derived from a sky model.

With respect to this fit, we note that in both sky-based and redundant calibration one is ultimately interested in the cross-correlations of the antenna voltages, for which the effect of absolute gain phase offsets common between antennas cancels out. As such, it is not necessary to constrain $\psi$  to calibrate the visibility data (this is also evident from the fact that the degeneracy in $V^\mathrm{red}_{\alpha}$ is independent of $\psi$). Thus, while the additional amplitude and phase constraints on the reference antennas, described above, fix all four degeneracies of the redundant gain parameters in \autoref{Eq:DegenerateParameterSubstitution}, relating these solutions to the general direction-independent gain parameters of interest (equal to those solved for in \autoref{Sec:ConstructingASkyBasedCalibrationLikelihood}) simplifies to solving for the appropriate overall amplitude and tip-tilt phase offset, per frequency and per time, required to fit the maximum likelihood redundant visibility model to a simulated model of the true sky visibilities.

\subsubsection{Constructing an absolute calibration likelihood for redundant calibration}
\label{Sec:ConstructingAnAbsoluteCalibrationLikelihood}

To derive absolute calibration solutions from our relative calibration solutions calculated in \autoref{Sec:ConstructingASkyBasedCalibrationLikelihood}, we start by defining $\hat{h}_{p}^\mathrm{m}$ as our maximum likelihood redundant gain parameters derived when prescribing a fixed gain amplitude to a reference antenna and fixed gain phases to two reference antennas. We can then define an absolutely calibrated model for the data as,
\begin{equation}
\label{Eq:PureSkyVisModel}
\bm{V}^\mathrm{model} = \hat{\mathbfss{H}}^\mathrm{m}\mathbfss{D}^\mathrm{m}\bm{V}^\mathrm{sim} \ .
\end{equation}
Here, we have defined the degenerate gain matrix $\mathbfss{D}^\mathrm{m}$, which is a diagonal matrix that for a single time integration and frequency channel has elements, $D_{ij} = \delta_{ij}A\mr{e}^{i(\bm{b}_{i}^{T} \sPhi)}$, $\hat{\mathbfss{H}}^\mathrm{m}$ is the diagonal matrix containing the maximum likelihood redundant gain parameters solved for in \autoref{Sec:ConstructingASkyBasedCalibrationLikelihood} and the elements of our model vector are given by,
\begin{equation}
\label{Eq:MEqSingePolShorthandRedundantModelNonDegenerate} 
V^\mathrm{model}_{\alpha,j} = \hat{h}_{p}^\mathrm{m}\hat{h}_{q}^{\mathrm{m}*}A\mr{e}^{i(\bm{b}_{i}^{T} \sPhi)}V^\mathrm{sim}_{\alpha}\, ,
\end{equation}
where $V^\mathrm{sim}_{\alpha}$ is a simulated model for the true visibility on the $\alpha$th unique baseline.
 
We can thus write a general likelihood for our model of the data as a function of amplitude and tip-tilt gain phase parameters, $\bm{A}$, $\bm{\mathit{\Phi}}_{l}$ and $\bm{\mathit{\Phi}}_{m}$, encoded in the degenerate gain matrix, $\mathbfss{D}^\mathrm{m}$, as,
\begin{multline}
\label{Eq:AbsCalVisLike}
\mathrm{Pr}(\bm{V}^\mathrm{obs} \;|\; \bm{A}, \bm{\mathit{\Phi}}_{l}, \bm{\mathit{\Phi}}_{m}) = \frac{1}{\pi^{N_{\mathrm{vis}}}\mathrm{det}(\mathbfss{N})} \\
\times\exp\left[-\left(\bm{V}^\mathrm{obs} - \hat{\mathbfss{H}}^\mathrm{m}\mathbfss{D}^\mathrm{m}\bm{V}^\mathrm{sim}\right)^\dagger\mathbfss{N}^{-1}\left(\bm{V}^\mathrm{obs} - \hat{\mathbfss{H}}^\mathrm{m}\mathbfss{D}^\mathrm{m}\bm{V}^\mathrm{sim}\right)\right] \ .
\end{multline} 
The amplitude and tip-tilt gain phase parameters (one of each parameter per channel) can then be estimated using the same approaches as for \autoref{Eq:BasicVisLike}. 

In principle, one can simply fit for the degenerate amplitude and tip-tilt gain phase parameters jointly with the redundant gain parameters by sampling from $\mathrm{Pr}(\bm{V}^\mathrm{obs} \;|\; \bm{A}, \bm{\mathit{\Phi}}_{l}, \bm{\mathit{\Phi}}_{m}, \bm{h}, \bm{V}^\mathrm{red,set}) = \mathrm{Pr}(\bm{V}^\mathrm{obs} \;|\; \bm{A}, \bm{\mathit{\Phi}}_{l}, \bm{\mathit{\Phi}}_{m})\, \mathrm{Pr}(\bm{V}^\mathrm{obs} \;|\; \bm{h}, \bm{V}^\mathrm{red,set})$. However, the larger parameter space associated with $\mathrm{Pr}(\bm{V}^\mathrm{obs} \;|\; \bm{A}, \bm{\mathit{\Phi}}_{l}, \bm{\mathit{\Phi}}_{m}, \bm{h}, \bm{V}^\mathrm{red,set})$ (i.e. the curse of dimensionality), means that, for a fully redundant array, it is computationally preferable to perform the relative and absolute calibration stages independently.

In comparison to \autoref{Eq:BasicVisLike}, deriving our calibration solutions by solving \autoref{Eq:AbsCalVisLike} has the advantage of greatly reducing the number of parameters that are solved for with an imperfect forward model of the true visibilities, $\bm{V}^\mathrm{sim}$. Nevertheless, since $\bm{A}$, $\bm{\mathit{\Phi}}_{l}$, $\bm{\mathit{\Phi}}_{m}$ are all frequency dependent, their maximum likelihood solutions will still absorb chromatic errors in $\bm{V}^\mathrm{sim}$ relative to $\bm{V}^\mathrm{true}$ as spurious spectral structure, in an analogous manner to standard sky based calibration. Indeed, within this framework, \citet{2019ApJ...875...70B} have shown that model incompleteness can still couple a prohibitive level of spectral structure into the gain solutions for 21 cm cosmology applications. However, in the next section, we will address this calibration model limitation by extending this approach to absolute calibration of relatively calibrated visibilities to incorporate a fitted statistical model for the emission components that, due to incomplete knowledge of the brightness distribution of the sky, are omitted in the calculation of $\bm{V}^\mathrm{sim}$.

%%%%%%%%%%%%%%%%%%%%%%%%%%%%%%%%%%%%%%%%%%%%%%%%%%%%%%%%%
\section{\textsc{BayesCal}: joint estimation of the calibration solutions and a parametrised sky model}
\label{Sec:BayesCal}
%%%%%%%%%%%%%%%%%%%%%%%%%%%%%%%%%%%%%%%%%%%%%%%%%%%%%%%%%

To extend the approach to absolute calibration of redundantly calibrated visibilities described in \autoref{Sec:ConstructingAnAbsoluteCalibrationLikelihood} to account for sky-model incompleteness, we subdivide our calibration model into two components, \begin{enumerate*}\item a simulated component, $\bm{V}^\mathrm{sim}$, derived using our a priori knowledge of the brightness distribution of known Galactic emission and extragalactic sources in the region of sky under observation and \item a fitted component, $\bm{V}^\mathrm{fit}$, derived using a parametrised model for the contribution to $\bm{V}^\mathrm{true}$ missing in $\bm{V}^\mathrm{sim}$ and constrained by a prior on the two-dimensional spatial power spectrum of the emission. 
\end{enumerate*}

In \autoref{Sec:BayesCalVisibilityModel}, we begin by describing the \textsc{BayesCal} visibility model for the absolute calibration of redundantly calibrated visibilities. In \autoref{Sec:BayesianInference}, we outline the principles of Bayesian inference on which our \textsc{BayesCal} calibration procedure is based. In \autoref{Sec:InstrumentalUncertainties}, we briefly describe the challenge to recovery of unbiased calibration solutions posed by instrumental uncertainties and outline an approach with which this can be addressed within the \textsc{BayesCal} framework.
In \autoref{Sec:FittedImageDomainSkyModel}, we describe the image domain sky model from which we derive $\bm{V}^\mathrm{fit}$. In \autoref{Sec:TemporalModelling}, we describe the use of temporal priors when calibrating multiple time integrations. We discuss approaches to determining the optimal choice of spectral model complexity of the fitted sky model in \autoref{Sec:SpectralModelComplexity} and describe our constraint on the brightness distribution of the fitted sky model in \autoref{Sec:PowerSpectralPriors}. In \autoref{Sec:DegenerateGainModelParametrisation}, we describe a parametrisation of the redundant calibration degeneracy function to enable non-sky-based constraints on the gain solutions, such as electric and electromagnetic co-simulation and reflectometry measurements of the receiver system, to be incorporated into the \textsc{BayesCal} calibration solutions. In \autoref{Sec:AnalyticMarginalisationOverTheSkyModelParameters}, we analytically marginalise over the parameters of our fitted sky model in order to sample directly from the posterior for the calibration solutions and, in \autoref{Sec:AnalyticMarginalisationOverPointSourceUncertainties}, we describe how a similar approach can be used to marginalise over uncertainties in catalogued point sources in $\bm{V}^\mathrm{sim}$. In \autoref{Sec:BayesCalForSkyBasedCalibration}, we describe how \textsc{BayesCal} can be applied to sky-based, as well as, redundant calibration.

\subsection{\textsc{BayesCal} visibility model}
\label{Sec:BayesCalVisibilityModel}

As described in \autoref{Sec:SkyBasedCalibration}, in this paper we focus on the calibration of individual components of the polarised visibility vector. For concreteness, we describe calibration of a given correlation of the polarised visibility vector and we use $V^\mathrm{nn}$ for this purpose, which corresponds to the correlation of signals from the northward oriented antenna feeds of an orthogonal dual-feed system. However, in general, the calibration approach we describe can be used to individually calibrate any of the four components of the polarised visibility vector in \autoref{Eq:PolarisedVisibilityVector} (see \citet{2018MNRAS.477.5670D} for the degenerate gain parameters associated with redundant calibration of a multiple-correlation visibility data set). We plan to generalise the approach to joint estimation of the gain parameters associated with the four correlation states of the visibility matrix, within the \textsc{BayesCal} framework, in future work.

In this regime, \autoref{Eq:MEqSingePolShorthand} describes the observed visibilities, $\bm{V}^\mathrm{obs}$. It will be convenient to define $\bm{V}^\mathrm{fit}$ in terms of a set of image-space parameters encoding the direction-dependent brightness temperature of the component of the sky-emission missing from $\bm{V}^\mathrm{sim}$. Therefore, we start by re-writing \autoref{Eq:MEqSingePolShorthand} in terms of a set of discretised models for the polarised sky brightness temperature\footnote{The brightness temperature, $T$, is related to the spectral brightness distribution via the Rayleigh--Jeans law, $T(\nu) = \dfrac{c^2 B}{2\nu^2k_\mathrm{B}}$, with $c$ and $k_\mathrm{B}$ the speed of light and Boltzmann constant, respectively.} in Stokes I, Q, U and V
observed by the interferometer, vectorised over angle, frequency and time: $\bm{T}_{I}$, $\bm{T}_{Q}$, $\bm{T}_{U}$ and $\bm{T}_{V}$, respectively. To do this, we define an interferometric fringe matrix, with elements $\mathbfss{F}_{\mathrm{fr}, jk}  = \gamma \mathrm{e}^{ -2\pi i (\bm{u} \otimes \bm{l} + \bm{v} \otimes \bm{m} + \bm{w} \otimes \bm{n})_{jk}}$, which encodes the transformation from the image domain to the $uvw$-coordinates sampled by the interferometric array, and polarised primary beam matrices, $\mathbfss{P}_{\mathrm{nn},I}$, $\mathbfss{P}_{\mathrm{nn},Q}$, $\mathbfss{P}_{\mathrm{nn},U}$ and $\mathbfss{P}_{\mathrm{nn},V}$, which encode the polarisation, frequency and direction-dependent primary beam profiles (see \autoref{Eq:PolarisedPrimaryBeamMatrixElements}). 

To convert the contribution to the visibilities from diffuse and point source emission to a common unit, when $\mathbfss{F}_\mathrm{fr}$ operates on a vector of sky brightness temperatures in $\mathrm{K}$, we perform the visibility unit conversion between $\mathrm{K~sr}$ and $\mathrm{Jy}$ via the Rayleigh-Jeans law, which yields: $\gamma = 2\times10^{26}\nu^2k_\mathrm{B}\Delta\Omega/c^2$, with $\Delta\Omega$ the image domain pixel area. When $\mathbfss{F}_\mathrm{fr}$ operates on a vector of source flux densities in $\mathrm{Jy}$, $\gamma = 1$.

By writing $\bm{V}^\mathrm{true}$ in terms of these matrices as, 
\begin{equation}
\label{Eq:VtrueExpanded}
\bm{V}^\mathrm{true} = \mathbfss{F}_\mathrm{fr}[\mathbfss{P}_{\mathrm{nn},I}\bm{T}_{I} + \mathbfss{P}_{\mathrm{nn},Q}\bm{T}_{Q} + \mathbfss{P}_{\mathrm{nn},U}\bm{T}_{U} + \mathbfss{P}_{\mathrm{nn},V}\bm{T}_{V}] \ ,
\end{equation}
we can rewrite \autoref{Eq:MEqSingePolShorthand},
\begin{equation}
\label{Eq:MEqSingePolShorthandExpanded}
\bm{V}^\mathrm{obs} = \bm{n} + \mathbfss{G}\mathbfss{F}_\mathrm{fr}[\mathbfss{P}_{\mathrm{nn},I}\bm{T}_{I} + \mathbfss{P}_{\mathrm{nn},Q}\bm{T}_{Q} + \mathbfss{P}_{\mathrm{nn},U}\bm{T}_{U} + \mathbfss{P}_{\mathrm{nn},V}\bm{T}_{V}] \ .
\end{equation}
\autoref{Eq:MEqSingePolShorthandExpanded} is a generic discretised form of the expression for the interferometric visibilities formed by correlating voltages induced in south-north oriented feeds, within the approximations associated with \autoref{Eq:MEqSingePol}. 

As described in \autoref{Sec:RedundantCalibration}, when an array has redundancy in its antenna layout, such that one has visibilities derived from the cross-correlation of the voltage responses from sets of antennas with identical baseline separations and beam patterns, the antenna gain parameters encoded in $\mathbfss{G}$ can be reparametrised in terms of a set of redundant gain parameters encoded in $\mathbfss{H}$ and a set of degeneracy parameters encoded in $\mathbfss{D}$. With this reparametrisation, we can rewrite \autoref{Eq:MEqSingePolShorthandExpanded} as, 
\begin{equation}
\label{Eq:MEqSingePolShorthandExpandedRedundant}
\bm{V}^\mathrm{obs} = \bm{n} + \mathbfss{H}\mathbfss{D}\mathbfss{F}_\mathrm{fr}[\mathbfss{P}_{\mathrm{nn},I}\bm{T}_{I} + \mathbfss{P}_{\mathrm{nn},Q}\bm{T}_{Q} + \mathbfss{P}_{\mathrm{nn},U}\bm{T}_{U} + \mathbfss{P}_{\mathrm{nn},V}\bm{T}_{V}] \ .
\end{equation}

When applying redundant and absolute calibration with an incomplete sky model, \autoref{Eq:PureSkyVisModel} is used to model $\bm{V}^\mathrm{true}$. However, the incompleteness of the sky model imparts spurious spectral structure to the recovered gain solutions that prevents robust separation of the foregrounds and 21 cm signal in the data via their intrinsically distinct spectral structures.

To minimise or eliminate this spurious spectral structure imparted to the model gains during calibration, we introduce an additional model for the contribution to the measured visibilities from emission not accounted for in $\bm{V}^\mathrm{sim}$, which we denote $\bm{V}^\mathrm{fit} (\sTheta)$, with $\sTheta$ a set of parameters of the fitted model. In this case, we can write our model visibilites as the sum of these components,
\begin{equation}
\label{Eq:SkyPlusFittedVisModel}
\bm{V}^\mathrm{model} = \hat{\mathbfss{H}}^\mathrm{m}\mathbfss{D}^\mathrm{m}(\bm{V}^\mathrm{sim} + \bm{V}^\mathrm{fit}(\sTheta)) \ .
\end{equation}
Here, for succinctness, we have written $\bm{V}^\mathrm{model}$ in terms of the visibilites $\bm{V}^\mathrm{sim}$ and $\bm{V}^\mathrm{fit}(\sTheta)$, rather than using an explicit image domain representation similar to \autoref{Eq:VtrueExpanded}. In principle, $\sTheta$, could encode parameters defined in the visibility domain; however, in \autoref{Sec:FittedImageDomainSkyModel} we will describe why, rather than taking this approach, it is advantageous to define $\sTheta$ in the image domain and then propagate the parameters through an instrumental forward model to the visibilities. In that case, $\bm{V}^\mathrm{fit}(\sTheta)$ will take on a form similar to \autoref{Eq:VtrueExpanded} but with the polarised sky brightness distributions replaced with a fitted image-cube, parameterised by $\sTheta$, which models the component of the sky brightness distribution that has been incompletely or imperfectly modelled in $\bm{V}^\mathrm{sim}$.

\subsection{Bayesian inference}
\label{Sec:BayesianInference}

Bayesian inference provides a consistent approach to estimate a set of parameters, $\sTheta$, of a model, $M$, given a set of data, $\bm{D}$, and, through the use of the Bayesian evidence $\mathrm{Pr}(\bm{D}\vert M_{i})\equiv\mathcal{Z}_{i}$, to select from a set of models the ones that best describe the data. Bayes' theorem states that,
\begin{equation}
\label{Eq:BayesEqn}
\mathrm{Pr}(\sTheta\vert\bm{D},M) = \dfrac{\mathrm{Pr}(\bm{D}\vert\sTheta,M)\ \mathrm{Pr}(\sTheta\vert M)}{\mathrm{\mathrm{Pr}}(\bm{D}\vert M)} = \dfrac{\mathcal{L}(\sTheta)\pi(\sTheta)}{\mathcal{Z}} \ , 
\end{equation}
where $\mathrm{Pr}(\sTheta\vert\bm{D},M)$ is the posterior probability distribution of the parameters, $\mathrm{Pr}(\bm{D}\vert\sTheta,M) \equiv \mathcal{L}(\sTheta)$ is the likelihood and $\mathrm{Pr}(\sTheta\vert M) \equiv \pi(\sTheta)$ is the prior probability distribution of the parameters.

The Bayesian evidence (the factor required to normalise the posterior over the parameters), is given by,
\begin{equation}
\label{Eq:Evidence}
\mathcal{Z}=\int\mathcal{L}(\sTheta)\pi(\sTheta)\mr{d}^{n}\sTheta \ ,
\end{equation}
where $n$ is the dimensionality of the parameter space. Comparison of the evidence for different models enables a statistically robust selection of a preferred model for the data. As the average of the likelihood over the prior, the evidence is larger for a model if more of its parameter space is likely and smaller for a model where large areas of its parameter space have low likelihood values, even if the likelihood function is very highly peaked.  Thus, the evidence automatically implements Occam's razor: a simpler theory with a compact parameter space will have a larger evidence than a more complicated one, unless the latter is significantly better at explaining the data.

\subsection{Instrumental uncertainties}
\label{Sec:InstrumentalUncertainties}

In the remainder of \autoref{Sec:BayesCal}, we derive the posterior probability distribution for calibration and sky-model parameters assuming uncertainty-free a priori knowledge of the voltage beam patterns and positions of the antennas that define the interferometric fringe and polarised primary beam matrices in Equations \ref{Eq:VtrueExpanded} -- \ref{Eq:SkyPlusFittedVisModel}. In practice, these quantities are known with finite precision. At the cost of increased dimensionality, these model uncertainties can also be accounted for in the \textsc{BayesCal} framework by introducing a set of model parameters that describe perturbations to the instrument forward model and which can be constrained by priors derived from the measurement errors corresponding to the precision with which the instrument has been characterised. These instrumental parameters can then be jointly sampled from during calibration in order to address this potential source of bias. We leave further exploration of this to future work.

\subsection{Fitted image domain sky model}
\label{Sec:FittedImageDomainSkyModel}

Since the incompleteness of our simulated visibilities, $\bm{V}^\mathrm{sim}$, derives from unaccounted-for emission in the image domain, this provides a natural space in which to define the parameters of our fitted visibility model, $\bm{V}^\mathrm{fit}$. Furthermore, defining our model parameters in the image domain and forward modelling to the visibilities ensures that $\bm{V}^\mathrm{fit}$ is automatically restricted to a model space in which the model visibilities are correlated, via the instrumental forward model, in a manner that best reflects our expectation of the instrumental correlations in the data.

In \autoref{Sec:FittedStokesISkyModel}, we begin by describing our model for the Stokes I brightness temperature contribution to $\bm{V}^\mathrm{true}$ that is unaccounted for in $\bm{V}^\mathrm{sim}$. We describe how we adapt this model to account for incompleteness in the remaining Stokes parameters in \autoref{Sec:FittedStokesQUVSkyModel}, and describe the resulting data likelihood for the model in \autoref{Sec:DataLikelihood}.

\subsubsection{Fitted Stokes $I$ sky model}
\label{Sec:FittedStokesISkyModel}

We parameterise our model for the Stokes I brightness temperature of the component of emission in $\bm{V}^\mathrm{true}$ that is unaccounted for in $\bm{V}^\mathrm{sim}$ using a set of amplitude parameters, $\bm{\varepsilon}$, defined on a \textsc{HEALPix} grid (\citealt{2005ApJ...622..759G}) at a reference frequency, $\nu_{0}$. The $N_\mathrm{pix}$ pixels of the \textsc{HEALPix} grid cover the full sky; however, the number of pixels relevant to calibration of a single short duration integration is at most $\sim N_\mathrm{pix}/2$ per polarisation, since sky emission below the horizon will not contribute to the measured visibilities\footnote{Ionospheric refraction can extend the radio horizon of the antenna below the geometric horizon (e.g. \citealt{2014MNRAS.437.1056V}); however, for antennas with narrow or moderate fields-of-view, emission at large zenith angles, including that between the radio and geometric horizon, is greatly downweighted by the beam and its contribution to the visibilities is small.}. Furthermore, if the primary beam of a visibility downweights emission at large zenith angles, a subset of the pixels within the observer's horizon may be sufficient to model the visibilities. To take advantage of this and reduce the computational cost of the calibration when applying \textsc{BayesCal}, we define our set of parameters, $\bm{\varepsilon}$, such that they represent the amplitudes in the $N_\mathrm{pix, s}$ length subset of pixels defined by the union of pixels that fall within a fixed zenith angle, $\theta_\mathrm{cut}$, such that, within this region, the primary beam lies above a given threshold weight, at each central local sidereal time (LST) of the integrations over which the calibration solutions are being jointly estimated.

To construct a model image cube for the contribution from sky emission missing in $\bm{V}^\mathrm{sim}$ from our \textsc{HEALPix} grid amplitudes, one must define a model for the expected spectral structure of the emission, $f_{\nu}$. 

The spectral structure of the dominant astrophysical emission components at radio frequencies is well described by power laws, thus, making a power law-based model a natural choice in this frequency range. However, the diffuse synchrotron, free-free emission and point source synchrotron emission components that dominate the sky brightness distribution at these frequencies are characterised by distinct mean temperature spectral indices $\beta$, falling in the approximate range $2 \lesssim \beta \lesssim 3$. In principle, this could motivate one to jointly estimate multiple spectral components, with each component tailored to a specific astrophysical foreground. In practice, a more complex spectral model will lead to increased correlation between $\bm{V}^\mathrm{fit}$ and our gain parameters; thus, a compromise between model accuracy and recovered gain parameter uncertainties is necessary\footnote{In practice, this will be dependent on the relative brightness and complexity of the foregrounds. This, in turn, will be dependent on a range of variables, including the angular scales probed by the instrument, which will affect the relative dominance of the different foreground components, as well as the frequency range of the observations and the choice of calibration-field, which will both influence the total brightness of the foregrounds and, correspondingly, the precision with which they must be modelled.}.

If we restrict our focus to emission on large angular scales probed by short baselines, Galactic diffuse synchroton emission (GDSE) accounts for a large fraction of the power in the visibilities and, owing to the level of uncertainty associated with models for this emission, it is also the dominant source of model incompleteness in fields covered by high sensitivity point-source surveys in the frequency range relevant to 21 cosmology during CD and the EoR ($\nu \lesssim 200~\mathrm{MHz}$),  such as GLEAM (\citealt{2015PASA...32...25W, 2017MNRAS.464.1146H}) and LoTSS (\citealt{2019A&A...622A...1S}). In paper II it will be shown that, when calibrating data in this regime, a single power law with power law index $\beta_\mathrm{m} = \langle \beta \rangle_\mathrm{GDSE}$, with ${\langle \beta \rangle_\mathrm{GDSE}}$ the mean spectral index of the a priori known component of the GDSE brightness temperature distribution, is sufficient to suppress spurious spectral fluctuations in the calibration solutions recovered with \textsc{BayesCal} in low and moderate incompleteness calibration model scenarios to a level where power in foreground systematics is significantly subdominant relative to a fiducial EoR signal.

To construct our Stokes $I$ vectorised model image cube at a fixed LST, $t_{i}$,  we define an $N_{\mathrm{pix, s}, t_{i}}N_{\nu} \times N_{\mathrm{pix, s}, t_{i}}$ matrix, $\mathbfss{S}_{t_{i}}$, encoding one's choice of spectral model\footnote{In \autoref{Eq:FittedImageCubeSingleLST}, the spectrum along each line of sight is characterised by a shape specified by $\mathbfss{S}_{t_{i}}$ and reference amplitude set by $\bm{\varepsilon}_{t_{i}}$. If one wishes to fit for $n>1$ spectral components, each associated with their own characteristic reference amplitude, $\mathbfss{S}_{t_{i}}$ will be replaced by a block diagonal matrix in which each block describes the spectral structure of the $j$th emission component, where $j$ runs from 1 to $n$ and $\bm{\varepsilon}_{t_{i}}$ is replaced with a concatenation over reference amplitude vectors, where the $j$th reference amplitude vector encodes the reference amplitudes of the $j$th spectral component.} and apply this matrix to the subset of pixel amplitude parameters, with $\theta(t_{i}) \le \theta_\mathrm{cut}$, used to construct our visibilities,
\begin{equation}
\label{Eq:FittedImageCubeSingleLST}
\bm{T}_{t_{i}}^\mathrm{model} =  \mathbfss{S}_{t_{i}}\bm{\varepsilon}_{t_{i}} \ .
\end{equation}
Here, $\bm{\varepsilon}_{t_{i}}$ are the length $N_{\mathrm{pix, s}, t_{i}}$ subset of image-domain amplitude parameters, $\bm{\varepsilon}$, that fall within the zenith angle $\theta_\mathrm{cut}$ in the $i$th time integration.
Correspondingly, we derive the contribution from $\bm{T}_{t_{i}}^\mathrm{model}$ to our fitted visibility model at time $t_{i}$, as,
\begin{equation}
\label{Eq:FittedVisibilitiesSingleLST}
\bm{V}^\mathrm{fit}_{t_{i}} = \mathbfss{F}_{\mathrm{fr}, t_{i}}\mathbfss{P}_{\mathrm{nn},I, t_i}\mathbfss{S}_{t_{i}}\bm{\varepsilon}_{t_{i}}\ .
\end{equation}
We define our total vectorised visibility model over the $n$ LSTs in our data set as the concatenation of the visibility vectors defined at the central LSTs of the integrations comprising the data set: $\bm{V}^\mathrm{fit} = ((\bm{V}^\mathrm{fit}_{t_{0}})^{T}, (\bm{V}^\mathrm{fit}_{t_{1}})^{T}, \ldots (\bm{V}^\mathrm{fit}_{t_{n}})^{T})^{T}$. To construct $\bm{V}^\mathrm{fit}$, we define block diagonal matrices $\mathbfss{F}_\mathrm{fr}$, $\mathbfss{P}_{\mathrm{nn},I}$, and $\mathbfss{S}$, each comprised of matrix blocks given by $\mathbfss{F}_{\mathrm{fr}, t_{i}}$, $\mathbfss{P}_{\mathrm{nn},I, t_i}$, and $\mathbfss{S}_{t_{i}}$, respectively, where $i$ runs from 0 to $n$, and we replace $\bm{\varepsilon}_{t_{i}}$ with the concatenation of image domain parameters over the $n$ LSTs the calibration solutions are being jointly estimated over: $\bm{\varepsilon}_\mathrm{concat} = ((\bm{\varepsilon}_{t_{0}})^{T}, (\bm{\varepsilon}_{t_{1}})^{T}, \ldots (\bm{\varepsilon}_{t_{n}})^{T})^{T}$. 

In general, one wishes to calibrate interferometric data comprised of a number of narrowly spaced successive time integrations, in which case, the region of sky contributing to successive integrations is strongly correlated in time. $\bm{\varepsilon}$ represents the amplitudes of fixed LST-independent regions of the sky; thus, to derive $\bm{\varepsilon}_\mathrm{concat}$, we need to operate on $\bm{\varepsilon}$ with a matrix that selects the relevant parameter values from $\bm{\varepsilon}$ at each LST and concatenates them,
\begin{equation}
\label{Eq:GriddedImageParams}
\bm{\varepsilon}_\mathrm{concat} = \mathbfss{C}\bm{\varepsilon}\ .
\end{equation}
Here, $\mathbfss{C}$ is an $\sum_{i=1}^{N_\mathrm{t}} N_{\mathrm{pix, s}, t_{i}} \times N_\mathrm{pix, s}$ block matrix comprised of $N_\mathrm{t}$ blocks,
\begin{equation}
\label{Eq:GriddingMatrix}
\mathbfss{C} =
\begin{pmatrix}
\mathbfss{C}^{0}  \\
\mathbfss{C}^{1}  \\
\vdots  \\
\mathbfss{C}^{N_\mathrm{t}}
\end{pmatrix} \ ,
\end{equation}
where the $i$th block, $\mathbfss{C}^{i}$ has shape $N_{\mathrm{pix, s}, t_{i}} \times N_\mathrm{pix, s}$ and is comprised of $N_{\mathrm{pix, s}, t_{i}}$ row vectors, one for each element of $\bm{\varepsilon}$ for which $\Theta^\mathrm{za}_{in} \le \theta_\mathrm{cut}$, and the $j$th row vector, corresponding to the $n$th element of $\bm{\varepsilon}$, has the form,
\begin{equation}
\label{Eq:GriddingMatrixBlockRow}
\mathbfss{C}^{i}_{jk} = \delta_{kn}
\ .
\end{equation}
Here, $\Theta^\mathrm{za}_{in}$ is the zenith angle of the \textsc{HEALPix} pixel associated with amplitude parameter $\varepsilon_{n}$ at LST $t_{i}$.

With these definitions, we can write the Stokes $I$ contribution to our fitted visibility model over all LSTs of our calibration data set as,
\begin{equation}
\label{Eq:FittedVisibilitiesFullModel}
\bm{V}^\mathrm{fit} = \mathbfss{F}_\mathrm{fr}\mathbfss{P}_{\mathrm{nn},I}\mathbfss{S}\mathbfss{C}\bm{\varepsilon}\ .
\end{equation}

\subsubsection{Fitted Stokes $Q,U,V$ sky models}
\label{Sec:FittedStokesQUVSkyModel}

In principle, one could individually account for model incompleteness on a per-polarisation basis by jointly fitting for the Stokes $Q$, $U$ and $V$ contributions to our fitted visibility model over all LSTs of our calibration data set in an analogous manner to \autoref{Eq:FittedVisibilitiesFullModel}. However, for a fixed spectral structure model over Stokes parameters, the contributions of the Stokes sky models to a single instrumental visibility correlation and time are degenerate.

This degeneracy can be broken if one places additional constraints on the model, for example by fitting to the four instrumental correlations of the polarised visibility vector, rather than a single instrumental correlation, as considered here, or through the differing temporal evolution of $\bm{V}_{j}^\mathrm{fit}$, with $j \in [I,Q,U,V]$, when one calibrates a sufficiently long duration data set.

Alternatively, one can make use of the degeneracy between the polarised sky-model parameters to reduce the number of sky-model parameters required to accurately model incompleteness in the intensity distribution used to construct $\bm{V}^\mathrm{fit}$ when calibrating a short duration data set. This second approach greatly reduces the computational complexity associated with jointly sampling from instrumental gains and the parameters of $\bm{V}^\mathrm{fit}$ but has the disadvantage that $\bm{V}^\mathrm{fit}$ will not model the temporal evolution of the contribution to $\bm{V}^\mathrm{sim}$ of missing Stokes $Q$, $U$ and $V$ brightness (if present). However, at low radio frequencies relevant to 21 cm cosmology, the sky brightness and fractional contribution to sky-model incompleteness is dominated by Stokes $I$ (Stokes $V$ is negligible and linearly polarised emission has been found to account for $\mathcal{O}(1\%)$ of the emission, e.g. \citealt{2016ApJ...830...38L}) and, thus, the sky-model incompleteness incurred by neglecting explicit modelling of Stokes $Q$, $U$ and $V$ in $\bm{V}^\mathrm{fit}$ is expected to be small relative to the total sky-model incompleteness. As such, here, we consider the case of jointly estimating the gain solutions and sky-model incompleteness as modelled by \autoref{Eq:FittedVisibilitiesFullModel}. We plan to address full Stokes modelling of sky-model incompleteness in more detail in future work.

\subsubsection{Data likelihood}
\label{Sec:DataLikelihood}

Our likelihood at this stage becomes, 
\begin{multline}
\label{Eq:BayesCalVisLike}
\mathrm{Pr}(\bm{V}^\mathrm{obs} \;|\; \bm{A}, \bm{\mathit{\Phi}}_{l}, \bm{\mathit{\Phi}}_{m}, \bm{\varepsilon}) =\\ \frac{1}{\pi^{N_{\mathrm{vis}}}\mathrm{det}(\mathbfss{N})} \exp\left[-\left(\bm{V}^\mathrm{obs} - \hat{\mathbfss{H}}^\mathrm{m}\mathbfss{D}^\mathrm{m}(\bm{V}^\mathrm{sim} + \bm{V}^\mathrm{fit})\right)^\dagger \right.\\ \left. \mathbfss{N}^{-1} 
 \left(\bm{V}^\mathrm{obs} - \hat{\mathbfss{H}}^\mathrm{m}\mathbfss{D}^\mathrm{m}(\bm{V}^\mathrm{sim} + \bm{V}^\mathrm{fit})\right)\right] \ .
 \end{multline}
In standard sky-based and redundancy-based calibration (sections \ref{Sec:SkyBasedCalibration} and \ref{Sec:RedundantCalibration}), the gain solutions at each frequency can be solved for independently. In the \textsc{BayesCal} framework, the introduction of the fitted frequency dependent visibility model, $\bm{V}^\mathrm{fit}$, means the instrumental gains are now correlated in frequency and, correspondingly, the gain parameters of each channel of our data set are fit for jointly. Thus, we redefine our model degenerate gain parameter matrix in the context of \textsc{BayesCal}, such that, for a single time integration, $\mathbfss{D}^\mathrm{m}$ is a block diagonal matrix where each block, $\mathbfss{D}^\mathrm{m}_{r}$, is a diagonal matrix with elements $D_{r, ij} = \delta_{ij}A_{r}\mr{e}^{i\bm{b}_{i}^{T} \sPhi_{r}}$ and the subscript $r$ runs over the $N_\nu$ channels of the data set (a generalisation of $\mathbfss{D}^\mathrm{m}$ to simultaneous calibration of multiple times is described in \autoref{Sec:TemporalModelling}). We also redefine $\hat{\mathbfss{H}}^\mathrm{m}$ in an analogous manner to now encode the maximum likelihood redundant gain parameters for each channel of the data set.

For calibrated data, the noise level on a visibility resulting from a pair of identical antennas individually experiencing equal system noise is given by (e.g. \citealt{1999ASPC..180.....T}),
\begin{equation}
\label{Eq:VisabilityNoise}
\sigma_{V}=\dfrac{1}{\eta_{s}\eta_{a}}\dfrac{2k_{B}T_{\mathrm{sys}}}{A\sqrt{2\Delta\nu\tau}} \ ,
\end{equation}
where $k_{B}=1.3806\times10^{-23}~\mathrm{JK^{-1}}$ is Boltzmann's constant, $T_{\mathrm{sys}}$ is the system noise temperature, $A$ is the antenna area, $\Delta\nu$ is the channel width, $\tau$ is the integration time and $\eta_{s}$ and $\eta_{a}$ are the system and antenna efficiencies, respectively.

To take advantage of \autoref{Eq:VisabilityNoise}, we can rewrite the covariance matrix as,
\begin{equation}
\label{Eq:CovarianceCalVis}
\mathbfss{N} = (\mathbfss{H}\mathbfss{D})^{\dagger}\mathbfss{N}^{\prime}(\mathbfss{H}\mathbfss{D}) \ .
\end{equation}
Here, we have assumed that the noise on the raw visibilities, in the frequency range of interest, is dominated by components whose amplitude scales with antenna gain. This includes sky noise, spillover from ground noise and receiver noise injected prior to or by the first low noise amplifier (LNA) in the signal chain\footnote{In the case that noise introduced into the signal chain after the first low noise amplifier is significant relative to the amplified noise injected prior to or by the first LNA, \autoref{Eq:CovarianceCalVis} can be generalised to include a second component to model this additional noise contribution. In this case, the appropriate covariance matrix is of the form $\mathbfss{N} = (\mathbfss{H}\mathbfss{D})^{\dagger}\mathbfss{N}^{\prime}(\mathbfss{H}\mathbfss{D}) + \mathbfss{N}^{\prime}_\mathrm{rec}$, where $\mathbfss{N}^{\prime}_\mathrm{rec}$ describes the noise deriving from electronics on the signal path between the first LNA and the correlator. This can be either calculated or estimated jointly with the calibration and sky-model parameters.}. $\mathbfss{N}^{\prime}$ is the covariance matrix of the noise on the calibrated data. The diagonal elements of $\mathbfss{N}^{\prime}$ are given by,
\begin{equation}
\label{Eq:CalibratedCovarianceElements}
N_{jj}^{\prime} = \left< n_j n_j^{*}\right> = \sigma_{j}^{2} \ ,
\end{equation}
where $\left< .. \right>$ represents the expectation value and $\sigma_{j}$ is the rms value of the noise term for visibility $j$, as given by \autoref{Eq:VisabilityNoise}. In the weak source limit, when the ground and instrument noise temperature exceeds the sky noise temperature, $\mathbfss{N}^{\prime}$ is well approximated as being diagonal. In the strong source limit, self-noise becomes an important consideration. In this case, the off-diagonal elements are non-zero, with the specific form of the expected covariance between two baselines a function of the source distribution and dependent on whether or not the baselines share an antenna (e.g. \citealt{1989AJ.....98.1112K}).

\subsection{Temporal priors}
\label{Sec:TemporalModelling}

In addition to redefining the model degenerate gain parameter matrix to account for the spectral correlation between the calibration parameters in the general case described in \autoref{Sec:FittedImageDomainSkyModel}, simultaneous calibration of multiple time integrations enables one to take advantage of a priori knowledge constraining the temporal evolution of the gain solutions. To account for this, a full model degenerate gain parameter matrix, for all time integrations in the data set being calibrated, must be defined. In this case, $\mathbfss{D}^\mathrm{m}$ is a block diagonal matrix where each block, $\mathbfss{D}^\mathrm{m}_{s}$, is itself a block diagonal matrix comprised of blocks, $\mathbfss{D}^\mathrm{m}_{r}$, which are diagonal matrices with elements $D_{s,r, ij} = \delta_{ij}A_{s,r}\mr{e}^{i\bm{b}_{i}^{T} \sPhi_{s,r}}$, where the subscripts $r$ and $s$ run over the $N_\nu$ channels and $N_\mathrm{t}$ integrations in the data set, respectively. $\hat{\mathbfss{H}}^\mathrm{m}$ will also be redefined analogously to now encode the maximum likelihood redundant gain parameters for all integrations in the data set.

When multiple time integrations are calibrated simultaneously, the total duration over which calibration solutions are simultaneously estimated has important implications. Variations in the environmental conditions of the receiver and signal chain, such as changing ambient temperature, will impart temporal fluctuations in the antenna gains on the temporal scales of those variations. On short time scales relative to the characteristic timescales of those variations, the calibration solutions are expected to be relatively stable. However, on longer time scales, this approximation will be less accurate. To address this, a model can be introduced for the temporal evolution of the calibration solutions (e.g. \citealt{2021MNRAS.506.4578G}). The parameters of this model can be constrained by their own priors. In this case one can write,
\begin{align}
\label{Eq:TemporalModel}
A_{s,r} &= f_{A}(\sTheta_{A,\mathrm{t}})_{s,r} \\ \nonumber
\sPhi_{s,r} &= f_{\sPhi}(\sTheta_{\sPhi,\mathrm{t}})_{s,r} \ ,
\end{align}
where $\sTheta_{A,\mathrm{t}}$ and $\sTheta_{\sPhi,\mathrm{t}}$ are the parameters of our temporal model for the degenerate amplitudes and tip-tilt phases. Defining $\sTheta_\mathrm{t} = [\sTheta_{A,\mathrm{t}}, \sTheta_{\sPhi,\mathrm{t}}]^{T}$ as the full set of parameters of our temporal model for the gain solutions, one now samples from the joint probability density of the image-space model coefficients and $\sTheta_\mathrm{t}$,
\begin{multline}
\label{Eq:ProbCalTemporal}
\mathrm{Pr}(\bm{\varepsilon}, \sTheta_\mathrm{t} \;|\; \bm{V}^\mathrm{obs}) \; \propto \; \mathrm{Pr}(\bm{V}^\mathrm{obs} \;|\; \bm{\varepsilon}, \sTheta_\mathrm{t}) \; \mathrm{Pr}(\bm{\varepsilon}) \; \mathrm{Pr}(\sTheta_\mathrm{t}) \ ,
\end{multline}
where $\mathrm{Pr}(\bm{\varepsilon})$ and $\mathrm{Pr}(\sTheta_\mathrm{t})$ are our priors on the sky-model parameters and the parameters of our temporal model for the gain solutions, respectively.

If one wished to jointly estimate the parameters of $\mathbfss{H}^\mathrm{m}$ and $\mathbfss{D}^\mathrm{m}$, or if applying the \textsc{BayesCal} framework to sky-based calibration (see \autoref{Sec:BayesCalForSkyBasedCalibration}), temporal priors would be applied to $\mathbfss{H}^\mathrm{m}$ or $\mathbfss{G}^\mathrm{m}$, respectively, in an analogous manner.

\subsection{Spectral model complexity}
\label{Sec:SpectralModelComplexity}

In the general case, when the optimal spectral model for the incomplete component of the calibration model is not known a priori, rather than using, for example, a single power law with a fixed power law index, as proposed for use on short baselines in \autoref{Sec:FittedStokesISkyModel}, the power law index of the model can be fit for as a free parameter in the analysis.

Additionally, as alluded to in \autoref{Sec:FittedImageDomainSkyModel}, an arbitrarily  more complex foreground model, defined by a set of foreground parameters $\sTheta_\mathrm{f}$, could be substituted for this spectral model. In this case, the fitted visibility model can be written as,
\begin{equation}
\label{Eq:FittedVisibilitiesFullModelGeneral}
\bm{V}^\mathrm{fit} = \mathbfss{F}_\mathrm{fr}\mathbfss{P}\mathbfss{S}(\sTheta_\mathrm{f})\mathbfss{C}\bm{\varepsilon}\ .
\end{equation}
The joint probability density of the image-space amplitude coefficients, spectral model parameters and calibration parameters is then sampled from, and the posterior probability is given by,
\begin{multline}
\label{Eq:ProbCalSkySpectrum}
\mathrm{Pr}(\bm{A}, \bm{\mathit{\Phi}}_{l}, \bm{\mathit{\Phi}}_{m}, \bm{\varepsilon}, \sTheta_\mathrm{f} \;|\; \bm{V}^\mathrm{obs}) \; \propto \\
\; \mathrm{Pr}(\bm{V}^\mathrm{obs} \;|\; \bm{A}, \bm{\mathit{\Phi}}_{l}, \bm{\mathit{\Phi}}_{m}, \bm{\varepsilon}, \sTheta_\mathrm{f}) \; \mathrm{Pr}(\bm{A}) \; \mathrm{Pr}(\bm{\mathit{\Phi}}_{l}) \; \mathrm{Pr}(\bm{\mathit{\Phi}}_{m}) \; \mathrm{Pr}(\bm{\varepsilon}) \; \mathrm{Pr}(\sTheta_\mathrm{f}) \ ,
\end{multline}
where $\mathrm{Pr}(\sTheta_\mathrm{f})$ is a prior on the foreground parameters\footnote{For a power law spectral model, such as that used in this paper, a Gaussian prior on the spectral index, informed by GDSE measurements in the FoV being calibrated, could be used. To calibrate visibilities with a significant spread in baseline lengths, such that GDSE and point source components of the emission are each dominant in subsets of the data, it may be preferable to use a more complex spectral model (a double power law, for example) constrained by priors derived from measurements of the respective emission distributions.} and we have not assumed any a priori correlation between $\bm{A}$, $\bm{\mathit{\Phi}}_{l}$, $\bm{\mathit{\Phi}}_{m}$, $\bm{\varepsilon}$ and $\sTheta_\mathrm{f}$.

In this case, the Bayesian evidence can be used to determine whether such a model is preferred and to select an optimal model, from a set of models for the data (see \citet{2020MNRAS.492...22S} for an example of this approach applied to foreground modelling in global 21 cm experiments).

\subsection{Fitted calibration model power priors}
\label{Sec:PowerSpectralPriors}

When fitting \autoref{Eq:BayesCalVisLike}, we aim to recover solutions of the form 
$\hat{\mathbfss{H}}^\mathrm{m}\mathbfss{D}^\mathrm{m}(\bm{V}^\mathrm{sim} + \bm{V}^\mathrm{fit}(\bm{\varepsilon})) = \mathbfss{H}\mathbfss{D}\bm{V}^\mathrm{true}$, with
$\bm{V}^\mathrm{fit} + \bm{V}^\mathrm{sim} = \bm{V}^\mathrm{true}$ and $\hat{\mathbfss{H}}^\mathrm{m}\mathbfss{D}^\mathrm{m} = \mathbfss{H}\mathbfss{D}$.
However, without additional constraints, correlation between the calibration parameters and $\bm{V}^\mathrm{fit}$ limits the confidence with which the calibration parameters can be recovered. 

The impact of correlation between calibration parameters and $\bm{V}^\mathrm{fit}$ can be seen most simply when the fixed simulated component of the calibration model is excluded (setting $\bm{V}^\mathrm{sim} = 0$
in \autoref{Eq:SkyPlusFittedVisModel}). In this case, the desired solution to \autoref{Eq:BayesCalVisLike} is of the form 
$\hat{\mathbfss{H}}^\mathrm{m}\mathbfss{D}^\mathrm{m} \bm{V}^\mathrm{fit} = \mathbfss{H}\mathbfss{D}\bm{V}^\mathrm{true}$, with $\bm{V}^\mathrm{fit} = \bm{V}^\mathrm{true}$ and $\hat{\mathbfss{H}}^\mathrm{m}\mathbfss{D}^\mathrm{m} = \mathbfss{H}\mathbfss{D}$; however, this is just one of a degenerate set of solutions with the form $\hat{\mathbfss{H}}^\mathrm{m}\mathbfss{D}^{\mathrm{m} \prime} \bm{V}^\mathrm{fit \prime} = \mathbfss{H}\mathbfss{D}\bm{V}^\mathrm{true}$ with $\hat{\mathbfss{H}}^\mathrm{m}\mathbfss{D}^{\mathrm{m} \prime} \ne \mathbfss{H}\mathbfss{D}$, where $\mathbfss{D}^{\mathrm{m} \prime} = \mathbfss{D}^{\mathrm{m}}\mathbfss{D}^{\prime}$ and  $\bm{V}^{\mathrm{fit} \prime} = \bm{V}^\mathrm{fit}(\bm{\varepsilon}^{\prime})) =  (\mathbfss{D}^{\prime})^{-1}\bm{V}^\mathrm{fit}(\bm{\varepsilon}))$.
Here, $\mathbfss{D}^{\prime}$ is a diagonal degeneracy matrix which for a single frequency channel and time integration has elements,
\begin{equation}
\label{Eq:DVfitCorrelation2}
D^{\prime}_{ij} = \delta_{ij}A^{\prime}\mr{e}^{i\bm{u}_{i}^{T} \sPhi^{\prime}} \ .
\end{equation}
Unlike the degeneracy matrices associated with relative calibration for which the degenerate gain amplitude and phase can, in principle, have arbitrarily complex spectral structure, the degeneracy described by $\mathbfss{D}^{\prime}$ is present only for values of $A^{\prime}$ and $\sPhi^{\prime}$ with spectral structure that is modellable with $\bm{V}^\mathrm{fit}$ (which, in turn, is a function of the chromatic structure associated with the fringe, primary beam and intrinsic spectral structure matrices from which $\bm{V}^\mathrm{fit}$ is constructed). 

If one includes a non-zero simulated component of the calibration model, in the limit that the spectral structure of $\bm{V}^\mathrm{sim}$ and $\bm{V}^\mathrm{fit}$ is identical and one limits $\bm{V}^\mathrm{sim}$ to modelling diffuse emission that is exactly describable using $\bm{V}^\mathrm{fit}$, one can rewrite $\bm{V}^\mathrm{fit}$ as,
\begin{equation}
\label{Eq:DVfitCorrelation1}
\bm{V}^\mathrm{fit} = -\bm{V}^\mathrm{sim} + \bm{V}^{\mathrm{fit} \prime} \ ,
\end{equation}
where $\bm{V}^\mathrm{fit \prime} = \mathbfss{F}_\mathrm{fr}\mathbfss{P}\mathbfss{S}(\sTheta_\mathrm{f})\mathbfss{C}\bm{\varepsilon}^{\prime}$.
Substituting \autoref{Eq:DVfitCorrelation1} into \autoref{Eq:SkyPlusFittedVisModel}, one recovers a data model, $\bm{V}^\mathrm{model} = \hat{\mathbfss{H}}^\mathrm{m}\mathbfss{D}^\mathrm{m}\bm{V}^{\mathrm{fit} \prime}$, with the same form as in the degenerate case with $\bm{V}^\mathrm{sim} = 0$ described above, and a degeneracy of the same form is present.

In the absence of noise, this degeneracy is broken in the more realistic scenario that the spectral structures of $\bm{V}^\mathrm{sim}$ and $\bm{V}^\mathrm{fit}$ are similar but not identical, in which case \autoref{Eq:DVfitCorrelation1} no longer holds exactly. However, it persists when the  spectral structure of $\bm{V}^\mathrm{sim}$ and $\bm{V}^\mathrm{fit}$ are sufficiently similar to construct $\bm{V}^\mathrm{fit}$ models that can destructively interfere with $\bm{V}^\mathrm{sim}$ to a residual level comparable or smaller than the noise level.

Solutions to \autoref{Eq:BayesCalVisLike} with fitted visibilities either approximately or exactly described by \autoref{Eq:DVfitCorrelation1} require that the power in $\bm{V}^\mathrm{fit \prime}$ exceeds that in $\bm{V}^\mathrm{sim}$ and greatly exceeds the expected power in $\bm{V}^\mathrm{fit}$, which for a reasonably complete and accurate simulated calibration model will be a sub-dominant component of the total power in $\bm{V}^\mathrm{model}$. Thus, models of $\bm{V}^\mathrm{fit}$ of this form, and the corresponding degeneracy in solutions to \autoref{Eq:BayesCalVisLike} that they introduce, are possible only if a priori knowledge of the expected power in the calibration model components is neglected\footnote{In practice, we expect $\bm{V}^\mathrm{sim}$ to be a sufficiently good model of the true visibilites to account for the majority of the power in the calibration visibility model. However, if this were not the case, this degeneracy can further be suppressed by placing priors on the gain parameters (see \autoref{Sec:DegenerateGainModelParametrisation}). In \textsc{BayesCal}, we combine both priors, thereby strengthening the preference for high fidelity calibration solutions of the form $\bm{V}^\mathrm{fit} + \bm{V}^\mathrm{sim} = \bm{V}^\mathrm{true}$ and $\hat{\mathbfss{H}}^\mathrm{m}\mathbfss{D}^\mathrm{m} = \mathbfss{H}\mathbfss{D}$.}, and they can be ruled out by incorporating in the calibration framework one's a priori knowledge of the expected level of power in $\bm{\varepsilon}$ or $\bm{V}^\mathrm{fit}$.

While the specific brightness distribution of the sky emission modelled by $\bm{\varepsilon}$ in the region of sky contributing to the visibilities to be calibrated is unknown (otherwise we would simply incorporate it into $\bm{V}^\mathrm{sim}$), the less stringent requirement that we have an estimate of the power in this emission is more easily met. Such an estimate can be made in the $uv$- or image-domain. In the former case, a prior on the expected power in the fitted calibration model can be formulated in terms of the two-dimensional spatial power spectrum of $\bm{V}^\mathrm{fit}$. In the later, a prior can be formulated in terms of the per-element squared amplitude of $\bm{\varepsilon}$ at a particular angular scale, or as the variance of $\bm{\varepsilon}$, in the limit that the temperature distribution in the incomplete component of the sky model is reasonably approximated as zero-mean and homoscedastic. The former approach has the benefit of encoding the expected spatial dependence of the power in $\bm{\varepsilon}$; however, if one wishes to calibrate a dataset with sparse $uv$-coverage, it places no constraint on fitted sky models in the null-space of the fitted visibility model. In contrast, constraining the per-element squared amplitude of $\bm{\varepsilon}$ at a particular angular scale constrains all choices of fitted sky-model brightness distribution; additionally, it provides a simple means to encode in the prior heteroscedastic uncertainty estimates, which are likely to be appropriate for diffuse emission models. In paper II, compared to placing a prior on the two-dimensional spatial power spectrum of $\bm{V}^\mathrm{fit}$, we find that a prior on the per-element squared amplitude of $\bm{\varepsilon}$ yields improved stability of the matrix inversion associated with marginalising out $\bm{\varepsilon}$ to sample directly from the posterior for the calibration parameters. However, for generality, here we consider both approaches.

\subsubsection{Power spectral priors}
\label{Sec:SubsecPSP}

The expected power in emission missing from $\bm{V}^\mathrm{sim}$, and that we seek to model with $\bm{\varepsilon}$, can be divided into two components:
\begin{enumerate}
 \item power in uncatalogued point sources below the completeness limit of the available source catalogues used in the construction of $\bm{V}^\mathrm{sim}$,
 \item power in unmodelled or mismodelled diffuse emission, which derives from the discrepancy between our model for the full-Stokes diffuse emission and the true full-Stokes brightness distribution.
\end{enumerate}

Assuming a uniform uncertainty of the diffuse emission model in the field being calibrated, the two-dimensional spatial power spectrum of this emission can be reasonably approximated as proportional to the two-dimensional spatial power spectrum of the a priori known component of the diffuse emission,
\begin{equation}
\label{Eq:DiffusePower}
\bm{P}_{uv,\delta T} = f^{2} \bm{P}_{uv,T} \ ,
\end{equation}
with $f$ the fractional uncertainty on our diffuse emission model. Here, $\bm{P}_{uv,T}$ can be estimated as the variance of the diffuse component of $\bm{V}^\mathrm{sim}$, binned over the baseline lengths in the array. 

While the second component incorporates the faint unresolved sea of point sources included in the first, when calibrating using intermediate and long baseline lengths, which resolve out large spatial scale diffuse emission, it is useful to be able to separately account for the power in point sources with flux densities $S<S_\mathrm{min}$, described in the first category, which are the dominant source of model incompleteness on these baseline lengths. Here, $S_\mathrm{min}$ is the minimum flux density of sources that we include in $\bm{V}^\mathrm{sim}$. In the limit of Poisson distributed point sources, this is given by (e.g. \citealt{2002ApJ...564..576D}),
\begin{equation}
\label{Eq:PoissonPointSourcePower}
\bm{P}_{uv,\mathrm{ps},\mathrm{Poisson}} = \int_{0}^{S_\mathrm{min}} S^{2}\dfrac{\mathrm{d}N}{\mathrm{d}S}\mathrm{d}S \ ,
\end{equation}
with $\mathrm{d}N/\mathrm{d}S$ the differential source count distribution.
For clustered sources, assuming a power-law
angular correlation function, $w(\theta)=(\theta/\theta_{0})^{-\beta}$,  with Legendre transform $\widetilde{w}$, the power spectrum has an additional contribution of the form (e.g., \citealt{1999A&A...346....1S}),
\begin{equation}
\label{Eq:ClusteredPointSourcePower}
\bm{P}_{uv,\mathrm{ps},\mathrm{clustered}} = \widetilde{w}\left[\int_{0}^{S_\mathrm{min}} S\dfrac{\mathrm{d}N}{\mathrm{d}S}\mathrm{d}S \right]^2 \ .
\end{equation}

To construct a prior on the two-dimensional spatial power spectrum of $\bm{V}^\mathrm{fit}$, assuming the $i$th visibility is an element of redundant baseline group $\alpha$, we define a covariance matrix $\bm{\sSigma}_{\delta T}$ of $\bm{V}^\mathrm{fit}$ with elements,
\begin{equation}
\label{Eq:PowerPrior}
\sSigma_{uv, \delta T, ij} = f^{2}\left< (V^\mathrm{sim}_{\alpha}) 
(V^\mathrm{sim}_{\beta})^{*} \right> = P_{uv,\delta T,j}\delta_{ij} \ .
\end{equation}
In addition to constraining the expected variance of $\bm{V}^\mathrm{fit}$ as a function of spatial scale, $\bm{\varepsilon}$ can additionally be constrained by encoding in $\bm{\sSigma}_{uv, \delta T}$ the expected covariance structure of the visibilites, when known. However, constraining the variance alone is sufficient to break the degeneracy described at the start of this section. 

Using \autoref{Eq:FittedVisibilitiesFullModel}, we can thus write our power spectral prior as,
\begin{multline}
\label{Eq:ProbObsPSofEpsilon}
\mathrm{Pr}(\bm{\varepsilon} \;|\; \bm{P}_{uv,\delta T}) \; \propto \; \frac{1}{\mathrm{det}(\bm{\sSigma}_{uv, \delta T})} \\
\times \exp\left[-~\bm{\varepsilon}^{T}\mathbfss{C}^{\dagger}\mathbfss{S}^{\dagger}\mathbfss{P}^{\dagger}(\mathbfss{F}_\mathrm{fr})^{\dagger}\bm{\sSigma}_{uv, \delta T}^{-1}\mathbfss{F}_\mathrm{fr}\mathbfss{P}\mathbfss{S}\mathbfss{C}\bm{\varepsilon}\right] \ .
\end{multline}
Incorporating this power spectral prior on the parameters of our fitted visibility model, we can write the joint probability density of our calibration parameters and our image-space model coefficients conditioned on their two-dimensional spatial power spectrum and the raw data as,
\begin{multline}
\label{Eq:ProbCalSkyGivenPS}
\mathrm{Pr}(\bm{A}, \bm{\mathit{\Phi}}_{l}, \bm{\mathit{\Phi}}_{m}, \bm{\varepsilon} \;|\; \bm{V}^\mathrm{obs}, \bm{P}_{uv,\delta T}) \; \propto
\; \mathrm{Pr}(\bm{V}^\mathrm{obs} \;|\; \bm{A}, \bm{\mathit{\Phi}}_{l}, \bm{\mathit{\Phi}}_{m}, \bm{\varepsilon}) \\ \times \mathrm{Pr}(\bm{\varepsilon} \;|\; \bm{P}_{uv,\delta T}) \; \mathrm{Pr}(\bm{A}) \; \mathrm{Pr}(\bm{\mathit{\Phi}}_{l}) \; \mathrm{Pr}(\bm{\mathit{\Phi}}_{m}) \ . 
\end{multline}
Here, we have assumed $\bm{P}_{uv,\delta T}$ is known a priori. However, in a scenario where the uncertainty on $f$ was assumed to be significant, rather than using a fixed estimate of the two-dimensional spatial power spectrum, one could instead jointly sample from a parametrised model for the power in $\bm{V}^\mathrm{fit}$, our image-space model coefficients and the calibration parameters as,
\begin{multline}
\label{Eq:ProbCalSkyPS}
\mathrm{Pr}(\bm{A}, \bm{\mathit{\Phi}}_{l}, \bm{\mathit{\Phi}}_{m}, \bm{\varepsilon}, \bm{P}_{uv,\delta T} \;|\; \bm{V}^\mathrm{obs}) \; \propto
\; \mathrm{Pr}(\bm{V}^\mathrm{obs} \;|\; \bm{A}, \bm{\mathit{\Phi}}_{l}, \bm{\mathit{\Phi}}_{m}, \bm{\varepsilon}) \\ \times \mathrm{Pr}(\bm{\varepsilon} \;|\; \bm{P}_{uv,\delta T}) \; \mathrm{Pr}(\bm{P}_{uv,\delta T}) \; \mathrm{Pr}(\bm{A}) \; \mathrm{Pr}(\bm{\mathit{\Phi}}_{l}) \; \mathrm{Pr}(\bm{\mathit{\Phi}}_{m}) \ , 
\end{multline}
where $\mathrm{Pr}(\bm{P}_{uv,\delta T})$ is our prior on $\bm{P}_{uv,\delta T}$ (see \citet{2016MNRAS.462.3069S} for an application of this approach in the context of estimating the three dimensional power spectrum of redshifted 21 cm emission from interferometric observations).

\subsubsection{Image-domain power priors}
\label{Sec:SubsecIDPP}

An image-domain prior on the expected power in the fitted calibration model can be formulated in a similar manner to that described in the previous section but now, rather than constraining the covariance structure of the visibilities that derive from $\bm{\varepsilon}$, one instead constrains the covariance structure of $\bm{\varepsilon}$ directly. In this case, one can define a Gaussian prior on $\bm{\varepsilon}$ of the from,
\begin{multline}
\label{Eq:ProbObsPSofEpsilonIm}
\mathrm{Pr}(\bm{\varepsilon} \;|\; \bm{\sSigma}_{lm, \delta T}) \; \propto \; \frac{1}{\sqrt{\mathrm{det}(\bm{\sSigma}_{lm, \delta T})}} \exp\left[-\frac{1}{2}~\bm{\varepsilon}^{T}\bm{\sSigma}_{lm, \delta T}^{-1}\bm{\varepsilon}\right] \ .
\end{multline}
Here, $\bm{\sSigma}_{lm, \delta T}$ is the expected covariance matrix of $\bm{\varepsilon}$. Incorporating this prior, we can write the joint probability density of our calibration parameters and our image-space model coefficients conditioned on our estimate of the covariance matrix of $\bm{\varepsilon}$ and the raw data as,
\begin{multline}
\label{Eq:ProbCalSkyGivenPSepsilon}
\mathrm{Pr}(\bm{A}, \bm{\mathit{\Phi}}_{l}, \bm{\mathit{\Phi}}_{m}, \bm{\varepsilon} \;|\; \bm{V}^\mathrm{obs}, \bm{\sSigma}_{lm, \delta T}) \; \propto
\; \mathrm{Pr}(\bm{V}^\mathrm{obs} \;|\; \bm{A}, \bm{\mathit{\Phi}}_{l}, \bm{\mathit{\Phi}}_{m}, \bm{\varepsilon}) \\ 
\times \mathrm{Pr}(\bm{\varepsilon} \;|\; \bm{\sSigma}_{lm, \delta T}) \; \mathrm{Pr}(\bm{A}) \; \mathrm{Pr}(\bm{\mathit{\Phi}}_{l}) \; \mathrm{Pr}(\bm{\mathit{\Phi}}_{m}) \ . 
\end{multline}
\autoref{Eq:ProbCalSkyGivenPSepsilon} assumes $\bm{\sSigma}_{lm, \delta T}$ is known a priori. In principle, $\bm{\sSigma}_{lm, \delta T}$ can be estimated when producing radio sky-maps and would enable a highly informative prior on $\bm{\varepsilon}$. In practice, more approximate uncertainty estimates, such as the fractional uncertainty or estimated RMS error in a map, are more typical. The commonly used Global Sky Model (GSM; e.g.  \citealt{2008MNRAS.388..247D, 2017MNRAS.464.3486Z}) derived from publicly available, total power, large-area radio surveys does not currently provide error estimates but an updated version of the model including per-pixel error estimates is in preparation (A. Liu, private communication).

In the absence of a full covariance matrix of the diffuse emission model, one can use a diagonal approximation to $\bm{\sSigma}_{lm, \delta T}$ with diagonal elements,
\begin{equation}
\label{Eq:PowerPrior}
\sSigma_{lm, \delta T, ii} = \sigma_{\varepsilon, i}^{2} \ .
\end{equation}
Here, $\sigma_{\varepsilon, i}$ is the estimated uncertainty in pixel $i$ of the sky model used to construct $\bm{V}^\mathrm{sim}$. If only the average error is available, this can be further simplifed to $\sSigma_{lm, \delta T, ii} = \sigma_{\varepsilon}^{2}$, where $\sigma_{\varepsilon}$ is the expected RMS error on the sky model used to construct $\bm{V}^\mathrm{sim}$. While these approximations are less informative, in paper II we will show that, in the cases considered, constraining the average power in the incomplete or uncertain component of the calibration sky model alone is sufficient to enable recovery of high fidelity calibration solutions with the \textsc{BayesCal} calibration framework. In this case, we can write the joint probability density of our calibration parameters and our image-space model coefficients conditioned on the average power in the incomplete or uncertain component of the calibration sky model and the raw data as,
\begin{multline}
\label{Eq:ProbCalSkySigma2epsilon}
\mathrm{Pr}(\bm{A}, \bm{\mathit{\Phi}}_{l}, \bm{\mathit{\Phi}}_{m}, \bm{\varepsilon} \;|\; \bm{V}^\mathrm{obs}, \sigma_{\varepsilon}^{2}) \; \propto
\; \mathrm{Pr}(\bm{V}^\mathrm{obs} \;|\; \bm{A}, \bm{\mathit{\Phi}}_{l}, \bm{\mathit{\Phi}}_{m}, \bm{\varepsilon}) \\ 
\times \mathrm{Pr}(\bm{\varepsilon} \;|\; \sigma_{\varepsilon}^{2}) \; \mathrm{Pr}(\bm{A}) \; \mathrm{Pr}(\bm{\mathit{\Phi}}_{l}) \; \mathrm{Pr}(\bm{\mathit{\Phi}}_{m}) \ . 
\end{multline}

In the context of calibration exclusively on long baselines for which the contribution from diffuse emission is small, $\sigma_{\varepsilon}^2$ can be replaced with the estimated power in faint point sources missing from $\bm{V}^\mathrm{sim}$ given in \autoref{Sec:SubsecPSP}.

In a similar manner to that described in \autoref{Sec:SubsecPSP}, in a scenario where there is significant uncertainty on the level of power in the incomplete or uncertain component of the calibration sky model, rather than using a fixed estimate, one could instead jointly sample from a parametrised model for the power in the incomplete or uncertain component of the calibration sky model, our image-space model coefficients and the calibration parameters. In the limit that we place our prior on the average power in the incomplete or uncertain component of the calibration sky model, the corresponding posterior is given by,
\begin{multline}
\label{Eq:ProbCalSkyGivenSigma2epsilon}
\mathrm{Pr}(\bm{A}, \bm{\mathit{\Phi}}_{l}, \bm{\mathit{\Phi}}_{m}, \bm{\varepsilon}, \sigma_{\varepsilon}^{2} \;|\; \bm{V}^\mathrm{obs}) \; \propto
\; \mathrm{Pr}(\bm{V}^\mathrm{obs} \;|\; \bm{A}, \bm{\mathit{\Phi}}_{l}, \bm{\mathit{\Phi}}_{m}, \bm{\varepsilon}) \\ \times \mathrm{Pr}(\bm{\varepsilon} \;|\; \sigma_{\varepsilon}^{2}) \; \mathrm{Pr}(\sigma_{\varepsilon}^{2}) \; \mathrm{Pr}(\bm{A}) \; \mathrm{Pr}(\bm{\mathit{\Phi}}_{l}) \; \mathrm{Pr}(\bm{\mathit{\Phi}}_{m}) \ .
\end{multline}
Here $\mathrm{Pr}(\sigma_{\varepsilon}^{2})$ is our prior on $\sigma_{\varepsilon}^{2}$.

\subsection{Degenerate gain model parametrisation and spectral smoothness priors}
\label{Sec:DegenerateGainModelParametrisation}

In addition to sky-based and redundant calibration, which use an astrophysical sky model to constrain the antenna gain solutions exclusively or in part, respectively, the antenna gains can also be constrained using alternate techniques that are independent of an astrophysical sky model. Techniques in this category include:
\begin{enumerate}
 \item calibration with an artificial sky signal generated by a satellite (e.g. \citealt{2016ApJ...826..199N}) or drone (e.g. \citealt{2017PASP..129c5002J}), 
 \item electric and electromagnetic co-simulations of the receiver system (e.g. \citealt{2021MNRAS.500.1232F}),
 \item reflectometry measurements of the feed-dish system (e.g. \citealt{2018ExA....45..177P}).
 \item three-position-switch calibration of individual antennas (e.g. \citealt{2017ApJ...835...49M, 2020arXiv201114052R}; Murray et al. in prep.).
\end{enumerate}

While current applications of the first three of these techniques may not have sufficient accuracy to calibrate interferometric arrays for precision 21 cm cosmology applications, they can, nevertheless, provide valuable constraints on the expected level of fluctuations associated with the antenna gain amplitudes (see e.g. \citealt{2018ExA....45..177P, 2021MNRAS.500.1232F}). We are not aware of published applications of the fourth technique to interferometric calibrations, but future constraints on the calibration parameters derived from three-position-switch calibration can be incorporated in \textsc{BayesCal}, in a similar manner to the first three, as described below.

In general, to leverage this a priori knowledge, one can incorporate these constraints as spectral priors on the gain solutions. For absolute calibration of a redundantly calibrated data set, this corresponds to placing priors on the spectral dependence of the redundant gain degeneracy function $f_\mathrm{vis}(A, \Phi_{l}, \Phi_{m})$. In this work, we focus on placing priors on the power in amplitude fluctuations of the  redundant gain degeneracy function; however, one could also place priors on the tip and tilt phases of the redundant gain degeneracy function in an analogous manner, given an equivalent set of constraints on these parameters. 

The power spectrum of amplitude fluctuations of the redundant gain degeneracy function $P_{A}(\eta)$ (with $\eta$ the Fourier conjugate to frequency $\nu$) is given by the Fourier transform of the redundant gain two-point correlation function,  
\begin{equation}
\label{Eq:GainAmplitudePS} 
\left\langle \widetilde{A}(\eta)\widetilde{A}^{*}(\eta^\prime) \right\rangle(\Delta\nu)^{2} \equiv \delta_\mathrm{D}(\eta-\eta^\prime)P_{A}(\eta) \ ,
\end{equation}
with $\widetilde{A}(\eta)$ the Fourier transform of $A(\nu)$ and $\delta_D$ the Dirac delta function. This follows from the Einstein--Wiener--Khinchin theorem, which states that the autocorrelation function of a wide-sense-stationary random process has a spectral decomposition given by the power spectrum of that process (e.g. \citealt{EWKtheorem}). Strictly speaking, the angular brackets denote an ensemble average; however, since we are estimating the power in amplitude fluctuations of the redundant gain degeneracy function specific to a given observing setup, we expect the power to be reasonably approximated as constant on short timescales (see \autoref{Sec:TemporalModelling}). Thus, we approximate the ensemble distribution as a delta function.

Given this, a Fourier parametrisation of the degenerate amplitudes provides a natural space to encode priors on the power in these gain fluctuations as a function of frequency. Although, here, we do not impose an informative prior on the distribution of the tip-tilt phases, for mathematical convenience we nevertheless apply the same reparametrisation to them. We, therefore, rewrite our degeneracy function parameters $\bm{A}(\nu)$, $\bm{\mathit{\Phi}}_{l}(\nu)$ and $\bm{\mathit{\Phi}}_{m}(\nu)$ using a new set of parameters that describe the amplitudes of their Fourier decompositions, $\bm{A}_\mathrm{F}(\eta)$, $\bm{\mathit{\Phi}}_{l,\mathrm{F}}(\eta)$ and $\bm{\mathit{\Phi}}_{m,\mathrm{F}}(\eta)$, with, 
\begin{align}
\label{Eq:DegeneracyFunctionFourierParameters}
\bm{A} &= \mathbfss{F}^{-1}\bm{A}_\mathrm{F} \\ \nonumber
\bm{\mathit{\Phi}}_{l} &= \mathbfss{F}^{-1}\bm{\mathit{\Phi}}_{l,\mathrm{F}} \\ \nonumber
\bm{\mathit{\Phi}}_{m} &= \mathbfss{F}^{-1}\bm{\mathit{\Phi}}_{m,\mathrm{F}} \ .
\end{align}
Here, $\mathbfss{F}^{-1}$ is a one dimensional inverse Fourier transform matrix mapping from the 'halfcomplex' representation\footnote{Here, we adopt the terminology used by \textsc{fftw3} (\citealt{FFTW3}), with 'halfcomplex' referring to the non-redundant half of the complex output for the one-dimensional discrete Fourier transform of the real frequency-domain gain parameters.} of the $\eta$-domain of the gain parameters to their frequency-domain representations, and $\bm{A}_\mathrm{F}$, $\bm{\mathit{\Phi}}_{l,\mathrm{F}}$ and $\bm{\mathit{\Phi}}_{m,\mathrm{F}}$ are $N_{\nu} \times 1$ real column vectors encoding the 'halfcomplex' representation of Fourier degenerate gain amplitude, tip- and tilt-phase parameters, respectively. This formulation enables one to sample from a $3\times N_{\nu}$ dimensional space of real numbers when calibrating the degenerate gain parameters and enforces that $\bm{A}$, $\bm{\mathit{\Phi}}_{l}$ and $\bm{\mathit{\Phi}}_{m}$ are real for arbitrary, real $\bm{A}_\mathrm{F}$, $\bm{\mathit{\Phi}}_{l,\mathrm{F}}$ and $\bm{\mathit{\Phi}}_{m,\mathrm{F}}$. When converting between the degenerate gain parameters and their Fourier transforms, we assume the \textsc{numpy} (e.g. \citealt{2020Natur.585..357H}) `backwards' discrete Fourier transform normalisation convention, such that the inverse transform differs from the forward transform by the sign of the exponential argument and by a $1/n_{\eta}$ normalisation prefactor, and $\mathbfss{F}^{-1}\mathbfss{F} = \mathbfss{I}$ with $\mathbfss{I}$ the identity matrix. 

We assume a uniform prior on the parameters defining the Fourier decomposition of the tip-tilt phases such that $\mathrm{Pr}(\mathit{\Phi}_{l,\mathrm{F}, j}) = 1/(\mathit{\Phi}_{l,\mathrm{F}, j,\mathrm{max}}-\mathit{\Phi}_{l,\mathrm{F}, j,\mathrm{min}}), \mathrm{Pr}(\mathit{\Phi}_{m,\mathrm{F}, j})= 1/(\mathit{\Phi}_{m,\mathrm{F}, j,\mathrm{max}}-\mathit{\Phi}_{m,\mathrm{F}, j,\mathrm{min}})$, where $j$ indexes over the $N_\nu$ tip-tilt phases and $\mathit{\Phi}_{l,\mathrm{F}, j,\mathrm{min}}$, $\mathit{\Phi}_{m,\mathrm{F}, j,\mathrm{min}}$ and $\mathit{\Phi}_{l,\mathrm{F}, j,\mathrm{max}}$, $\mathit{\Phi}_{m,\mathrm{F}, j,\mathrm{max}}$ are the minimum and maximum values of the $j$th Fourier tip phase and tilt phases, respectively, and we impose a Gaussian prior encoding the power in amplitude fluctuations of the redundant gain degeneracy function as,
\begin{multline}
\label{Eq:ProbAgivenSigmaA}
\mathrm{Pr}(\bm{A}_\mathrm{F} \;|\; \bm{\sigma}_{A_\mathrm{F}}^{2}) \; \propto \; \frac{1}{\sqrt{\mathrm{det}(\bm{\sSigma}_{A_\mathrm{F}})}} \\
\times \exp\left[-\frac{1}{2}(\bm{A}_\mathrm{F} -\bar{\bm{A}}_\mathrm{F})^{T}\bm{\sSigma}_{A_\mathrm{F}}^{-1}(\bm{A}_\mathrm{F} -\bar{\bm{A}}_\mathrm{F})\right] \ .
\end{multline}
Here, $\bar{\bm{A}}_\mathrm{F}$ is a vector of amplitudes defining the expectation value of the Fourier decomposition of $\bm{A}$, $\bm{\sigma}_{A_\mathrm{F}}^{2}$ is the vector of expected variances of $\bm{A}_\mathrm{F}$ and $\bm{\sSigma}_{A_\mathrm{F}}$ is a diagonal matrix with elements $\sSigma_{A_\mathrm{F},ij} = \delta_{ij}\sigma_{A_\mathrm{F}}^{2}$.

Designing antennas with minimal spectral structure in their gains is one way in which existing 21 cm cosmology experiments aim to mitigate spurious spectral structure in calibration solutions derived from sky-based or redundant calibration with an incomplete sky model. This approach leverages the fact that, if the true level of the fluctuating spectral structure in the antenna gains is known a priori to be much smaller than the ratio of the power between the foregrounds and 21 cm signal, fitting a simple flat bandpass will be sufficient to obtain calibration solutions that do not impart statistically significant foreground systematics.  

However, in practice, a range of effects, including dish-feed reflections, micro-reflections along the cable connecting the antenna to the correlator, and reflections at the ends of those cables due to imperfect impedance matching, make meeting this exacting specification challenging (e.g. \citealt{2018ExA....45..177P, 2021MNRAS.500.1232F}). If this specification is not met, fitting a smooth bandpass solution that is unable to model the true gain fluctuations in the bandpass will introduce similar systematic effects, in analyses trying to isolate the 21 cm signal from the foregrounds with approaches relying on the spectral smoothness of the foregrounds in the data, to those introduced when fitting the data with the product of a general gain model and an incomplete sky model.

At this stage, the joint probability density of our image-space model coefficients and our Fourier space calibration parameters, conditioned on the raw data, the expected power in the fitted calibration model and the power in fluctuations in the degenerate gain amplitude, is given by,
\begin{multline}
\label{Eq:ProbCalSkyGivenPSVNA}
\mathrm{Pr}(\bm{A}_\mathrm{F}, \bm{\mathit{\Phi}}_{l,\mathrm{F}}, \bm{\mathit{\Phi}}_{m,\mathrm{F}}, \bm{\varepsilon} \;|\; \bm{V}^\mathrm{obs}, \sigma_{\varepsilon}^{2}, \bm{\sigma}_{A_\mathrm{F}}^{2}) \; \propto \\
\; \mathrm{Pr}(\bm{V}^\mathrm{obs} \;|\; \bm{A}_\mathrm{F}, \bm{\mathit{\Phi}}_{l,\mathrm{F}}, \bm{\mathit{\Phi}}_{m,\mathrm{F}}, \bm{\varepsilon}) \; \mathrm{Pr}(\bm{\varepsilon} \;|\; \sigma_{\varepsilon}^{2}) \;  \\
\times \mathrm{Pr}(\bm{A}_\mathrm{F} \;|\; \bm{\sigma}_{A_\mathrm{F}}^{2}) \; \mathrm{Pr}(\bm{\mathit{\Phi}}_{l,\mathrm{F}}) \; \mathrm{Pr}(\bm{\mathit{\Phi}}_{m,\mathrm{F}}) \ ,
\end{multline}
where $\mathrm{Pr}(\bm{V}^\mathrm{obs} \;|\; \bm{A}_\mathrm{F}, \bm{\mathit{\Phi}}_{l,\mathrm{F}}, \bm{\mathit{\Phi}}_{m,\mathrm{F}}, \bm{\varepsilon})$ is given by \autoref{Eq:BayesCalVisLike} reparametrised in terms of the Fourier gain parameters of \autoref{Eq:DegeneracyFunctionFourierParameters}. Here, we have assumed $\bm{\sigma}_{A_\mathrm{F}}^{2}$ is known a priori. If the uncertainties associated with estimates of $\bm{\sigma}_{A_\mathrm{F}}^{2}$ are significant, one can, instead, jointly sample from $\bm{\sigma}_{A_\mathrm{F}}$ with $\bm{A}_\mathrm{F}$, $\bm{\mathit{\Phi}}_{l,\mathrm{F}}$, $\bm{\mathit{\Phi}}_{m,\mathrm{F}}$ and $\bm{\varepsilon}$, in an analogous manner to that outlined for the high power spectral uncertainty regime at the end of \autoref{Sec:PowerSpectralPriors}. In that case, the joint probability density of our image-space model coefficients, Fourier space calibration parameters and power in fluctuations in the degenerate gain amplitude, conditioned on the raw data and the expected power in the fitted calibration model, is given by,
\begin{multline}
\label{Eq:ProbCalSkyVNAGivenPS}
\mathrm{Pr}(\bm{A}_\mathrm{F}, \bm{\mathit{\Phi}}_{l,\mathrm{F}}, \bm{\mathit{\Phi}}_{m,\mathrm{F}}, \bm{\varepsilon}, \bm{\sigma}_{A_\mathrm{F}}^{2} \;|\; \bm{V}^\mathrm{obs}, \sigma_{\varepsilon}^{2}) \; \propto \\
\; \mathrm{Pr}(\bm{V}^\mathrm{obs} \;|\; \bm{A}_\mathrm{F}, \bm{\mathit{\Phi}}_{l,\mathrm{F}}, \bm{\mathit{\Phi}}_{m,\mathrm{F}}, \bm{\varepsilon}) \; \mathrm{Pr}(\bm{\varepsilon} \;|\; \sigma_{\varepsilon}^{2}) \;  \\
\times \mathrm{Pr}(\bm{A}_\mathrm{F} \;|\; \bm{\sigma}_{A_\mathrm{F}}^{2}) \; \mathrm{Pr}(\bm{\sigma}_{A_\mathrm{F}}^{2}) \; \mathrm{Pr}(\bm{\mathit{\Phi}}_{l,\mathrm{F}}) \; \mathrm{Pr}(\bm{\mathit{\Phi}}_{m,\mathrm{F}}) \ ,
\end{multline}
where $\mathrm{Pr}(\bm{\sigma}_{A_\mathrm{F}}^{2})$ is our prior on $\bm{\sigma}_{A_\mathrm{F}}^{2}$.

\subsection{Analytic marginalisation over the sky-model parameters}
\label{Sec:AnalyticMarginalisationOverTheSkyModelParameters}

In principle, one could sample from $\mathrm{Pr}(\bm{A}_\mathrm{F}, \bm{\mathit{\Phi}}_{l,\mathrm{F}}, \bm{\mathit{\Phi}}_{m,\mathrm{F}}, \bm{\varepsilon} \;|\; \bm{V}^\mathrm{obs}, \sigma_{\varepsilon}^{2}, \bm{\sigma}_{A_\mathrm{F}}^{2})$ in \autoref{Eq:ProbCalSkyGivenPSVNA} in the case of redundant calibration, or an equivalent set of parameters for sky-based calibration (see \autoref{Sec:SkyBasedCalibration}), using a sampling algorithm that is efficient in a high dimensional space, such as Hamiltonian Monte Carlo (e.g. \citealt{2008MNRAS.389.1284T}). Then, the marginal distribution for the calibration parameters can be derived by numerical marginalisation over the sky-model parameters. However, by analytically marginalising over the sky-model parameters and directly sampling from the posterior probability distribution for the calibration parameters, one can greatly reduce the dimensionality of the sampling problem.

When calibrating a redundant interferometric array, in particular, where the majority of the calibration parameters can be derived using the constraints imposed by the requirement of internal consistency within redundant baseline groups, the total number of parameters to be sampled from can be dominated by the sky-model amplitude parameters.

In general, the number of sky-model parameters is proportional to the square of the longest baseline length of the visibilites being calibrated (which determines the image-domain angular scale resolved by the measurements), and the number of calibration parameters is proportional to the number of frequency channels to be jointly calibrated. In paper II, 10 channels are jointly calibrated, with each channel having three redundant calibration parameters. In contrast, the resolution of the baselines and the FoV of the sky model we use for calibration yield $\mathcal{O}(10^2)$ sky-model amplitude parameters. In this case, analytic marginalisation results in a greater than $90\%$ reduction in parameters. If longer baselines were calibrated, this fraction would be larger still. 

Analytically marginalising over these parameters significantly decreases the number of samples that must be drawn from the posterior for reliable inference. Depending on the specific choice of sampling algorithm, this can translate to a multiple order of magnitude reduction in the number of samples required to explore the joint posterior of the paramters (e.g. \citealt{2015MNRAS.453.4384H}). Additionally, it facilitates exploration of the parameter space with nested sampling techniques which yield the Bayesian model evidence as a primary output, thus providing a statistically robust means to optimize the calibration model for the data set and instrument under consideration. 

In order to perform the marginalisation over the sky-model amplitude parameters, we first simplify our notation by defining the residual visibility vector $\delta\bm{V}^\mathrm{obs} \equiv \bm{V}^\mathrm{obs} - \hat{\mathbfss{H}}^\mathrm{m}\mathbfss{D}^\mathrm{m}\bm{V}^\mathrm{sim}$, its weighted gridded projection on the parameter space of the fitted visibility model, $\overline{\delta \bm{V}}^\mathrm{obs} \equiv \bm{\upLambda}^{\dagger}\mathbfss{N}^{-1}\delta\bm{V}^\mathrm{obs}$, the system matrix mapping from our image domain parameters to the fitted model visibilities, $\bm{\upLambda} \equiv \hat{\mathbfss{H}}^\mathrm{m}\mathbfss{D}^\mathrm{m}\mathbfss{F}_\mathrm{fr}\mathbfss{P}\mathbfss{S}\mathbfss{C}$, and the covariance matrix of $\overline{\delta \bm{V}}^\mathrm{obs}$, which, assuming one defines their prior on the expected power in the fitted calibration model in the image domain, is given by $\bm{\upUpsilon} \equiv \bm{\upLambda}^{\dagger}\mathbfss{N}^{-1}\bm{\upLambda} + \frac{1}{2} \bm{\sSigma}_{lm, \delta T}^{-1}$ and, if instead it is defined in the $uv$-domain, is given by $\bm{\upUpsilon} \equiv \bm{\upLambda}^{\dagger}\mathbfss{N}^{-1}\bm{\upLambda} + \mathbfss{C}^{\dagger}\mathbfss{S}^{\dagger}\mathbfss{P}^{\dagger}(\mathbfss{F}_\mathrm{fr})^{\dagger}\bm{\sSigma}_{uv, \delta T}^{-1}\mathbfss{F}_\mathrm{fr}\mathbfss{P}\mathbfss{S}\mathbfss{C}$. 
In the remainder of this section we assume the prior on the expected power in the fitted calibration model is defined in the image domain; however, the derivation proceeds similarly if the prior on the expected power in the fitted calibration model is defined in the $uv$-domain. In that case, one simply substitutes: \begin{enumerate*} \item the appropriate definition of $\bm{\upUpsilon}$,  \item $\bm{\sSigma}_{lm, \delta T}$ for $\bm{\sSigma}_{uv, \delta T}$, and  \item the conditional parameters for expected power in the fitted calibration model defined in the image-domain for those defined in the $uv$-domain \end{enumerate*}.

Using the above definitions, we can write the log of the joint posterior in \autoref{Eq:ProbCalSkyGivenPSVNA} as,
\begin{multline}
\label{Eq:LogP}
\log(\mathrm{Pr}(\bm{A}_\mathrm{F}, \bm{\mathit{\Phi}}_{l,\mathrm{F}}, \bm{\mathit{\Phi}}_{m,\mathrm{F}}, \bm{\varepsilon} \;|\; \bm{V}^\mathrm{obs}, \sigma_{\varepsilon}^{2}, \bm{\sigma}_{A_\mathrm{F}}^{2})) = \\
-(\delta\bm{V}^\mathrm{obs})^{\dagger}\mathbfss{N}^{-1}(\delta\bm{V}^\mathrm{obs}) \\
 \; -\bm{\varepsilon}^T\bm{\upUpsilon}\bm{\varepsilon} + 2\bm{\varepsilon}^T\overline{\delta \bm{V}}^\mathrm{obs}
-\log(\mathrm{det}(\mathbfss{N})) -\frac{1}{2}\log(\mathrm{det}(\bm{\sSigma}_{lm, \delta T})) \\
-\frac{1}{2}\log(\mathrm{det}(\bm{\sSigma}_{A_\mathrm{F}})) -\frac{1}{2}(\bm{A}_\mathrm{F} -\bar{\bm{A}}_\mathrm{F})^{T}\bm{\sSigma}_{A_\mathrm{F}}^{-1}(\bm{A}_\mathrm{F} -\bar{\bm{A}}_\mathrm{F}) + k\ . 
\end{multline}
Here, $k$ is a normalisation factor. The posterior probability distribution of the parameters is independent of $k$ and, for brevity, we omit it going forward. 

Taking the derivative with respect to $\bm{\varepsilon}$ of $\log(\mathrm{Pr}(\bm{A}_\mathrm{F}, \bm{\mathit{\Phi}}_{l,\mathrm{F}}, \bm{\mathit{\Phi}}_{m,\mathrm{F}}, \bm{\varepsilon} \;|\; \bm{V}^\mathrm{obs}, \sigma_{\varepsilon}^{2}, \bm{\sigma}_{A_\mathrm{F}}^{2}))$ gives us, 
\begin{equation}
\label{Eq:Gradb}
\frac{\partial \log(\mathrm{Pr}(\bm{A}_\mathrm{F}, \bm{\mathit{\Phi}}_{l,\mathrm{F}}, \bm{\mathit{\Phi}}_{m,\mathrm{F}}, \bm{\varepsilon} \;|\; \bm{V}^\mathrm{obs}, \sigma_{\varepsilon}^{2}, \bm{\sigma}_{A_\mathrm{F}}^{2}))}{\partial \bm{\varepsilon}} =  -2\bm{\upUpsilon}\bm{\varepsilon} + 2\overline{\delta \bm{V}}^\mathrm{obs} \ ,
\end{equation}
which can be solved to give the maximum likelihood vector of coefficients $\hat{\bm{\varepsilon}}$,
\begin{equation}
\label{Eq:epsilonmax}
\hat{\bm{\varepsilon}} = \bm{\upUpsilon}^{-1}\overline{\delta \bm{V}}^\mathrm{obs} \ .
\end{equation}
Re-expressing \autoref{Eq:LogP} in terms of $\hat{\bm{\varepsilon}}$ yields,
\begin{multline}
\log(\mathrm{Pr}(\bm{A}_\mathrm{F}, \bm{\mathit{\Phi}}_{l,\mathrm{F}}, \bm{\mathit{\Phi}}_{m,\mathrm{F}}, \bm{\varepsilon} \;|\; \bm{V}^\mathrm{obs}, \sigma_{\varepsilon}^{2}, \bm{\sigma}_{A_\mathrm{F}}^{2})) = \\ 
-(\delta\bm{V}^\mathrm{obs})^{\dagger}\mathbfss{N}^{-1}(\delta\bm{V}^\mathrm{obs}) + \hat{\bm{\varepsilon}}^T\bm{\upUpsilon}\hat{\bm{\varepsilon}} -(\bm{\varepsilon} - \hat{\bm{\varepsilon}})^T\bm{\upUpsilon}(\bm{\varepsilon} - \hat{\bm{\varepsilon}}) \\
-\log(\mathrm{det}(\mathbfss{N})) -\frac{1}{2}\log(\mathrm{det}(\bm{\sSigma}_{lm, \delta T})) -\frac{1}{2}\log(\mathrm{det}(\bm{\sSigma}_{A_\mathrm{F}})) 
\\ -\frac{1}{2}(\bm{A}_\mathrm{F} -\bar{\bm{A}}_\mathrm{F})^{T}\bm{\sSigma}_{A_\mathrm{F}}^{-1}(\bm{A}_\mathrm{F} -\bar{\bm{A}}_\mathrm{F}) \ .
\end{multline}
The 3rd term in this expression can then be integrated with respect to the $N_\mathrm{pix, s}$ elements of $\bm{\varepsilon}$ to give,
\begin{eqnarray}
\label{Eq:RealGaussianIntegral}
I &=& \int_{-\infty}^{+\infty}\mathrm{d}\bm{\varepsilon}\exp\left[-(\bm{\varepsilon} - \hat{\bm{\varepsilon}})^T\bm{\upUpsilon}(\bm{\varepsilon} - \hat{\bm{\varepsilon}})\right] \nonumber \\
&=& (2\pi)^{-N_\mathrm{pix, s}/2}~\mathrm{det}(\bm{\upUpsilon})^{-\frac{1}{2}}.
\end{eqnarray}
Our marginalised probability distribution for the calibration parameters is thus given by,
\begin{multline}
\label{Eq:Margin}
\mathrm{Pr}(\bm{A}_\mathrm{F}, \bm{\mathit{\Phi}}_{l,\mathrm{F}}, \bm{\mathit{\Phi}}_{m,\mathrm{F}} \;|\; \bm{V}^\mathrm{obs}, \sigma_{\varepsilon}^{2}, \bm{\sigma}_{A_\mathrm{F}}^{2}) \propto \\
\frac{\mathrm{det}(\mathbfss{N})^{-1}}{\sqrt{\mathrm{det} (\bm{\upUpsilon}) ~ \mathrm{det}(\bm{\sSigma}_{lm, \delta T}) ~ \mathrm{det}(\bm{\sSigma}_{A_\mathrm{F}})}} \\
\times \exp \biggl[-(\delta\bm{V}^\mathrm{obs})^{\dagger}\mathbfss{N}^{-1}(\delta\bm{V}^\mathrm{obs}) + (\overline{\delta \bm{V}}^\mathrm{obs})^{T} \bm{\upUpsilon}^{-1}(\overline{\delta \bm{V}}^\mathrm{obs}) \\
 - \frac{1}{2}(\bm{A}_\mathrm{F} -\bar{\bm{A}}_\mathrm{F})^{T}\bm{\sSigma}_{A_\mathrm{F}}^{-1}(\bm{A}_\mathrm{F} -\bar{\bm{A}}_\mathrm{F} )\biggr].
\end{multline}

\subsection{Marginalising over uncertainties in point source flux densities}
\label{Sec:AnalyticMarginalisationOverPointSourceUncertainties}

Up to this point, we have assumed that the uncertainties on the catalogued point sources contributing to $\bm{V}^\mathrm{sim}$ are small, such that the contribution of residual source emission to the measured visibilities, relative to the contribution from unmodelled sources, is also small. In this case, the residuals can be simply subsumed into our fitted visibility model $\bm{V}^\mathrm{fit}$. 

However, if the uncertainties on the flux densities of the sources contributing to $\bm{V}^\mathrm{sim}$ are significant, one can make better use of one's a priori knowledge of the positions of the residual source emission by extending the image domain model to include an additional set of parameters, defined at the coordinates of the catalogued sources, which model the misestimated contribution to the visibilities associated with their uncertainties.

In this case, assuming there are $N_{\delta S}$ catalogued point sources with uncertain flux-densities contributing to $\bm{V}^\mathrm{sim}$, the existing diffuse sky-model parameters, $\bm{\varepsilon}$, can be augmented with an additional set of model parameters, $\bm{\varepsilon}_{\delta S}$, of length $N_{\delta S}$, which describe the difference between the true and catalogued flux densities of the sources included in $\bm{V}^\mathrm{fit}$. We construct our prior on the expected power in the fitted calibration model in the $uv$-domain by writing as $\bm{P}_{uv, \delta S}$ the two-dimensional spatial power associated with catalogued source uncertainties, as $\bm{\varepsilon}_{\delta T, \delta S}$ our augmented image-space parameter vector and as $\bm{P}_{uv,\delta T, \delta S} = \bm{P}_{uv,\delta T} + \bm{P}_{uv, \delta S}$ the total two-dimensional spatial power due to incompleteness and uncertainties associated with emission contributing to $\bm{V}^\mathrm{sim}$; thus, we can define an updated power spectral prior incorporating this contribution as,
\begin{multline}
\label{Eq:ProbObsPSofEpsilonPlusSources}
\mathrm{Pr}(\bm{\varepsilon}_{\delta T, \delta S} \;|\; \bm{P}_{uv,\delta T, \delta S}) \; \propto \; \frac{1}{\mathrm{det}(\bm{\sSigma}_{uv, \delta T, \delta S})} \\
\times \exp\left[-~\bm{\varepsilon}_{\delta T, \delta S}^{T}\bm{T}_{\delta T, \delta S}^{\dagger}\bm{\sSigma}_{uv, \delta T, \delta S}^{-1}\bm{T}_{\delta T, \delta S}\bm{\varepsilon}_{\delta T, \delta S}\right] \ .
\end{multline}
Here, $\bm{\sSigma}_{uv, \delta T, \delta S, ij} = \delta_{ij}P_{uv,\delta T, \delta S, j}$. $\bm{T}_{\delta T, \delta S} = [\bm{T}, \bm{T}_{\delta S}]$ is a block matrix with $\bm{T} = \mathbfss{F}_\mathrm{fr}\mathbfss{P}\mathbfss{S}\mathbfss{C}$, and  $\bm{T}_{\delta S} = \mathbfss{F}_\mathrm{fr, \delta S}^{-1}\mathbfss{P}_{\delta S}\mathbfss{S}_{\delta S}$. $\mathbfss{F}_\mathrm{fr, \delta S}^{-1}$ and $\mathbfss{P}_{\delta S}$ describe interferometric fringe and primary beam matrices, respectively, evaluated at the $lmn$ coordinates of the source residuals. $\mathbfss{S}_{\delta S}$ is a matrix encoding the expected spectral structure of the source residuals with elements $S_{\delta S,ij} = (\nu_{j}/\nu_{0})^{-\beta_{\delta S,j}}$, where $i$ runs over the set of source residuals being fit for, $\nu_{0}$ is a reference frequency and $\beta_{\delta S,j}$ is a model for the $j$th source spectral index.

Assuming the errors on the point source flux densities are Gaussian distributed\footnote{While it is reasonable to assume that the errors on the point source flux densities follow a Gaussian distribution, the approach described for deriving a prior for the point source contribution to the residuals is also applicable for sampling from an arbitrary set of point source flux-densities error distributions and, thus, it is not a requirement.}, with standard deviation $\sigma_{\delta S, j}$ on source $j$, a prior on $\bm{P}_{uv, \delta S}$ can be constructed computationally from samples of $\bm{T}_{\delta S} \delta \bm{S}$, where the $j$th element of $\delta \bm{S}$ is drawn from the   Gaussian distribution $N(0,\sigma^{2}_{\delta S, j})$ describing the amplitude of the flux-density residual on the $j$th cataloged point source. For a narrow prior on $\bm{P}_{uv, \delta S}$, such that the spread in power is small relative to the expected power, one can reasonably approximate it as fixed at its expectation value, in a similar manner to $\bm{\sSigma}_{lm, \delta T}$. More generally, one can instead jointly sample from the posterior for the calibration and sky-model parameters and the power spectrum, in an analogous manner to that described at the conclusion of \autoref{Sec:PowerSpectralPriors}.

In order to sample directly from the posterior probability distributions of the calibration parameters, one can analytically marginalise over these additional model parameters in a similar manner to that described in \autoref{Sec:AnalyticMarginalisationOverTheSkyModelParameters}. To achieve this, we update the vector and matrix definitions in \autoref{Sec:AnalyticMarginalisationOverTheSkyModelParameters} to account for our additional set of sky-model parameters: $\bm{\upUpsilon}_{uv, \delta T, \delta S} \equiv \bm{\upLambda}_{\delta T, \delta S}^{\dagger}\mathbfss{N}^{-1}\bm{\upLambda}_{\delta T, \delta S} + \bm{T}_{\delta T, \delta S}^{\dagger}\bm{\sSigma}_{uv, \delta T, \delta S}^{-1}\bm{T}_{\delta T, \delta S}$ and $\overline{\delta \bm{V}}_{\delta T, \delta S}^\mathrm{obs} \equiv \bm{\upLambda}_{\delta T, \delta S}^{\dagger}\mathbfss{N}^{-1}\delta\bm{V}^\mathrm{obs}$, where $\bm{\upLambda}_{\delta T, \delta S} \equiv \hat{\mathbfss{H}}^\mathrm{m}\mathbfss{D}^\mathrm{m}\bm{T}_{\delta T, \delta S}$. We then proceed in an identical manner to derive the marginalised probability distribution for the calibration parameters given the observed visibilities, the two-dimensional spatial power spectrum of the emission missing from $\bm{V}^\mathrm{sim}$ and the flux density uncertainties on the catalogued point sources included in $\bm{V}^\mathrm{sim}$, and the power spectrum of fluctuations in the degenerate gain amplitude as,
\begin{multline}
\label{Eq:MarginPlusSources}
\mathrm{Pr}(\bm{A}_\mathrm{F}, \bm{\mathit{\Phi}}_{l,\mathrm{F}}, \bm{\mathit{\Phi}}_{m,\mathrm{F}} \;|\; \bm{V}^\mathrm{obs}, \bm{P}_{uv, \delta T, \delta S}, \bm{\sigma}_{A_\mathrm{F}}^{2}) \propto \\
\frac{\mathrm{det}(\mathbfss{N})^{-1} ~ \mathrm{det}(\bm{\sSigma}_{uv, \delta T, \delta S})^{-1}}{\sqrt{\mathrm{det}(\bm{\upUpsilon}_{uv, \delta T, \delta S}) ~ \mathrm{det}(\bm{\sSigma}_{A_\mathrm{F}})}} \\
\times \exp \biggl[-(\delta\bm{V}^\mathrm{obs})^{\dagger}\mathbfss{N}^{-1}(\delta\bm{V}^\mathrm{obs}) + (\overline{\delta \bm{V}}_{\delta T, \delta S}^\mathrm{obs})^{T} \bm{\upUpsilon}_{uv, \delta T, \delta S}^{-1}(\overline{\delta \bm{V}}_{\delta T, \delta S}^\mathrm{obs}) \\
 - \frac{1}{2}(\bm{A}_\mathrm{F} -\bar{\bm{A}}_\mathrm{F})^{T}\bm{\sSigma}_{A_\mathrm{F}}^{-1}(\bm{A}_\mathrm{F} -\bar{\bm{A}}_\mathrm{F})\biggr].
\end{multline}

Alternatively, a similar marginalisation can also be performed if we define both our priors on the power in both our fitted diffuse and point source uncertainty models in the image-domain. In this case, we can write our prior on the expected power in the fitted calibration model as, 
\begin{multline}
\label{Eq:ProbObsPSofEpsilonPlusSourcesImDomain}
\mathrm{Pr}(\bm{\varepsilon}_{\delta T, \delta S} \;|\; \bm{\sigma}_{lm,\delta T, \delta S}) \; \propto \; \frac{1}{\sqrt{\mathrm{det}(\bm{\sSigma}_{lm, \delta T, \delta S})}} \\
\times \exp\left[-\frac{1}{2}\bm{\varepsilon}_{\delta T, \delta S}^{T}\bm{\sSigma}_{lm, \delta T, \delta S}^{-1}\bm{\varepsilon}_{\delta T, \delta S}\right] \ .
\end{multline}
Here, $\bm{\sSigma}_{lm, \delta T, \delta S}$ is a diagonal matrix with elements given by $\sSigma_{lm, \delta T, \delta S, ij} = \delta_{ij}\sigma^{2}_{lm,\delta T, \delta S, j}$, where $\sigma_{lm,\delta T, \delta S, j}$ is the uncertainty on the $j$-th sky-model parameter and $j$ runs from $1$ to $N_{\mathrm{pix, s}} + N_{\delta S}$. Defining $\bm{\upUpsilon}_{lm, \delta T, \delta S} \equiv \bm{\upLambda}_{\delta T, \delta S}^{\dagger}\mathbfss{N}^{-1}\bm{\upLambda}_{\delta T, \delta S} + \frac{1}{2}\bm{\sSigma}_{lm, \delta T, \delta S}^{-1}$, the marginalised probability distribution for the calibration parameters given the observed visibilities, the amplitude of uncertainties in our fitted diffuse and point source uncertainty models, and the power spectrum of fluctuations in the degenerate gain amplitude is given by,
\begin{multline}
\label{Eq:MarginPlusSourcesImDomain}
\mathrm{Pr}(\bm{A}_\mathrm{F}, \bm{\mathit{\Phi}}_{l,\mathrm{F}}, \bm{\mathit{\Phi}}_{m,\mathrm{F}} \;|\; \bm{V}^\mathrm{obs}, \bm{\sigma}_{lm,\delta T, \delta S}^{2}, \bm{\sigma}_{A_\mathrm{F}}^{2}) \propto \\
\frac{\mathrm{det}(\mathbfss{N})^{-1}}{\sqrt{\mathrm{det}\bm{\upUpsilon}_{lm, \delta T, \delta S} ~ \mathrm{det}(\bm{\sSigma}_{lm, \delta T, \delta S}) ~ \mathrm{det}(\bm{\sSigma}_{A_\mathrm{F}})}} \\
\times \exp \biggl[-(\delta\bm{V}^\mathrm{obs})^{\dagger}\mathbfss{N}^{-1}(\delta\bm{V}^\mathrm{obs}) + (\overline{\delta \bm{V}}_{\delta T, \delta S}^\mathrm{obs})^{T} \bm{\upUpsilon}_{lm, \delta T, \delta S}^{-1}(\overline{\delta \bm{V}}_{\delta T, \delta S}^\mathrm{obs}) \\
 - \frac{1}{2}(\bm{A}_\mathrm{F} -\bar{\bm{A}}_\mathrm{F})^{T}\bm{\sSigma}_{A_\mathrm{F}}^{-1}(\bm{A}_\mathrm{F} -\bar{\bm{A}}_\mathrm{F})\biggr].
\end{multline}

\subsection{\textsc{BayesCal} for sky-based calibration}
\label{Sec:BayesCalForSkyBasedCalibration}

In the preceding subsections, we derived the data likelihood and the posterior probability distributions when incorporating a range of priors in the \textsc{BayesCal} data model, in the context of redundant calibration, because this minimises the number of calibration parameters one must jointly estimate with the sky-model parameters. This, correspondingly, improves the computational efficiency of the analysis. In paper II, we show that \textsc{BayesCal} enables recovery of calibration solutions that are sufficiently high fidelity to remove calibration-based systematics as a limiting factor in 21 cm signal estimation. However, these results assume that the array is redundant and, correspondingly, that the relative gain solutions derived using constraints based on the redundancy of the array are robust. If the array is assumed to be redundant when in reality it not, this has been shown to introduce errors into the relative gain solutions (e.g. \citealt{2019MNRAS.487..537O, 2021MNRAS.506.2066C}) which will not be corrected by \textsc{BayesCal} if calibration parameter estimation is limited to the degenerate gain parameters of a redundant array.

However, there is nothing that fundamentally restricts the \textsc{BayesCal} approach of jointly estimating calibration parameters and the parameters of a statistical model for the flux missing from $\bm{V}^\mathrm{sim}$ to calibrating redundant arrays. Instead, for a non-redundant array, \textsc{BayesCal} can be used to jointly estimate the antenna based gains of purely sky-based calibration (see \autoref{Sec:SkyBasedCalibration}) with the parameters of the statistical model for the flux missing from $\bm{V}^\mathrm{sim}$.

Derivation of a data likelihood for sky-based calibration using the \textsc{BayesCal} framework proceeds in an identical manner to that of \autoref{Sec:FittedImageDomainSkyModel}, up to \autoref{Eq:BayesCalVisLike}, at which point, we instead calculate the likelihood with respect to the full set of antenna gains,
\begin{multline}
\label{Eq:BayesCalVisLikeSkyBased}
\mathrm{Pr}(\bm{V}^\mathrm{obs} \;|\; \bm{g}^\mathrm{m}, \bm{\varepsilon}) =\\ \frac{1}{\pi^{N_{\mathrm{vis}}}\mathrm{det}(\mathbfss{N})} \exp\biggl[-\left(\bm{V}^\mathrm{obs} - \mathbfss{G}^\mathrm{m}(\bm{V}^\mathrm{sim} + \bm{V}^\mathrm{fit})\right)^\dagger \\  \mathbfss{N}^{-1} 
 \left(\bm{V}^\mathrm{obs} - \mathbfss{G}^\mathrm{m}(\bm{V}^\mathrm{sim} + \bm{V}^\mathrm{fit})\right)\biggr] \ .
 \end{multline}
 
\autoref{Eq:BayesCalVisLikeSkyBased} can be used, in combination with the priors discussed in Sections \ref{Sec:TemporalModelling}, \ref{Sec:PowerSpectralPriors}, and \ref{Sec:DegenerateGainModelParametrisation}, by both redundant and non-redundant arrays to recover high fidelity calibration solutions; however, the larger number of gain parameters that must be jointly estimated makes it more computationally expensive than absolute calibration of redundantly calibrated visibilities. Thus, if one is calibrating data from a truly redundant array, it is preferable to perform relative and absolute calibration independently.

Additionally, if one is calibrating a partially redundant array, the likelihood can be supplemented with a prior on the parameters derived from the partial redundancy of the visibilities (see \citealt{2017arXiv170101860S} and \citealt{2021MNRAS.503.2457B} for a discussion of this approach). Since \autoref{Eq:BayesCalVisLikeSkyBased} is designed to address the sky-model incompleteness problem, for a sufficiently accurate $\bm{V}^\mathrm{fit}$, one expects to recover high fidelity gain solutions without requiring the additional constraints derived from the partial redundancy of the array. However, for a sufficiently redundant array, supplementing sky-based \textsc{BayesCal} calibration with additional constraints derived from the array redundancy should further reduce the uncertainties on the recovered parameters. We leave further investigation of this to future work.

%%%%%%%%%%%%%%%%%%%%%%%%%%%%%%%%%%%%%%%%%%%%%%%%%%%%%%%%%
\section{Summary \& Conclusions}
\label{Sec:SummaryAndConclusions}
%%%%%%%%%%%%%%%%%%%%%%%%%%%%%%%%%%%%%%%%%%%%%%%%%%%%%%%%%

Existing approaches to radio interferometric calibration have been shown to produce biased calibration solutions containing spurious spectral structure if the sky model is incomplete (e.g. \citealt{2016MNRAS.461.3135B, 2019ApJ...875...70B}). In 21 cm cosmology, a unifying theme amongst approaches to separating the cosmological signal from the foregrounds is the use of a priori knowledge of the spectral smoothness of the foreground emission relative to the cosmological signal, which is expected to fluctuate more rapidly due to inhomogeneities in the ionization, temperature and density of hydrogen in the IGM along the line of sight. However, this foundational assumption is violated when spurious spectral structure in the calibration solutions imparts spectral fluctuations to the bright foreground emission in the calibrated data. When this occurs, the spectrally fluctuating component of the data will contain calibration-derived foreground systematics on the spectral scales of the spurious fluctuations in the calibration solutions that will prevent unbiased estimation of the 21 cm signal on these scales by spectral means.

In this work, we have introduced a Bayesian framework for interferometric calibration, \textsc{BayesCal}, which is designed to address this problem by mitigating sky-model incompleteness in the calibration model, enabling recovery of higher fidelity gain solutions. To achieve this, the \textsc{BayesCal} calibration model is comprised of \begin{enumerate*}\item a simulated visibility model, $\bm{V}^\mathrm{sim}$, which describes the expected contribution to the observed visibilities associated with a priori known sky emission (this model component is standard in sky-referenced calibration) and \item a fitted visibility model, $\bm{V}^\mathrm{fit}$, that models the contribution to the observed visibilities associated with emission missing from $\bm{V}^\mathrm{sim}$ due to the incompleteness and uncertainties associated with our limited a priori knowledge of the brightness distribution of the sky.\end{enumerate*}

We have addressed calibration of visibilities corresponding to a single instrumental correlation. When deriving $\bm{V}^\mathrm{fit}$, we focus on modelling the contribution to the observed visibilities associated with Stokes $I$ emission missing from $\bm{V}^\mathrm{sim}$ due to the incompleteness and uncertainties associated with our a priori knowledge of the brightness distribution of the sky. When calibrating short duration data sets, the flexibility of the intrinsic sky model enables $\bm{V}^\mathrm{fit}$ to approximately account for sky-model incompleteness in polarised Stokes emission. We describe how $\bm{V}^\mathrm{fit}$ can be extended to model incompleteness in full-Stokes emission. However, this corresponds to an increased dimensionality of the sky-model parameter space and, for a fixed spectral structure model over Stokes parameters, the contributions of the Stokes sky models to a single instrumental visibility correlation and time are degenerate. In principle, both of these obstacles can be tackled by placing additional constraints on the sky-model parameters of $\bm{V}^\mathrm{fit}$, for example, by simultaneously fitting visibilities corresponding to the four instrumental correlations and by jointly fitting for calibration parameters over an extended time interval. We intend to explore this approach in more detail in future work.

From the starting point of defining a calibration model in which the parameters of a fitted visibility model and instrumental calibration parameters are jointly estimated, we have incorporated a number of novel features that enhance the fidelity of the recovered calibration solutions and improve the computational efficiency of the algorithm:
\begin{itemize}
 \item We have demonstrated how $\bm{V}^\mathrm{fit}$ can be constrained using a Gaussian prior encoding either: \begin{enumerate*}\item its expected two-dimensional spatial power spectrum or \item the expected power in the image-domain parameters of the fitted calibration model \end{enumerate*}. This eliminates degeneracy between the jointly estimated calibration and sky-model parameters and enables more stringent constraints on each.
 \item In general, sampling jointly from the parameters of $\bm{V}^\mathrm{fit}$ and the instrumental calibration parameters significantly increases the dimensionality of the optimization problem that must be solved to calibrate the instrument. However, we have shown how this limitation can be overcome by analytically marginalising over the sky-model parameters of $\bm{V}^\mathrm{fit}$, allowing one to sample directly from the marginal posterior probability distribution of the calibration parameters. 
 \item We have demonstrated how the above approach can also be extended to marginalise over flux-density uncertainties in catalogued point sources included in $\bm{V}^\mathrm{sim}$.
 \item We have described how priors on the spectral smoothness of the instrumental gains can be incorporated in the \textsc{BayesCal} framework, informed by theoretical and measurement-based constraints, such as electric and electromagnetic co-simulation and reflectometry measurements of the receiver system. We have also shown how such priors can be parametrised such that calibration solutions requiring spurious levels of spectral structure are disfavoured in a statistically principled manner, without preventing any true low-level structure in the instrumental gains on small spectral scales from being calibrated.
 \item When defining $\bm{V}^\mathrm{fit}$, one must specify a model for its spectral structure. We have described how Bayesian model selection can be used to determine whether a simple physically motivated model parametrisation, such as a power law, is sufficient for modelling a given data set to be calibrated (and, if so, to determine the optimal power law index) or whether an alternate or more complex spectral model is preferred.
\item If one wishes to simultaneously calibrate multiple time integrations of a data set, we have shown how a temporal model and priors are simply incorporated in the \textsc{BayesCal} framework, enabling one to take advantage of a priori knowledge constraining the temporal evolution of the gain solutions. In general, one expects the temporal stability of signal propagation effects and the total duration over which calibration solutions are simultaneously estimated to determine the necessary complexity of the temporal model for the evolution of the gain solutions. Bayesian model selection can again be used to select between such models in a rigorous manner.
\end{itemize}

We have derived the posterior probability distribution for the absolute calibration parameters of a redundantly calibrated array in the \textsc{BayesCal} calibration framework as our primary case study. In addition to enabling construction of a theoretically complete sky model and thus mitigating or removing (for a sufficiently accurate model for $\bm{V}^\mathrm{fit}$) spurious spectral structure in the calibration solutions due to sky-model incompleteness, the framework naturally accounts for and covariantly propagates through to the calibration parameter estimates the non-uniform uncertainty on the raw visibilities, both due to the varying redundancy of the baseline groups and to the effect of the antenna gains on the noise in the data. This enables correct propagation of these uncertainties through to the calibrated data, which, in turn, is necessary for accurately accounting for the uncertainties on astrophysical parameters derived from the calibrated data. 

In this paper, when deriving the posterior probability distribution for the absolute calibration parameters of a redundantly calibrated array, we assume that nominally redundant baselines have perfect redundancy (i.e. that the visibility data measured by baselines in a redundant group is identical barring the noise). In practice, deviations from this assumption will occur because of small errors in antenna positions and differences in antenna voltage patterns across the array due to perturbations in feed positions and orientations between antennas and varying levels of mutual coupling between neighbouring antennas in different parts of the array. Here, we have additionally described the generalisation of the \textsc{BayesCal} calibration framework to fully sky based calibration, which circumvents this problem if the parameters of the array are perfectly known; however, the increased dimensionality of the calibration parameter space in this regime will mean a corresponding increase in computational complexity.

Even if array redundancy is neglected and sky based calibration is used to solve for all of the calibration degrees of freedom, one still requires precise a priori knowledge of the voltage beam patterns and positions of the antennas to avoid introducing spurious spectral structure into the gain solutions. In practice, these quantities are known with finite precision. However, at the cost of increased dimensionality, these model uncertainties can also be accounted for in the \textsc{BayesCal} framework by jointly sampling from the calibration parameters, sky-model parameters and an additional set of model parameters that describe perturbations to the instrument forward model and which can be constrained by priors derived from the measurement errors corresponding to the precision with which the instrument has been characterised. Exploring this approach is an interesting direction for future analysis.

In addition to redundant plus absolute calibration of perfectly redundant arrays and purely sky-based calibration of non-redundant arrays, frameworks for calibrating partially redundant arrays have been explored in detail in \citet{2017arXiv170101860S} and \citet{2021MNRAS.503.2457B}. This approach to bridging sky-based and redundant calibration is complementary to the \textsc{BayesCal} framework developed in this paper, which focuses on explicitly addressing the sky-model incompleteness problem. Combining these two developments provides another exciting direction for future work. 

In paper II, we demonstrate the application of \textsc{BayesCal} to simulated observations in the context of absolute calibration of data from redundant interferometric arrays. In that paper we show that this enables recovery of significantly higher fidelity calibration solutions relative to standard calibration approaches (up to four orders of magnitude suppression in spurious spectral fluctuations in the calibration solutions), which is sufficient to reduce calibration-based foreground systematics to a sub-EoR level on all Nyquist sampled spectral scales in the data when sky-model incompleteness is moderate or low.

%%%%%%%%%%%%%%%%%%%%%%%%%%%%%%%%%%%%%%%%%%%%%%%%%%%%%%%%%
\section*{Acknowledgements}
%%%%%%%%%%%%%%%%%%%%%%%%%%%%%%%%%%%%%%%%%%%%%%%%%%%%%%%%%

We thank the anonymous reviewer for valuable comments that improved the manuscript. PHS and JCP acknowledge support from NSF Award \#1907777 and Brown University's Richard B. Salomon Faculty Research Award Fund. JLS and PHS acknowledge funding from the Canada 150 Research Chairs Program. PHS was supported in part by a McGill Space Institute fellowship. This research was conducted using computational resources and services at the Center for Computation and Visualization, Brown University. PHS thanks James Aguirre, Adrian Liu and Irina Stefan for valuable discussions and Irina Stefan and Adrian Liu for helpful comments on a draft of this manuscript.

%%%%%%%%%%%%%%%%%%%%%%%%%%%%%%%%%%%%%%%%%%%%%%%%%%%%%%%%%
\section*{Data Availability}
%%%%%%%%%%%%%%%%%%%%%%%%%%%%%%%%%%%%%%%%%%%%%%%%%%%%%%%%%

No new data were generated or analysed in support of this research.

%%%%%%%%%%%%%%%%%%%%%%%%%%%%%%%%%%%%%%%%%%%%%%%%%%%%%%%%%%%%%%%%%%%%%%%%%%%%%%%%%%%%%%%%%%%%%%%%%%%
%%%%%%%%%%%%%%%%%%%%%%%%%%%%%%%%%%%%%%%%%%%%%%%%%%%%%%%%%%%%%%%%%%%%%%%%%%%%%%%%%%%%%%%%%%%%%%%%%%%

%%%%%%%%%%%%%%%%%%%%%%%%%%%%%%%%%%%%%%%%%%%%%%%%%%%%%%%%%%%%%%%%%%%%%%%%%%%%%%%%%%%%%%%%%%%%%%%%%%%
%%%%%%%%%%%%%%%%%%%%%%%%%%%%%%%%%%%%%%%%%%%%%%%%%%%%%%%%%%%%%%%%%%%%%%%%%%%%%%%%%%%%%%%%%%%%%%%%%%%

\appendix

%%%%%%%%%%%%%%%%%%%%%%%%%%%%%%%%%%%%%%%%%%%%%%%%%%%%%%%%%
\section{Summary of variables}
\label{Sec:SummaryOfVariables}
%%%%%%%%%%%%%%%%%%%%%%%%%%%%%%%%%%%%%%%%%%%%%%%%%%%%%%%%%

\onecolumn

\begin{table}
    \centering
    \begin{tabular}{p{2.6cm} p{10cm} c c}
    \toprule
         Quantity &  Description & Size & First Appearance\\
    \midrule
        $\bm{v}$ & {\color{black}Vector of voltages induced in the antenna feeds by the vector electric field due to the source brightness distribution, evaluated at the antenna}  & $2 \times 1$ & \autoref{Eq:AntennaVoltage} \\
        $\mathbfss{J}$ & {\color{black}Jones matrix describing the antenna- and, in general, direction-dependent cumulative product of all propagation effects along the signal path between the emission sources, antenna and the correlator} & $2 \times 2$ & \autoref{Eq:AntennaVoltage} \\
        $\bm{E}$ & {\color{black}Vector electric field due to the source brightness distribution} & $2 \times 1$ & \autoref{Eq:AntennaVoltage} \\
        $\bm{V}_{pq}$ & Polarised visibility vector, for a pair of dual-polarised antennas $p$ and $q$, encoding the four correlations of the two voltage signals per antenna induced by the incident electric field at the antennas  & $4 \times 1$ & \autoref{Eq:MEq} \\
        $\bm{n}_{pq}$ & Polarised noise vector  & $4 \times 1$ & \autoref{Eq:MEq} \\
        $l$, $m$ and $n$ & Direction cosines  & & \autoref{Eq:MEq} \\
        $\bm{\hat{l}}$ & Unit vector of direction cosines  & $3 \times 1$  & \autoref{Eq:MEq} \\
        $\bm{C}$ & Coherency vector encoding the four correlations of the two components of the vector electric field incident on the antennas, assuming a direction-dependent orthogonal 2D basis with basis vectors perpendicular to $\bm{\hat{l}}$.  & $4 \times 1$ & \autoref{Eq:MEq} \\
        $\mathbfss{J}_{pq}$ & Matrix describing signal propagation effects on the voltage correlations measured by the interferometric baseline between antennas $p$ and $q$  & $4 \times 4$ & \autoref{Eq:MEq} \\
        $\mathbfss{J}_{p}$ & Jones matrix describing the cumulative product of all propagation effects along the signal paths of antennas $p$ to the correlator   & $2 \times 2$ & \autoref{Eq:Jpq} \\
        $E_{\delta}$ and $E_{\alpha}$ & Component of vector electric field in the directions of $\hat{\bm{e}}_{\delta}(\bm{\hat{l}})$ and $\hat{\bm{e}}_{\alpha}(\bm{\hat{l}})$, respectively  & & \autoref{Eq:CoherencyVector} \\
        $\hat{\bm{e}}_{\delta}$ and $\hat{\bm{e}}_{\alpha}$  & Direction dependent Dec and RA equatorial basis vectors   & $2 \times 1$ & \autoref{Eq:CoherencyVector} \\
        $I$, $Q$, $U$ and $V$  & Stokes parameters  & & \autoref{Eq:CoherencyVector} \\
        $\mathbfss{J}^\mathrm{ant}_{p}$ & Jones matrix describing the cumulative product of antenna-based signal propagation effects between the feeds of antenna $p$ and the correlator & $2 \times 2$ & \autoref{Eq:JonesDecomposition} \\
        $\mathbfss{J}^\mathrm{sky}_{p}$ & Jones matrix describing the cumulative product of sky-based signal propagation effects between the source and antenna $p$  & $2 \times 2$ & \autoref{Eq:JonesDecomposition} \\
        $K_{p}$ & Geometrical phase delay associated with antenna $p$, relative to a reference position, as a function of direction on the sky   & & \autoref{Eq:JonesDecomposition} \\
        $\mathbfss{G}_{p}$  & Polarised gain matrix associated with antenna $p$ and encoding direction-independent antenna-based signal propagation effects   & $2 \times 2$ & \autoref{Eq:JonesDecomposition} \\
        $\mathbfss{E}_{p}$ & Polarised voltage beam pattern of antenna $p$   & $2 \times 2$ & \autoref{Eq:JonesDecomposition} \\
        $\bm{x}_{p}$ & Position vector of antenna $p$   & $3 \times 1$ & \autoref{Eq:JonesDecomposition} \\
        $\lambda$ & Wavelength   & & \autoref{Eq:JonesDecomposition} \\
        $\mathbfss{G}_{pq}$ & Diagonal (assuming no polarisation leakage between feeds) polarised gain matrix describing direction independent signal propagation effects on the voltage correlations measured by the interferometric baseline between antennas $p$ and $q$  & $4 \times 4$ & \autoref{Eq:MEqexpanded2} \\
        $\mathbfss{P}_{pq}$ & Polarised primary beam matrix encoding the direction dependent primary beam response of the interferometric baseline between antennas $p$ and $q$ to the sky brightness distribution expressed in terms of Stokes parameters  & $4 \times 4$ & \autoref{Eq:MEqexpanded2} \\
        $\bm{C}^\mathrm{S}$ & Stokes coherency vector  & $4 \times 1$ & \autoref{Eq:MEqexpanded2} \\
        $\bm{b}_{pq}$ & Baseline vector between antennas $p$ and $q$  & $3 \times 1$ & \autoref{Eq:MEqexpanded2} \\
        $\mathbfss{D}$ & D-term / feed-error matrix encoding cross-coupling effects & $2 \times 2$ & Below \autoref{Eq:MEqexpanded2} \\
        $\mathbfss{E}_{pq}$ & Polarised primary beam matrix associated with the interferometric baseline between antennas $p$ and $q$  & $4 \times 4$ & Below \autoref{Eq:MEqexpanded2} \\
        $\mathbfss{T}^\mathrm{S}$ & Stokes transformation matrix  & $4 \times 4$ & Below \autoref{Eq:MEqexpanded2} \\
        $\hat{\bm{e}}_\mathrm{n}$ and $\hat{\bm{e}}_\mathrm{e}$ & Time dependent north and east basis vectors  & $2 \times 1$ & \autoref{Eq:VoltageNE} \\
        $v_{p}^\mathrm{n}$ and $v_{p}^\mathrm{e}$ & Component of the voltage vector induced in the feeds aligned with $\hat{\bm{e}}_\mathrm{n}$ and $\hat{\bm{e}}_\mathrm{e}$ by the respective component of the vector electric field, due to the source brightness distribution, evaluated at the antenna  & & \autoref{Eq:VoltageNE} \\
        $V^\mathrm{nn}$,  $V^\mathrm{ne}$,  $V^\mathrm{en}$ \newline and $V^\mathrm{ee}$ & Visibilities resulting from north-south feed correlations, east-west with north-south cross-correlations, north-south with east-west cross-correlations, and east-west correlations, respectively  & & \autoref{Eq:PolarisedVisibilityVector} \\
        $\mathbfss{P}$ & See $\mathbfss{P}_{pq}$   & $4 \times 4$ & \autoref{Eq:PolarisedPrimaryBeamMatrixElements} \\
        $P_{i,j}$ & Polarised primary beam matrix element which, for the bases used in this work, encodes the effective primary beam of the baselines coupling Stokes parameter $j \in [I,Q,U,V]$, on the sky, to visibility correlation $i \in [\mathrm{nn}, \mathrm{ne}, \mathrm{en}, \mathrm{ee}]$  & & \autoref{Eq:PolarisedPrimaryBeamMatrixElements} \\
        $g_{p}^\mathrm{n}$ & Complex gain associated with north-south feed of antenna $p$  & & \autoref{Eq:MEqSingePol} \\
        $\bm{u}_{pq}$ & Vector of time-stationary $uvw$-coordinates $\bm{u}_{pq} = \bm{b}_{pq}/\lambda = (u_{pq},v_{pq},w_{pq})$ of the baseline between antennas $p$ and $q$.  & $3 \times 1$ & \autoref{Eq:MEqSingePol} \\
        $\bm{V}^\mathrm{obs}$ & Vector of observed visibility data  & $N_{\mathrm{vis}} \times 1$ & \autoref{Eq:MEqSingePolShorthand} \\
        $\bm{n}$ & Noise vector  & $N_{\mathrm{vis}} \times 1$ & \autoref{Eq:MEqSingePolShorthand} \\
        $\bm{V}^\mathrm{true}$ & Vector of `true' visibilities: those that would be observed in the zero-noise limit if instrumental gains were unity  & $N_{\mathrm{vis}} \times 1$ & \autoref{Eq:MEqSingePolShorthand} \\
        \bottomrule
    \end{tabular}
    \caption{Lists of variables and their descriptions.}
    \label{Tab:VariablesList1}
\end{table} 

\clearpage

\begin{table}
    \centering
    \begin{tabular}{p{2.6cm} p{10cm} p{1.5cm} c}
    \toprule
         Quantity &  Description & Size & First Appearance\\
    \midrule
        $g_{p}$ & Complex gain of antenna $p$  & & \autoref{Eq:MEqSingePolShorthandElement} \\
        $N_t$ & Number of time integrations in the data set & & Below \autoref{Eq:MEqSingePolShorthandElement} \\
        $N_\nu$ & Number of channels in the data set & & Below \autoref{Eq:MEqSingePolShorthandElement} \\
        $N_\mathrm{ant}$ & Number of antennas in the array & & Below \autoref{Eq:MEqSingePolShorthandElement} \\
        $N_\mathrm{rb}$ & Number of unique redundant baselines per channel in the array & &  \\
        $N_{\mathrm{vis}}$ & Number of visibilities in the data set: $N_\mathrm{vis} = N_t N_\nu N_\mathrm{ant}(N_\mathrm{ant}-1)/2$ and $N_\mathrm{vis} = N_t N_\nu N_\mathrm{rb}$, before and after redundant baseline averaging, respectively & & Below \autoref{Eq:MEqSingePolShorthandElement} \\
        $g_{p}^\mathrm{m}$ & Model complex gain of antenna $p$  & & Above \autoref{Eq:MEqSingePolShorthandModel} \\
        $\bm{V}^\mathrm{model}$ & Model for the signal component of the observed visibilities  & $N_{\mathrm{vis}} \times 1$ & \autoref{Eq:MEqSingePolShorthandModel} \\
        $\mathbfss{G}^\mathrm{m}$ & Diagonal matrix of model gain parameters  & $N_{\mathrm{vis}} \times N_{\mathrm{vis}}$ & \autoref{Eq:MEqSingePolShorthandModel} \\
        $\bm{V}^\mathrm{sim}$ & Vector of simulated visibilities encoding one's best estimate of the true (gain-free) visibilites derived from one's a priori knowledge of the antenna beam, array layout and emission sources in the sky under observation  & $N_{\mathrm{vis}} \times 1$ & \autoref{Eq:MEqSingePolShorthandModel} \\
        $\mathbfss{N}$ & Covariance matrix of noise on $\bm{V}^\mathrm{obs}$   & $N_{\mathrm{vis}} \times N_{\mathrm{vis}}$ & Above \autoref{Eq:DataCovarianceMatrix} \\
        $\sigma_{j}$ & Expected noise on the $j$th visibility  & & \autoref{Eq:DataCovarianceMatrix} \\
        $\bm{g}^\mathrm{m}$ & Vector of model gain parameters\tablefootnote{\label{TableNote1}Here, we list the sizes of the per-instrumental-correlation gain parameter vectors and related matrices for the case that the gain parameters are modelled as constant over the duration of the data set to be calibrated. If the gain parameters at each integration are modelled as being independent, these dimensionalities would increase by a factor $N_t$. With a temporal model appropriate for gain parameters with partial temporal correlation (i.e. the absolute value of the correlation coefficient of the gain parameters lies between 0 and 1) the dimensionality will lie between these two extremes.}  & $N_{\mathrm{ant}}N_\nu \times 1$ & \autoref{Eq:BasicVisLike} \\
        $\mathrm{Pr}(\bm{V}^\mathrm{obs} \;|\; \bm{g}^\mathrm{m})$ & Data likelihood for sky calibration  & & \autoref{Eq:BasicVisLike} \\
        $h_{p}^\mathrm{m}$ & Model for the redundant gain parameter of antenna $p$ & & \autoref{Eq:MEqSingePolShorthandRedundantModel} \\
        $V^\mathrm{model}_{\alpha,j}$ & Model for the $j$th redundant visibility in the set of visibilities with the $\alpha$th unique baseline & & \autoref{Eq:MEqSingePolShorthandRedundantModel} \\
        $\bm{h}^\mathrm{m}$ & Vector of model redundant gain parameters\textsuperscript{\ref{TableNote1}} & $N_{\mathrm{ant}}N_\nu \times 1$ & \autoref{Eq:RedundantVisLike} \\
        $\bm{V}^\mathrm{red}$ & Vector of redundant visibilities & $N_{\mathrm{vis}} \times 1$ & \autoref{Eq:RedundantVisLike} \\
        $\bm{V}^\mathrm{red,set}$ & Vector of unique redundant visibility parameters & $N_t N_\nu N_\mathrm{rb} \times 1$ & \autoref{Eq:RedundantVisLike} \\
        $\mathrm{Pr}(\bm{V}^\mathrm{obs} \;|\; \bm{h}^\mathrm{m}, \bm{V}^\mathrm{red,set})$ & Data likelihood for redundant calibration & & \autoref{Eq:RedundantVisLike} \\
        $\mathbfss{H}^\mathrm{m}$ & Diagonal matrix of redundant calibration parameters & $N_{\mathrm{vis}} \times N_{\mathrm{vis}}$ & \autoref{Eq:RedundantVisLike} \\
        $A$ & Degenerate gain amplitude for a given time, frequency and polarisation & & Above \autoref{Eq:DegenerateParameterDefinitions} \\
        $\phi_{l}$ and $\phi_{m}$ & Degenerate tip and tilt phase, respectively, for a given time, frequency and polarisation  & & Above \autoref{Eq:DegenerateParameterDefinitions} \\
        $\psi$ & Degenerate absolute phase for a given time, frequency and polarisation & & Above \autoref{Eq:DegenerateParameterDefinitions} \\
        $\sPhi$ & Vector of degenerate tip-tilt gain phases for a given time, frequency and polarisation & $2 \times 1$ & Above \autoref{Eq:DegenerateParameterDefinitions} \\
        $f_\mathrm{ant}(A, \sPhi, \psi)$ & Antenna degeneracy function & & Above \autoref{Eq:DegenerateParameterDefinitions} \\
        $f_\mathrm{vis}(A, \sPhi)$ & Visibility degeneracy function & & Above \autoref{Eq:DegenerateParameterDefinitions} \\
        $\mathbfss{D}^\mathrm{m}$ & Model degenerate gain matrix & $N_{\mathrm{vis}} \times N_{\mathrm{vis}}$ & \autoref{Eq:PureSkyVisModel} \\
        $\bm{A}$ & Vector of degenerate gain amplitude parameters\textsuperscript{\ref{TableNote1}} & $N_\nu \times 1$ & \autoref{Eq:AbsCalVisLike} \\
        $\bm{\mathit{\Phi}}_{l}$ and $\bm{\mathit{\Phi}}_{m}$ & Vectors of degenerate gain tip phase and tilt phase parameters\textsuperscript{\ref{TableNote1}}, respectively & $N_\nu \times 1$ & \autoref{Eq:AbsCalVisLike} \\
        $\bm{u}$, $\bm{v}$ and $\bm{w}$ & Vectors of $uvw$-space coordinates & $N_{\mathrm{vis}} \times 1$ & \autoref{Eq:VtrueExpanded} \\
        $N_{\mathrm{im}}$ & Number of image-space coordinates ($N_{\nu}\sum_{i=1}^{N_\mathrm{t}} N_{\mathrm{pix, s}, t_{i}}$ for diffuse emission or $N_{\nu}\sum_{i=1}^{N_\mathrm{t}} N_{\mathrm{source}, t_{i}}$, with $N_{\mathrm{source}, t_{i}}$ the number of sources within the FoV at LST $t_{i}$, for point sources) &  & - \\
        $\bm{l}$, $\bm{m}$ and $\bm{n}$ & Vectors of direction cosine coordinates & $1 \times N_{\mathrm{im}}$ & \autoref{Eq:VtrueExpanded} \\
        $\mathbfss{F}_\mathrm{fr}$ & Interferometric fringe matrix & $N_{\mathrm{vis}} \times N_{\mathrm{im}}$ & \autoref{Eq:VtrueExpanded} \\
        $\bm{T}_{I}$, $\bm{T}_{Q}$, $\bm{T}_{U}$ and $\bm{T}_{V}$ & Vectors of polarised sky brightness temperatures in Stokes $I$, $Q$, $U$ and $V$, respectively  & $N_{\mathrm{im}} \times 1$ & \autoref{Eq:VtrueExpanded} \\
        $\mathbfss{P}_{\mathrm{nn},I}$, $\mathbfss{P}_{\mathrm{nn},Q}$, $\mathbfss{P}_{\mathrm{nn},U}$ and $\mathbfss{P}_{\mathrm{nn},V}$ & Polarised primary beam matrices encoding the effective primary beam coupling polarised sky brightness temperatures in Stokes $I$, $Q$, $U$ and $V$, respectively, to north-south voltage correlations  & $N_{\mathrm{im}} \times N_{\mathrm{im}}$ & \autoref{Eq:VtrueExpanded} \\
        $\mathbfss{G}$ & Diagonal matrix encoding direction-independent instrumental gains  & $N_{\mathrm{vis}} \times N_{\mathrm{vis}}$ & \autoref{Eq:MEqSingePolShorthandExpanded} \\
        $\mathbfss{H}$ and $\mathbfss{D}$ & Diagonal matrices encoding redundant instrumental gains relative to reference antennas and degenerate instrumental gains, respectively & $N_{\mathrm{vis}} \times N_{\mathrm{vis}}$ & \autoref{Eq:MEqSingePolShorthandExpandedRedundant} \\
        $\hat{\mathbfss{H}}^\mathrm{m}$ & Diagonal matrix of maximum likelihood redundant calibration parameters & $N_{\mathrm{vis}} \times N_{\mathrm{vis}}$ & \autoref{Eq:SkyPlusFittedVisModel} \\
        $\sTheta$ & Set of visibility model parameters & & \autoref{Eq:SkyPlusFittedVisModel} \\
        $\bm{V}^\mathrm{fit}(\sTheta)$ & Fitted visibility model. & $N_{\mathrm{vis}} \times 1$ & \autoref{Eq:SkyPlusFittedVisModel} \\
        $\mathrm{Pr}(\sTheta\vert\bm{D},M)$ & Posterior probability distribution of the parameters, $\sTheta$, given data, $\bm{D}$, and a model, $M$ &  & \autoref{Eq:BayesEqn} \\
        $\mathrm{Pr}(\bm{D}\vert\sTheta,M) \equiv \mathcal{L}(\sTheta)$ & Likelihood of data $\bm{D}$, given $M$ and $\sTheta$  & & \autoref{Eq:BayesEqn} \\
        $\mathrm{Pr}(\sTheta\vert M) \equiv \pi(\sTheta)$ & Prior probability distribution of the parameters $\sTheta$ & & \autoref{Eq:BayesEqn} \\
        $\mathrm{Pr}(\sTheta\vert M) \equiv \mathcal{Z}$ & Bayesian evidence of the model for the data & & \autoref{Eq:BayesEqn} \\
        $N_\mathrm{pix}$ & Number of pixels on the \textsc{HEALPix} grid & & \autoref{Sec:FittedStokesISkyModel} \\
        $\theta_\mathrm{cut}$ & Zenith angle cut-off defining the  field-of-view of the model ($\theta \le \theta_\mathrm{cut}$), at a given LST  & & \autoref{Sec:FittedStokesISkyModel} \\
        \bottomrule
    \end{tabular}
    \contcaption{}
    \label{Tab:VariablesList2}
\end{table} 

\clearpage

\begin{table}
    \centering
    \begin{tabular}{p{5cm} p{8cm} p{1.5cm} c}
    \toprule
         Quantity &  Description & Size & First Appearance\\
    \midrule
        $N_{\mathrm{pix, s}, t_{i}}$ & Subset of \textsc{HEALPix} pixels within field-of-view at LST, $t_{i}$  & & \autoref{Sec:FittedStokesISkyModel} \\
        $N_\mathrm{pix, s}$ & \textsc{HEALPix} pixels included in the model -- defined as the union of pixels that fall within the snapshot field-of-view at each central LST of the integrations over which the calibration solutions are being jointly estimated  & & \autoref{Sec:FittedStokesISkyModel} \\
        $\mathbfss{S}_{t_{i}}$ & Spectral model matrix for LST $t_{i}$ & $N_{\mathrm{pix, s}, t_{i}}N_{\nu} \times N_{\mathrm{pix, s}, t_{i}}$ & \autoref{Eq:FittedImageCubeSingleLST} \\
        $\bm{T}_{t_{i}}^\mathrm{model}$ & Fitted residual\tablefootnote{\label{TableNote2}Residual, here, refers to the contribution omitted or mis-modelled in the fixed simulated calibration model and that is fitted with the additional fitted visibility model component of the calibration model introduced in \textsc{BayesCal}.} brightness temperature image cube model at LST $t_{i}$ & $N_{\mathrm{pix, s}, t_{i}}N_{\nu} \times 1$ & \autoref{Eq:FittedImageCubeSingleLST} \\
        $\bm{\varepsilon}_{t_{i}}$ & Fitted residual\textsuperscript{\ref{TableNote2}} sky brightness temperature map model at reference frequency $\nu_{0}$ and LST $t_{i}$ & $N_{\mathrm{pix, s}, t_{i}} \times 1$ & \autoref{Eq:FittedImageCubeSingleLST} \\
        $\bm{\varepsilon}_\mathrm{concat}$ & Concatenation of fitted residual\textsuperscript{\ref{TableNote2}} sky brightness temperature models as a function of LST & $\sum_{i=1}^{N_\mathrm{t}} N_{\mathrm{pix, s}, t_{i}} \times 1$ & \autoref{Eq:GriddedImageParams} \\
        $\mathbfss{C}$ &  Concatenation matrix over LST & $\sum_{i=1}^{N_\mathrm{t}} N_{\mathrm{pix, s}, t_{i}} \times N_\mathrm{pix, s}$ & \autoref{Eq:GriddedImageParams} \\
        $\bm{\varepsilon}$ & Vector of fitted residual\textsuperscript{\ref{TableNote2}} sky-model parameters & $N_\mathrm{pix, s} \times 1$ & \autoref{Eq:GriddedImageParams} \\
        $\bm{\sTheta}^\mathrm{za}$ & Zenith angle matrix used in the construction of $\mathbfss{C}$ & $N_\mathrm{t}N_{\mathrm{pix, s}} \times N_\mathrm{pix, s}$ & \autoref{Eq:GriddingMatrix} \\
        $\mathbfss{S}$ & Spectral model matrix for all LST of the data set to be calibrated & $N_{\mathrm{im}} \times \sum_{i=1}^{N_\mathrm{t}} N_{\mathrm{pix, s}, t_{i}}$ & \autoref{Eq:FittedVisibilitiesFullModel} \\
        $\sigma_{V}$ & Expected noise level on a calibrated visibility resulting from a pair of identical antennas individually experiencing equal system noise & & \autoref{Eq:VisabilityNoise} \\
        $\mathbfss{N}^{\prime}$ & Covariance matrix of the noise on the calibrated data & $N_{\mathrm{vis}} \times N_{\mathrm{vis}}$ & \autoref{Eq:CovarianceCalVis} \\
        $\sTheta_{A,\mathrm{t}}$ and $\sTheta_{\sPhi,\mathrm{t}}$ & Parameters of degenerate gain amplitude and tip-tilt phase temporal models, respectively &  & \autoref{Eq:TemporalModel} \\
        $f_{A}(\sTheta_{A,\mathrm{t}})$ and $f_{\sPhi}(\sTheta_{\sPhi,\mathrm{t}})$ & Degenerate gain amplitude and tip-tilt phase temporal models, respectively &  & \autoref{Eq:TemporalModel} \\
        $\mathrm{Pr}(\bm{\varepsilon}, \sTheta_\mathrm{t} \;|\; \bm{V}^\mathrm{obs})$ & Posterior probability distributions for $\bm{\varepsilon}$ and $\sTheta_\mathrm{t}$ given the raw data &  & \autoref{Eq:ProbCalTemporal} \\
        $\mathrm{Pr}(\bm{V}^\mathrm{obs} \;|\; \bm{\varepsilon}, \sTheta_\mathrm{t})$ & Likelihood of the raw data given $\bm{\varepsilon}$ and $\sTheta_\mathrm{t}$ &  & \autoref{Eq:ProbCalTemporal} \\
        $\mathrm{Pr}(\bm{\varepsilon})$ and $\mathrm{Pr}(\sTheta_\mathrm{t})$ & Prior probability distributions on sky- and temporal-model parameters, respectively &  & \autoref{Eq:ProbCalTemporal} \\
        $\sTheta_\mathrm{f}$ & Foreground spectral model parameters  &  & \autoref{Eq:FittedVisibilitiesFullModelGeneral} \\
        $\mathrm{Pr}(\bm{A}, \bm{\mathit{\Phi}}_{l}, \bm{\mathit{\Phi}}_{m}, \bm{\varepsilon}, \sTheta_\mathrm{f} \;|\; \bm{V}^\mathrm{obs})$ & Posterior probability distributions for $\bm{A}$, $\bm{\mathit{\Phi}}_{l}$, $\bm{\mathit{\Phi}}_{m}$, $\bm{\varepsilon}$ and $\sTheta_\mathrm{f}$ given the raw data  &  & \autoref{Eq:ProbCalSkySpectrum} \\
        $\mathrm{Pr}(\bm{V}^\mathrm{obs} \;|\; \bm{A}, \bm{\mathit{\Phi}}_{l}, \bm{\mathit{\Phi}}_{m}, \bm{\varepsilon}, \sTheta_\mathrm{f})$ & Likelihood of the raw data given $\bm{A}$, $\bm{\mathit{\Phi}}_{l}$, $\bm{\mathit{\Phi}}_{m}$, $\bm{\varepsilon}$ and $\sTheta_\mathrm{f}$  &  & \autoref{Eq:ProbCalSkySpectrum} \\
        $\mathrm{Pr}(\bm{A})$, $\mathrm{Pr}(\bm{\mathit{\Phi}}_{l})$, $\mathrm{Pr}(\bm{\mathit{\Phi}}_{m})$ and $ \mathrm{Pr}(\sTheta_\mathrm{f})$ & Prior probability distributions on degenerate gain amplitude, tip-tilt phase and foregrounds spectral model parameters, respectively  &  & \autoref{Eq:ProbCalSkySpectrum} \\
        $\bm{P}_{uv,\delta T}$ & Two dimensional spatial power spectrum of residual\textsuperscript{\ref{TableNote2}} diffuse emission & $N_{\mathrm{vis}} \times 1$ & \autoref{Eq:DiffusePower} \\
        $\bm{P}_{uv,T}$ & Two dimensional spatial power spectrum of the a priori known component of diffuse emission & $N_{\mathrm{vis}} \times 1$ & \autoref{Eq:DiffusePower} \\
        $\bm{P}_{uv,\mathrm{ps},\mathrm{Poisson}}$ & Two dimensional spatial power spectrum of residual\textsuperscript{\ref{TableNote2}} Poisson distributed point sources & $N_{\mathrm{vis}} \times 1$ & \autoref{Eq:PoissonPointSourcePower} \\
        $S$ & Point source flux-density &  & \autoref{Eq:PoissonPointSourcePower} \\
        $\dfrac{\mathrm{d}N}{\mathrm{d}S}$ & Differential source count &  & \autoref{Eq:PoissonPointSourcePower} \\
        $\bm{P}_{uv,\mathrm{ps},\mathrm{clustered}}$ & Two dimensional spatial power spectrum of residual\textsuperscript{\ref{TableNote2}} clustered point sources  & $N_{\mathrm{vis}} \times 1$ & \autoref{Eq:ClusteredPointSourcePower} \\
        $\widetilde{w}$ & Legendre transform of the point source angular correlation function  &  & \autoref{Eq:ClusteredPointSourcePower} \\
        $\bm{\sSigma}_{uv, \delta T}$ & Covariance matrix encoding the $\bm{P}_{uv,\delta T}$ estimate of the power spectrum of $\bm{V}^\mathrm{fit}$ & $N_{\mathrm{vis}} \times N_{\mathrm{vis}}$ & \autoref{Eq:ProbObsPSofEpsilon} \\
        $\mathrm{Pr}(\bm{\varepsilon} \;|\; \bm{P}_{uv,\delta T})$ & Prior on the image domain calibration model parameters given their expected two dimensional spatial power spectrum &  & \autoref{Eq:ProbObsPSofEpsilon} \\
        \bottomrule
    \end{tabular}
    \contcaption{}
    \label{Tab:VariablesList3}
\end{table} 

\clearpage

\begin{table}
    \centering
    \begin{tabular}{p{5cm} p{8cm} p{1.6cm} c}
    \toprule
         Quantity &  Description & Size & First Appearance\\
    \midrule
        $\mathrm{Pr}(\bm{A}, \bm{\mathit{\Phi}}_{l}, \bm{\mathit{\Phi}}_{m}, \bm{\varepsilon} \;|\; \bm{V}^\mathrm{obs}, \bm{P}_{uv,\delta T})$ & Posterior probability distributions for $\bm{A}$, $\bm{\mathit{\Phi}}_{l}$, $\bm{\mathit{\Phi}}_{m}$ and $\bm{\varepsilon}$ given the raw data and the two-dimensional spatial power spectrum of the residual\textsuperscript{\ref{TableNote2}} emission &  & \autoref{Eq:ProbCalSkyGivenPS} \\
        $\mathrm{Pr}(\bm{P}_{uv,\delta T})$ & Prior on the two-dimensional spatial power spectrum of the residual\textsuperscript{\ref{TableNote2}} emission &  & \autoref{Eq:ProbCalSkyPS} \\
        $\mathrm{Pr}(\bm{\varepsilon} \;|\; \bm{\sSigma}_{lm, \delta T})$ & Prior on the image domain calibration model parameters given their expected covariance matrix &  & \autoref{Eq:ProbObsPSofEpsilonIm} \\
        $\bm{\sSigma}_{lm, \delta T}$ & Expected covariance matrix of $\bm{\varepsilon}$ &  & \autoref{Eq:ProbObsPSofEpsilonIm} \\
        $\mathrm{Pr}(\bm{A}, \bm{\mathit{\Phi}}_{l}, \bm{\mathit{\Phi}}_{m}, \bm{\varepsilon} \;|\; \bm{V}^\mathrm{obs}, \bm{\sSigma}_{lm, \delta T})$ & Posterior probability distributions for $\bm{A}$, $\bm{\mathit{\Phi}}_{l}$, $\bm{\mathit{\Phi}}_{m}$ and $\bm{\varepsilon}$ given the raw data and the expected covariance matrix of $\bm{\varepsilon}$ &  & \autoref{Eq:ProbCalSkyGivenPSepsilon} \\
        $\sigma_{\varepsilon, i}^{2}$ & Variance corresponding to our uncertainty on the $i$th pixel of the a priori known component of our image domain calibration model &  & \autoref{Eq:PowerPrior} \\
        $\sigma_{\varepsilon}^{2}$ & Variance corresponding to the expected RMS uncertainty on the a priori known component of our image domain calibration model &  & Above \autoref{Eq:ProbCalSkySigma2epsilon} \\
        $\mathrm{Pr}(\bm{A}, \bm{\mathit{\Phi}}_{l}, \bm{\mathit{\Phi}}_{m}, \bm{\varepsilon} \;|\; \bm{V}^\mathrm{obs}, \sigma_{\varepsilon}^{2})$ & Posterior probability distributions for $\bm{A}$, $\bm{\mathit{\Phi}}_{l}$, $\bm{\mathit{\Phi}}_{m}$ and $\bm{\varepsilon}$ given the raw data and the expected RMS uncertainty on the a priori known component of our image domain calibration model &  & \autoref{Eq:ProbCalSkySigma2epsilon} \\
        $\mathrm{Pr}(\bm{\varepsilon} \;|\; \sigma_{\varepsilon}^{2})$ & Prior on the image domain calibration model parameters given the expected RMS uncertainty on the a priori known component of our image domain calibration model &  & \autoref{Eq:ProbCalSkySigma2epsilon} \\
        $\mathrm{Pr}(\bm{A}, \bm{\mathit{\Phi}}_{l}, \bm{\mathit{\Phi}}_{m}, \bm{\varepsilon}, \sigma_{\varepsilon}^{2} \;|\; \bm{V}^\mathrm{obs})$ & Posterior probability distributions for $\bm{A}$, $\bm{\mathit{\Phi}}_{l}$, $\bm{\mathit{\Phi}}_{m}$, $\bm{\varepsilon}$ and $\sigma_{\varepsilon}^{2}$ given the raw data &  & \autoref{Eq:ProbCalSkyGivenSigma2epsilon} \\
        $\mathrm{Pr}(\sigma_{\varepsilon}^{2})$ & Prior on $\sigma_{\varepsilon}^{2}$ &  & \autoref{Eq:ProbCalSkyGivenSigma2epsilon} \\
        $\eta$ & Fourier conjugate to frequency, $\nu$ &  & \autoref{Eq:GainAmplitudePS} \\
        $P_{A}(\eta)$ & One-dimensional power spectrum of amplitude fluctuations of the redundant gain degeneracy function &  & \autoref{Eq:GainAmplitudePS} \\
        $\mathbfss{F}$ & One-dimensional Fourier decomposition matrix\textsuperscript{\ref{TableNote1}} & $N_{\nu} \times N_{\nu}$ & \autoref{Eq:DegeneracyFunctionFourierParameters} \\
        $\bm{A}_\mathrm{F}$, $\bm{\mathit{\Phi}}_{l,\mathrm{F}}$ and $\bm{\mathit{\Phi}}_{m,\mathrm{F}}$ & Vectors of Fourier degenerate gain amplitude, tip- and tilt-phase parameters\textsuperscript{\ref{TableNote1}}, respectively & $N_{\nu} \times 1$ & \autoref{Eq:DegeneracyFunctionFourierParameters} \\
        $\bm{\sigma}_{A_\mathrm{F}}^{2}$ & Vector encoding the power spectrum of the degenerate gain amplitudes\textsuperscript{\ref{TableNote1}} & $N_{\nu} \times 1$ & \autoref{Eq:ProbAgivenSigmaA} \\
        $\bm{\sSigma}_{A_\mathrm{F}}$ & Covariance matrix of the degenerate gain amplitudes\textsuperscript{\ref{TableNote1}} & $N_{\nu} \times N_{\nu}$ & \autoref{Eq:ProbAgivenSigmaA} \\
        $\mathrm{Pr}(\bm{A}_\mathrm{F} \;|\; \bm{\sigma}_{A_\mathrm{F}}^{2})$ & Gaussian prior encoding the power in amplitude fluctuations of the redundant gain degeneracy function &  & \autoref{Eq:ProbAgivenSigmaA} \\
        $\mathrm{Pr}(\bm{A}_\mathrm{F}, \bm{\mathit{\Phi}}_{l,\mathrm{F}}, \bm{\mathit{\Phi}}_{m,\mathrm{F}}, \bm{\varepsilon} \;|\; \bm{V}^\mathrm{obs}, \sigma_{\varepsilon}^{2}, \bm{\sigma}_{A_\mathrm{F}}^{2})$ & Posterior probability distributions for $\bm{A}$, $\bm{\mathit{\Phi}}_{l,\mathrm{F}}$, $\bm{\mathit{\Phi}}_{m,\mathrm{F}}$ and $\bm{\varepsilon}$ given the raw data, expected uncertainty on the a priori known component of our image domain calibration model and the one-dimensional power spectrum of the degenerate gain amplitudes &  & \autoref{Eq:ProbCalSkyGivenPSVNA} \\
        $\mathrm{Pr}(\bm{\mathit{\Phi}}_{l,\mathrm{F}})$ and $\mathrm{Pr}(\bm{\mathit{\Phi}}_{m,\mathrm{F}})$ & Prior probability distributions of the Fourier degenerate tip and tilt phase parameters, respectively &  & \autoref{Eq:ProbCalSkyGivenPSVNA} \\
        $\mathrm{Pr}(\bm{\sigma}_{A_\mathrm{F}}^{2})$ & Prior probability distribution of the power in amplitude fluctuations of the redundant gain degeneracy function &  & \autoref{Eq:ProbCalSkyVNAGivenPS} \\
        $\delta\bm{V}^\mathrm{obs}$ & Residual\textsuperscript{\ref{TableNote2}} visibility vector & $N_{\mathrm{vis}} \times 1$ & \autoref{Eq:LogP} \\
        $\overline{\delta \bm{V}}^\mathrm{obs}$ & Weighted gridded projection of the residual\textsuperscript{\ref{TableNote2}} visibility vector onto the parameter space of the fitted visibility model & $N_{\mathrm{pix, s}} \times 1$ & \autoref{Eq:LogP} \\
        $\bm{\upLambda}$ & System matrix mapping from the image domain diffuse emission parameters to the fitted model visibilities & $N_{\mathrm{vis}} \times N_{\mathrm{pix, s}}$ & \autoref{Eq:LogP} \\
        $\bm{\upUpsilon}$ & Covariance matrix of $\overline{\delta \bm{V}}^\mathrm{obs}$ regularised by $\mathrm{Pr}(\bm{\varepsilon} \;|\; \bm{\sSigma}_{lm, \delta T})$ & $N_{\mathrm{pix, s}} \times N_{\mathrm{pix, s}}$ & \autoref{Eq:LogP} \\
        $\hat{\bm{\varepsilon}}$ & Maximum likelihood vector of image-space coefficients & $N_{\mathrm{pix, s}} \times 1$ & \autoref{Eq:epsilonmax} \\
        $\mathrm{Pr}(\bm{A}_\mathrm{F}, \bm{\mathit{\Phi}}_{l,\mathrm{F}}, \bm{\mathit{\Phi}}_{m,\mathrm{F}} \;|\; \bm{V}^\mathrm{obs}, \sigma_{\varepsilon}^{2}, \bm{\sigma}_{A_\mathrm{F}}^{2})$ & Marginal posterior probability distribution for the Fourier degenerate gain amplitude and tip-tilt phase parameters given the raw data, expected uncertainty on the a priori known component of our image domain calibration model and the one-dimensional power spectrum of the degenerate gain amplitudes &  & \autoref{Eq:Margin} \\
        \bottomrule
    \end{tabular}
    \contcaption{}
    \label{Tab:VariablesList4}
\end{table}

\clearpage

\begin{table}
    \centering
    \begin{tabular}{p{5cm} p{8cm} p{1.6cm} c}
    \toprule
         Quantity &  Description & Size & First Appearance\\
    \midrule
        $N_{\delta S}$ & Number of catalogued point sources with uncertain flux-densities contributing to $\bm{V}^\mathrm{sim}$ &  & \autoref{Sec:AnalyticMarginalisationOverPointSourceUncertainties} \\
        $\bm{\varepsilon}_{\delta S}$ & Vector of sky-model parameters associated with catalogued point sources contributing to $\bm{V}^\mathrm{sim}$ with uncertain flux-densities  & $N_{\delta S} \times 1$ & \autoref{Sec:AnalyticMarginalisationOverPointSourceUncertainties} \\
        $\bm{\varepsilon}_{\delta T, \delta S}$ & Vector of sky-model parameters used to model total residual\textsuperscript{\ref{TableNote2}} emission from sky-model incompleteness and uncertainties associated with both diffuse emission and with catalogued flux-densities of point sources included in $\bm{V}^\mathrm{sim}$ & $(N_{\mathrm{pix, s}}+N_{\delta S}) \times 1$ & \autoref{Sec:AnalyticMarginalisationOverPointSourceUncertainties} \\
        $\bm{P}_{uv, \delta S}$ & Vector encoding the two dimensional spatial power spectrum of residual\textsuperscript{\ref{TableNote2}} emission associated with uncertainties in the catalogued flux-densities of point sources included in $\bm{V}^\mathrm{sim}$ & $N_\mathrm{vis} \times 1$ & Above \autoref{Eq:ProbObsPSofEpsilonPlusSources} \\
        $\bm{P}_{uv,\delta T, \delta S}$ & Vector encoding the two dimensional spatial power spectrum associated with the total residual\textsuperscript{\ref{TableNote2}} emission from sky-model incompleteness and uncertainties associated with both diffuse emission and with catalogued flux-densities of point sources included in $\bm{V}^\mathrm{sim}$ & $N_\mathrm{vis} \times 1$ & Above \autoref{Eq:ProbObsPSofEpsilonPlusSources} \\
        $\mathrm{Pr}(\bm{\varepsilon}_{\delta T, \delta S} \;|\; \bm{P}_{uv,\delta T, \delta S})$ & Prior on the image domain and point source uncertainty parameters given their expected two dimensional spatial power spectrum &  & \autoref{Eq:ProbObsPSofEpsilonPlusSources} \\
        $\bm{\sSigma}_{uv, \delta T, \delta S}$ & Covariance matrix encoding the $\bm{P}_{uv,\delta T, \delta S}$ estimate of the power spectrum of $\bm{V}^\mathrm{fit}$ & $N_\mathrm{vis} \times N_\mathrm{vis}$ & \autoref{Eq:ProbObsPSofEpsilonPlusSources} \\
        $\bm{T}$ & Transformation matrix mapping from the image space coefficients of our fitted diffuse calibration model to the model true visibilities & $N_\mathrm{vis} \times N_\mathrm{pix, s}$ & \autoref{Eq:ProbObsPSofEpsilonPlusSources} \\
        $\bm{T}_{\delta S}$ & Transformation matrix mapping from the coefficients associated with the catalogued point sources with uncertain flux-densities to the model true visibilities & $N_\mathrm{vis} \times N_{\delta S}$ & \autoref{Eq:ProbObsPSofEpsilonPlusSources} \\
        $\bm{T}_{\delta T, \delta S}$ & Transformation matrix mapping from the image domain diffuse + uncertain catalogued point source parameters to the model true visibilities & $N_\mathrm{vis} \times N_{\delta S}$ & \autoref{Eq:ProbObsPSofEpsilonPlusSources} \\
        $\mathbfss{F}_\mathrm{fr, \delta S}^{-1}$ & Interferometric fringe matrix for catalogued point sources with uncertain flux-densities contributing to $\bm{V}^\mathrm{sim}$ & $N_{\mathrm{vis}} \times N_{\mathrm{im}}$ & Below \autoref{Eq:ProbObsPSofEpsilonPlusSources} \\
        $\mathbfss{P}_{\delta S}$ & Primary beam matrix encoding the effective primary beam coupling point source flux densities to north-south voltage correlations & $N_{\mathrm{im}} \times N_{\mathrm{im}}$ & Below \autoref{Eq:ProbObsPSofEpsilonPlusSources} \\
        $\mathbfss{S}_{\delta S}$ & Point source spectral model matrix & $N_{\mathrm{im}} \times N_{\delta S}$ & Below \autoref{Eq:ProbObsPSofEpsilonPlusSources} \\
        $\overline{\delta \bm{V}}_{\delta T, \delta S}^\mathrm{obs}$ & Weighted gridded projection of the residual\textsuperscript{\ref{TableNote2}} visibility vector onto the diffuse + uncertain catalogued point source parameters space of the augmented fitted visibility model & $(N_{\mathrm{pix, s}}+N_{\delta S}) \times 1$ & Above \autoref{Eq:MarginPlusSources} \\
        $\bm{\upUpsilon}_{uv, \delta T, \delta S}$ & Covariance matrix of $\overline{\delta \bm{V}}_{\delta T, \delta S}^\mathrm{obs}$ regularised by $\mathrm{Pr}(\bm{\varepsilon}_{\delta T, \delta S} \;|\; \bm{P}_{uv,\delta T, \delta S})$ & $(N_{\mathrm{pix, s}}+N_{\delta S}) \times (N_{\mathrm{pix, s}}+N_{\delta S})$ & Above \autoref{Eq:MarginPlusSources} \\
        $\bm{\upLambda}_{\delta T, \delta S}$ & System matrix mapping from the image domain diffuse + uncertain catalogued point source parameters to the fitted model visibilities & $N_\mathrm{vis} \times (N_{\mathrm{pix, s}}+N_{\delta S})$ & Above \autoref{Eq:MarginPlusSources} \\
        $\mathrm{Pr}(\bm{A}_\mathrm{F}, \bm{\mathit{\Phi}}_{l,\mathrm{F}}, \bm{\mathit{\Phi}}_{m,\mathrm{F}} \;|\; \bm{V}^\mathrm{obs}, \bm{P}_{uv, \delta T, \delta S}, \bm{\sigma}_{A_\mathrm{F}}^{2})$ & Marginal posterior probability distribution for the Fourier degenerate gain amplitude and tip-tilt phase parameters given the raw data, two dimensional spatial power spectrum associated with the expected total residual\textsuperscript{\ref{TableNote2}} emission from sky-model incompleteness and uncertainties associated with both diffuse emission and with catalogued flux-densities of point sources included in $\bm{V}^\mathrm{sim}$, and the one-dimensional power spectrum of the degenerate gain amplitudes &  & \autoref{Eq:MarginPlusSources} \\
        $\mathrm{Pr}(\bm{\varepsilon}_{\delta T, \delta S} \;|\; \bm{\sigma}_{lm,\delta T, \delta S})$ & Prior on the image domain diffuse and point source uncertainty parameters given their expected uncertainties &  & \autoref{Eq:ProbObsPSofEpsilonPlusSourcesImDomain} \\
        $\bm{\sSigma}_{lm, \delta T, \delta S}$ & Covariance matrix encoding the estimated uncertainties on the a priori known component of the diffuse and point source emission & $(N_{\mathrm{pix, s}}+N_{\delta S}) \times (N_{\mathrm{pix, s}}+N_{\delta S})$ & \autoref{Eq:ProbObsPSofEpsilonPlusSourcesImDomain} \\
        $\bm{\upUpsilon}_{lm, \delta T, \delta S}$ & Covariance matrix of $\overline{\delta \bm{V}}_{\delta T, \delta S}^\mathrm{obs}$ regularised by $\mathrm{Pr}(\bm{\varepsilon}_{\delta T, \delta S} \;|\; \bm{\sigma}_{lm,\delta T, \delta S})$ & $(N_{\mathrm{pix, s}}+N_{\delta S}) \times (N_{\mathrm{pix, s}}+N_{\delta S})$ & Below \autoref{Eq:ProbObsPSofEpsilonPlusSourcesImDomain} \\
        $\mathrm{Pr}(\bm{A}_\mathrm{F}, \bm{\mathit{\Phi}}_{l,\mathrm{F}}, \bm{\mathit{\Phi}}_{m,\mathrm{F}} \;|\; \bm{V}^\mathrm{obs}, \bm{\sigma}_{lm,\delta T, \delta S}^{2}, \bm{\sigma}_{A_\mathrm{F}}^{2})$ & Marginal posterior probability distribution for the Fourier degenerate gain amplitude and tip-tilt phase parameters given the raw data, expected uncertainty on the a priori known component of our image domain calibration model including catalogued point sources with uncertain flux-densities contributing to $\bm{V}^\mathrm{sim}$, and the one-dimensional power spectrum of the degenerate gain amplitudes &  & \autoref{Eq:MarginPlusSourcesImDomain} \\
        \bottomrule
    \end{tabular}
    \contcaption{}
    \label{Tab:VariablesList5}
\end{table}

\label{lastpage}

\end{document}